\RequirePackage{ifpdf}
\documentclass[hyper,12pt,letterpaper]{JHEP3}

\usepackage{comment}
\usepackage{graphicx}
\usepackage{latexsym,amsmath,amsfonts,amssymb}
\usepackage{mathrsfs}
\usepackage[makeroom]{cancel}
\usepackage{bbm}
\usepackage{bm}
\usepackage{subfigure}
\usepackage{paralist}
\usepackage{url}

\newcommand{\vol}{\mathrm{vol}}

\newcommand{\ie}{\textit{i.e.}}

\numberwithin{equation}{section}

\newcommand{\nn}{\nonumber}
\newcommand{\mat}[1]{\begin{pmatrix} #1 \end{pmatrix}}

\newcommand{\be}{\begin{equation}} 
\newcommand{\ee}{\end{equation}}
\newcommand{\bea}{\begin{equation} \begin{aligned}} \newcommand{\eea}{\end{aligned} \end{equation}}

\newcommand{\bit}{\begin{itemize}} 
\newcommand{\eit}{\end{itemize}}

\newcommand{\cA}{\mathcal{A}}

\newcommand{\cC}{\mathcal{C}}
\newcommand{\cD}{\mathcal{D}}

\newcommand{\cF}{\mathcal{F}}

\newcommand{\cH}{\mathcal{H}}

\newcommand{\cM}{\mathcal{M}}
\newcommand{\cN}{\mathcal{N}}

\newcommand{\cP}{\mathcal{P}}

\newcommand{\cS}{\mathcal{S}}

\newcommand{\cW}{\mathcal{W}}

\newcommand{\cZ}{\mathcal{Z}}

\newcommand{\Z}{\mathbb{Z}}
\newcommand{\C}{\mathbb{C}}
\newcommand{\R}{\mathbb{R}}

\renewcommand{\t}{\widetilde }
\renewcommand{\d}{\partial }

\renewcommand{\b}{\bar }

\newcommand{\half}{{1\over 2}}

\newcommand{\bz}{{\b z}}
\newcommand{\bw}{{\b w}}

\newcommand{\CA}{\mathcal{A}}

\newcommand{\CC}{\mathcal{C}}

\newcommand{\CE}{\mathcal{E}}
\newcommand{\CF}{\mathcal{F}}
\newcommand{\CG}{\mathcal{G}}
\newcommand{\CH}{\mathcal{H}}
\newcommand{\CI}{\mathcal{I}}
\newcommand{\CJ}{\mathcal{J}}
\newcommand{\CK}{\mathcal{K}}

\newcommand{\CM}{\mathcal{M}}
\newcommand{\CN}{\mathcal{N}}
\newcommand{\CO}{\mathcal{O}}

\newcommand{\CQ}{\mathcal{Q}}
\newcommand{\CR}{\mathcal{R}}
\newcommand{\CS}{\mathcal{S}}
\newcommand{\CT}{\mathcal{T}}

\newcommand{\CW}{\mathcal{W}}

\newcommand{\CZ}{\mathcal{Z}}

\newcommand{\FR}{\mathfrak{R}}
\newcommand{\Fg}{\mathfrak{g}}
\newcommand{\Fh}{\mathfrak{h}}

\newcommand{\GG}{\mathbf{G}}

\newcommand{\GH}{\mathbf{H}}

\newcommand{\rk}{{{\rm rk}(\Fg)}}

\newcommand{\m}{\mathfrak{m}}
\newcommand{\n}{\mathfrak{n}}

\newcommand{\ba}{{\b a}}
\newcommand{\bb}{{\bb b}}

\newcommand{\h}{\hat}

\newcommand{\oneloop}{\text{1-loop}}

\newcommand{\Mgp}{\CM_{g,p}}
\newcommand{\pif}{{\Pi}}

\DeclareMathOperator{\Tr}{Tr}

\newcommand{\ov}{\over}

\newcommand{\dilog}{{\text{Li}_2}}

\title{$\CN=1$ supersymmetric indices \\ and the four-dimensional A-model 
}

\author{Cyril~Closset,$^\flat$ Heeyeon~Kim$^\sharp$ and Brian Willett$^{\natural}$\\

{}$^{\flat}$Theory Department, CERN\\
CH-1211, Geneva 23, Switzerland\\
{}$^{\sharp}$ Perimeter Institute for Theoretical Physics\\
31 Caroline Street North, Waterloo, N2L 2Y5, Ontario, Canada\\
{}$^{\natural}$ Kavli Institute for Theoretical Physics\\
 University of California, Santa Barbara, CA 93106
}

\preprint{CERN-TH-2017-180}

\abstract{We compute the supersymmetric partition function of $\CN=1$ supersymmetric gauge theories with an $R$-symmetry  on $\mathcal{M}_4 \cong \mathcal{M}_{g,p}\times S^1$, a principal elliptic fiber bundle of degree $p$ over a genus-$g$ Riemann surface, $\Sigma_g$. Equivalently, we compute the generalized supersymmetric index $I_{\mathcal{M}_{g,p}}$, with the supersymmetric three-manifold ${\mathcal{M}_{g,p}}$ as the spatial slice. The ordinary $\CN=1$ supersymmetric index on the round three-sphere is recovered as a special case. We approach this computation from the point of view of a topological $A$-model for the abelianized gauge fields on the base $\Sigma_g$. This $A$-model---or $A$-twisted two-dimensional $\CN=(2,2)$ gauge theory---encodes all the information about the generalized indices, which are viewed as expectations values of some canonically-defined surface defects wrapped on $T^2$ inside $\Sigma_g \times T^2$. Being defined by compactification on the torus, the $A$-model also enjoys natural modular properties, governed by the four-dimensional 't Hooft anomalies. As an application of our results, we provide new tests of Seiberg duality. We also present a new evaluation formula for the three-sphere index as a sum over two-dimensional vacua.
}

\begin{document}

\tableofcontents

\section{Introduction}
The $\CN=1$ supersymmetric index \cite{Romelsberger:2005eg, Romelsberger:2007ec} provides us with an invaluable tool in the study of four-dimensional supersymmetric quantum field theories with an exact $R$-symmetry,  $U(1)_R$. It is defined as the Witten index  \cite{Witten:1981nf} of the theory quantized on $S^3$, in presence of various fugacities:
\be\label{index pq}
I_{S^3}({\bf p}, {\bf q}, y)=  \Tr_{S^3}\left[ (-1)^F {\bf p}^{J_3+J_3' +\half R} {\bf q}^{J_3-J_3' +\half R}  \prod_\alpha y_\alpha^{Q^\alpha}\right]~, 
\ee 
with $J_3, J_3'$  the generators of the Cartan of  $SO(4)\cong SU(2) \times SU(2)'$ rotations on $S^3$,  $R$ the conserved $U(1)_R$ charge, and $Q_\alpha$ any conserved charges that commute with supersymmetry.~\footnote{Unless otherwise stated, $\alpha$ denotes an index for the flavor symmetry group. This should not be confused with the spinor index for the supercharges $\CQ_\alpha, \t \CQ_{\dot \alpha}$.}
If the supersymmetric theory is also conformal, or flows to a non-trivial conformal fixed point in the infrared, the index \eqref{index pq}  computes the superconformal index \cite{Kinney:2005ej}---by the state-operator correspondence, it counts certain operators in short representations of the $\CN=1$ superconformal algebra.
More generally, the three-sphere index of any $R$-symmetric $\CN=1$ theory is obtained by quantizing the theory on $S^3\times \R$ with supergravity background fields turned on to preserve at least two supercharges \cite{Festuccia:2011ws, Dumitrescu:2012ha}.

A natural generalization of \eqref{index pq} considers any spatial manifold $\CM_3$ allowed by supersymmetry. The corresponding ``generalized index'' takes the form:
\be\label{def gen index}
I_{\CM_3}(q, y) =\Tr_{\CM_3}\left[ (-1)^F q^{2J+ R}  \prod_\alpha y_\alpha^{Q^\alpha}\right]~.
\ee
Such an index exists if and only if $\CM_3$ is a Seifert manifold \cite{Dumitrescu:2012ha, Klare:2012gn, Closset:2012ru}---that is, $\CM_3$ is an $S^1$ bundle over a two-dimensional orbifold $\h \Sigma_g$. The conserved charge $J$ in \eqref{def gen index} is the generator of the $S^1$ isometry along the Seifert fiber. Whenever the base $\h \Sigma_g$ of the Seifert manifold admits an isometry, we might also introduce an additional fugacity for it, which  corresponds to a certain ``squashing'' of the index \eqref{def gen index}. For instance, the $S^3$ index \eqref{index pq} has two ``geometric'' fugacities ${\bf p}$ and ${\bf q}$, which correspond to a certain squashing of the three-sphere---see {\it e.g.} \cite{Aharony:2013dha, Closset:2013vra}. In this work, we will only consider the ``round''---non-squashed---index \eqref{def gen index}. In particular, we will study the $S^3$ index with the specialization ${\bf p}={\bf q}=q$.

On general grounds, the $\CM_3$ index \eqref{def gen index} with $|q|= e^{- 2\pi \beta}$ can be computed as a supersymmetric path integral on $\CM_3 \times S^1$, with $S^1$ a circle of radius $\beta$  \cite{Cecotti:1981fu}.
The path integral and Hamiltonian computations must agree up to scheme-dependent local terms. However, it is sometimes convenient to factor out the contribution of the $\CM_3$ vacuum to \eqref{def gen index}, and to define a ``normalized index'' $\CI_{\CM_3}$ that doesn't include that vacuum contribution.
We define:
\be\label{ZM3S1 gen intro}
  \CZ_{\CM_3\times S^1}  =  I_{\CM_3} = q^{ E_{\CM_3}} \, \CI_{\CM_3}~,
\ee
with $\CZ_{\CM_3\times S^1}$ the supersymmetric partition function. The vacuum-contribution $E_{\CM_3}$ is the so-called supersymmetric Casimir energy \cite{Assel:2014paa, Lorenzen:2014pna, Assel:2015nca, Bobev:2015kza}. With this definition, the normalized index $\CI_{\CM_3}$ has an expansion in $q$:
\be\label{expansion index intro}
\CI_{\CM_3}(y; q) = \CI_{\CM_3}^{(0)}(y) + O(q)~,
\ee
with the  $O(q^0)$ vacuum contribution given by the first term. For the standard $S^3$ index, we have $\CI^{(0)}_{S^3}= 1$---in the superconformal case, it is simply the contribution from the unit operator.
For the generalized $\CM_3$ index, the first term $ \CI_{\CM_3}^{(0)}(y)$ in \eqref{expansion index intro} can itself be interpreted as the flavored Witten index \cite{Hori:2014tda} of a one-dimensional theory obtained in the $q \rightarrow 0$ limit, corresponding to sending the size of $\CM_3$ to zero.

 The explicit computation of \eqref{ZM3S1 gen intro} for any Seifert manifold $\CM_3$ remains an open challenge.  In addition to $\CM_3\cong S^3$ \cite{Romelsberger:2007ec, Assel:2014paa}, the case $\CM_3 \cong \Sigma_g \times S^1$, with $\Sigma_g$ a Riemann surface, has been computed in   \cite{Benini:2015noa, Honda:2015yha, Benini:2016hjo, Closset:2016arn}.
 In particular, for $\Sigma_0\cong S^2$, the  $S^2 \times T^2$ partition function has an interesting interpretation as a direct sum of elliptic genera---2d indices \cite{Gadde:2013ftv, Benini:2013nda, Benini:2013xpa}---for  two-dimensional $\CN=(0,2)$ supersymmetric theories obtained from the four-dimensional $\CN=1$ theory compactified on $S^2$ with a topological twist \cite{Closset:2013sxa, Honda:2015yha, Gadde:2015wta}.

In this work, we compute the generalized index \eqref{def gen index} in the case:
\be
\CM_3 \cong \Mgp~,
\ee
where $\Mgp$ is a principal $U(1)$ bundle of first Chern class $p\in \Z$ over the genus-$g$ (closed, orientable) Riemann surface $\Sigma_g$.
\be
S^1 \longrightarrow  \Mgp  \stackrel{\pi}\longrightarrow \Sigma_g~.
\ee
These generalized indices were first studied in \cite{Nishioka:2014zpa} using supersymmetric localization and we will expand on those results, albeit using different techniques.
The $\Mgp$ family includes the two important examples  mentioned above:
\be
\CM_{0,1}\cong S^3~, \qquad \qquad \CM_{g, 0}\cong \Sigma_g \times S^1~.
\ee
We will derive an explicit formula for the $\Mgp\times S^1$ supersymmetric partition function, valid for any asymptotically-free gauge theory with a semi-simple, simply-connected gauge group $\GG$. 
This can be done rather elegantly by studying a ``four-dimensional $A$-model'' on $\Sigma_g \times T^2$, following the recent approach of  \cite{Closset:2017zgf}. Note that, while the $\Mgp$ manifolds form a small subfamily in the set of all Seifert manifolds, we expect  that, using similar methods, one may also consider most allowed ``half-BPS'' $\CM_3$ backgrounds.

The $A$-model approach relates all the $\Mgp\times S^1$ partition functions amongst themselves.
For instance, we find that the $S^3 \times S^1$ partition function  \cite{Romelsberger:2007ec, Assel:2014paa} can be related to the $S^2 \times T^2$ partition function \cite{Benini:2015noa, Honda:2015yha} by:
\be\label{S3 to S2 rel intro}
\CZ_{S^3\times S^1}= \langle \CF \rangle_{S^2 \times T^2}~.
\ee
Here the insertion $\CF$ in the $S^2 \times T^2$ path integral is a particular surface operator wrapped over $T^2$, which we  call the {\it fibering operator}. Its insertion at any point on $S^2$ induces a non-trivial fibration of $T^2$ over $S^2$, leading to the $S^3 \times S^1$ topology.

\subsection*{The four-dimensional $A$-model}
Let us consider the compactification of a four-dimensional $\CN=1$ theory on $T^2$. 
Let $(z, w)$ be complex coordinates on $\R^2 \times T^2$. In terms of angular coordinates $x_1$, $x_2$ of period $2\pi$ on $T^2$, we have $w= x_1+ \tau x_2$.  The parameter $\tau$ is the modular parameter of $T^2$.
Any four-dimensional field has a Kaluza-Klein (KK) expansion on the torus:
\be
\phi = \sum_{n, m \in \Z}  \varphi_{n, m}(z, \bz) \, e^{i (n x_1 + m x_2)}~.
\ee
We can  view the four-dimensional theory as a two-dimensional theory with $\CN=(2,2)$ supersymmetry, with an infinite number of fields due to the KK decomposition. The $\CN=(2,2)$ superalgebra allows for  two distinct sectors of  half-BPS local operators.
The {\it chiral} operators commute with the supercharges~\footnote{The four-dimensional supercharges $\CQ_\alpha$, $\t \CQ_{\dot \alpha}$ become $\CQ_\pm$, $\t \CQ_\pm$ on $\R^2$, with $\alpha= \dot \alpha = \pm$.}
 $\t \CQ_+$ and $\t \CQ_-$. These operators descend from ordinary 4d $\CN=1$ chiral operators. 
The {\it twisted chiral} operators commute with the supercharges $\CQ_-$ and $\t \CQ_+$. This condition breaks four-dimensional Lorentz invariance in $\R^4$. The two-dimensional twisted chiral operators  on $\R^2$ descend from half-BPS $\CN=1$ surface operators wrapped over $T^2$.
These half-BPS operators form a ring---their OPE is non-singular up to $\CQ$-exact terms. The structure of the twisted chiral ring---the ring of parallel half-BPS surface operators---can be usefully isolated by the topological $A$-twist \cite{Witten:1988xj}. This corresponds to a supersymmetric compactification of $\R^2$ to $\Sigma_g$, a genus-$g$ Riemann surface.~\footnote{See for instance \cite{Closset:2014pda}, whose conventions we mostly follow.}

In the case of an $\CN=1$ {\it gauge} theory with gauge group $\GG$, the most important degrees of freedom, upon compactification to two dimensions, are the Wilson lines on $T^2$. We define the complex fields:
\be
u_a = \frac{\tau}{2 \pi} \int_{S^1_1} A^a_\mu dx^\mu - \frac{1}{2 \pi} \int_{S^1_2} A^a_\mu dx^\mu ~, \qquad\qquad a=1, \cdots, {\rm rk}(\GG)~,
\ee
for the abelianized gauge field $A^a_\mu$ in the Cartan of $\GG$,
\be
\GH \equiv \prod_{a=1}^{\rk}U(1)_a \subset \GG~.
\ee
The fields $u_a$ are Coulomb branch coordinates in the $\CN=(2,2)$ theory. Their higher-dimensional origin manifests itself by the periodic identifications:
\be
u_a \;\sim\; u_a +1\;\sim\; u_a +\tau~,
\ee
due to large gauge transformations on $T^2$. We similarly define the parameters $\nu_\alpha$ for the flavor group $\GG_F$, with $\alpha= 1, \cdots, {\rm rk}(\GG_F)$, corresponding to Wilson lines for  $U(1)_\alpha$ background gauge fields. Importantly, the fields $u_a$ and the background fields $\nu_\alpha$ are the lowest components of 2d $\CN=(2,2)$ twisted chiral multiplets.

Consider an $\CN=1$ supersymmetric gauge theory with gauge group $\GG$ and chiral multiplets $\Phi_i$ charged under the gauge group. For simplicity, we will assume that $\GG$ is semi-simple and simply-connected.    We define the {\it four-dimensional $A$-model} of this gauge theory as the low-energy effective theory on the Coulomb branch in two dimensions, subjected to the topological $A$-twist.  In favorable circumstances, this effective theory has isolated vacua---in many examples, this will be the case for generic-enough flavor parameters $\nu_\alpha$.
The four-dimensional $A$-model is fully determined in terms of two potentials:
\be\label{W and Om}
\CW(u, \nu; \tau)~, \qquad \qquad \Omega(u, \nu; \tau)~,
\ee
locally holomorphic in all variables.
The {\it effective twisted superpotential}, $\CW$, governs the dynamics of the low energy effective theory on $\R^2 \times T^2$, and the {\it effective dilaton}, $\Omega$, governs the coupling to curved space \cite{Witten:1993xi, Nekrasov:2014xaa}. The $A$-model vacua correspond to the solutions of the  {\it Bethe equations}  \cite{Nekrasov:2009uh},
\be
\exp\Big(2 \pi i  \d_{u_a} \CW(u, \nu; \tau)\Big) =1~, \qquad a=1, \cdots, \rk~,
\ee
which are not left invariant by the Weyl group. These so-called {\it Bethe vacua}, the two-dimensional vacua of the theory compactified on $T^2$,   will play a central role in this work.

\subsection*{Supersymmetric partition function from the $A$-model}
One can build a number of ``canonical'' $A$-model operators from $\CW$ and $\Omega$ \cite{Nekrasov:2014xaa, Closset:2017zgf}:
\bea
&\pif_\alpha &=&\; \exp\Big(2 \pi i {\d_{\nu_\alpha}\CW}\Big)~, \qquad\cr
&\CH &=&\; \exp\Big(2 \pi i {\Omega}\Big)\, \det_{ab}\Big(\d_{u_a}\d_{u_b}\CW\Big)~,
\qquad\cr
 &\CF  &=&\; \exp\Big(2 \pi i {\d_{\tau}\CW}\Big)~.
\eea
The operator $\pif_\alpha$ is the {\it flavor flux operator}, which inserts one unit of $U(1)_\alpha$ flux for a flavor background gauge field at a point on $\Sigma_g$. The operator $\CH$ is the {\it handle-gluing operator}, whose insertion at a point is equivalent to changing the topology of $\Sigma_g$ to $\Sigma_{g+1}$.  Finally, the {\it fibering operator} $\CF$ introduces a non-trivial fibration of $T^2$ over $\Sigma_g$. (More precisely, as we will discuss, there are two distinct fibering operators $\CF_1$ and $\CF_2$, related by a modular transformation of $T^2$. Here we chose $\CF=\CF_1$.) All these operators are local operators in the $A$-model---equivalently, they are half-BPS surface operators in the four dimensional $\CN=1$ theory.
By construction, we obtain:
\be\label{rel Mgp to S2T2 intro}
\CZ_{\Mgp\times S^1}= \Big\langle  \CH^{g} \, \CF^p \, \prod_\alpha \pif_\alpha^{\n_\alpha} \Big\rangle_{S^2 \times T^2}~,
\ee
for the $\Mgp\times S^1$ partition function with background fluxes $\n_\alpha$, generalizing \eqref{S3 to S2 rel intro}---the insertion of $\CH^g$ on $S^2$ changes the topology of the base to $\Sigma_g$, the fibering operator insertion $\CF^p$ changes the first Chern class of the principal circle bundle from $0$ to $p$, and the flavor flux operators $\pif_\alpha^{\n_\alpha}$ introduce background fluxes $\n_\alpha$. These operators can be inserted anywhere on $S^2$ since the $A$-model is topological in two dimensions.

The supersymmetric partition function can be computed explicitly as a sum over Bethe vacua:
\be\label{Z BE intro}
\CZ_{\Mgp\times S^1}(\nu; \tau) = \sum_{\h u \in \CS_{\rm BE}}  \CF(\h u,\nu; \tau)^{p}\, \CH(\h u,\nu; \tau)^{g-1} \,\prod_{\alpha} \pif_\alpha(\h u,\nu; \tau)^{\n_\alpha}~.
\ee
Here $\CS_{\rm BE}$ denotes the set of all Bethe vacua. The parameters $\nu$ and $\tau$ are related to the fugacities $y$ and $q$ in the index \eqref{def gen index} by:
\be
y= e^{2\pi i \nu}~, \qquad\qquad q= e^{2 \pi i \tau}~.
\ee
One can pull out a $\h u$-independent supersymmetric Casimir energy term from \eqref{Z BE intro}, like in \eqref{ZM3S1 gen intro}. This supersymmetric Casimir energy is determined entirely by the various 't Hooft anomalies of the theory, and it is therefore scheme-independent. 

The supersymmetric partition function  \eqref{Z BE intro} enjoys natural transformations properties under large transformations $\nu\sim \nu+1\sim \nu+\tau$ for the flavor parameters, and under modular transformations of the $T^2$ fiber. While  $\CZ_{\Mgp\times S^1}$ is not fully invariant under large gauge transformations for flavor background gauge fields, this lack of invariance is naturally expressed in terms of  `t Hooft anomalies. Incidentally, the gauge theory itself should, of course, be non-anomalous---all gauge and gauge-flavor anomalies must vanish for the $A$-model to be well-defined.

We also note that the $A$-model formalism naturally allows the insertion of  more general supersymmetric surface defects supported along the $T^2$ fiber of $\Mgp \times S^1$.  Their expectation value is computed by modifying the sum over Bethe vacua according to:
\be\label{Z BE intro surface}
\langle \cS \rangle = \sum_{\h u \in \CS_{\rm BE}}  \CF(\h u,\nu; \tau)^{p}\, \CH(\h u,\nu; \tau)^{g-1} \,\prod_{\alpha} \pif_\alpha(\h u,\nu; \tau)^{\n_\alpha}~ \cS(\h u,\nu,\tilde{\nu} ).
\ee
where $\cS(u, \nu,\tilde{\nu})$ is the $T^2$ partition function, or elliptic genus, of an $\cN=(0,2)$ surface defect theory, which may couple to the $4d$ gauge fields, as well as to a $2d$ flavor symmetry group with fugacities $\tilde{\nu}$.  
We leave a more detailed study of surface defects in $4d$ $\cN=1$ theories for future work.

\subsection*{A new evaluation formula for the three-sphere index}
The special case of $S^3\times S^1$ is worth discussing in more detail. One important property of the $A$-model is that the $R$-charges of all fundamental fields should be integer-quantized. This is so that the fields are well-defined on $\Sigma_g$ with the topological $A$-twist. From the point of view of curved-space supersymmetry \cite{Festuccia:2011ws}, we should view $\CM_4 \cong \Mgp\times S^1$ as a complex manifold. The $U(1)_R$ background gauge field is then the connection on a complex line bundle $L^{(R)}$ over $\CM_4$ \cite{Dumitrescu:2012ha}.  We have:
\be
c_1(L^{(R)})= g-1 \quad {\rm mod} \; p~,
\ee
its $\Z_p$-valued first Chern class, which generally imposes a Dirac quantization condition on the $R$-charge.  We refer to Appendix \ref{app:  SUSY backg} for a detailed exposition of the $\Mgp \times S^1$ supersymmetric background.

In the special case $\CM_4 \cong S^3 \times S^1$, however, the $R$-symmetry gauge field is topologically trivial, so that the $R$-charges can be taken real, not only integer-valued. More precisely, this is true in a $U(1)_R$ gauge for which the $R$-symmetry gauge fields vanishes along the three-sphere.~\footnote{Strictly speaking, that is only true for the round metric on $S^3$. See Appendix  \ref{app:  SUSY backg}.} This so-called ``physical gauge'' is related to the ``A-twist gauge'' used in most of this work by a large $U(1)_R$ gauge transformation along the Hopf fiber inside the $S^3$.

In the physical gauge, the $S^3\times S^1$ partition function has a well-known expression as an elliptic hypergeometric integral \cite{Romelsberger:2007ec, Assel:2014paa}:
\be\label{S3S1 expl}
\CZ_{S^3\times S^1}^{\rm phys} =q^{E_{S^3}}\; { (q; q)_\infty^{2{\rm rk}(\GG)}\ov |W_\GG|}\, \oint_{|x|=1} \prod_{a=1}^{{\rm rk}(\GG)}{dx_a \ov 2 \pi i x_a} \;  {\prod_i\prod_{\rho_i}  \Gamma_0\big(x^{\rho_i} q^{r_i-1}; q\big)\ov \prod_{\alpha\in \Fg} \Gamma_0\big(x^\alpha q^{-1}; q\big)}~,
\ee
where the integrand is given in terms of elliptic gamma-functions,~\footnote{Here we defined  $\Gamma_0(x; q)= \Gamma_e(x q; q, q)$, with $\Gamma_e(x; {\bf p},{\bf  q})$  the standard elliptic gamma function.}
with the numerator and denominator corresponding to the chiral and vector multiplets, respectively. The factor $E_{S^3}$ in front of \eqref{S3S1 expl} is the supersymmetric Casimir energy \cite{Assel:2015nca, Bobev:2015kza}. Note that  we suppressed all dependence on the flavor fugacities  $y_\alpha$   in \eqref{S3S1 expl}, to avoid clutter. Our results lead to an explicit evaluation formula for  \eqref{S3S1 expl} as a sum over the Bethe vacua of the schematic form:
\be\label{ZS3S1 ev form intro}
\CZ_{S^3\times S^1}^{\rm phys} = \sum_{\h u \in \CS_{\rm BE}^{\rm phys}}  \CG^{\rm phys}(\h u)~,
\ee
completely analogously, and closely related, to \eqref{Z BE intro}.   We will explain the precise meaning of \eqref{ZS3S1 ev form intro}   in section \ref{sec: free R}.

\subsection*{A new test of Seiberg duality}
Arguably, the most striking application of the supersymmetric index is that it provides highly non-trivial tests of supersymmetric dualities, such as Seiberg duality \cite{Seiberg:1994pq}. This is possible because the index---or the supersymmetric partition function---is renormalization group (RG) invariant. This allows us to  easily compute it for any weakly-coupled theory in the ultraviolet (UV) in order to deduce properties of the strongly-coupled infrared (IR). 

In the case of the $S^3$ index, Seiberg duality manifests itself as rather formidable identities between elliptic hypergeometric integrals like \eqref{S3S1 expl}   for dual theories \cite{Dolan:2008qi}. Such identities were discovered by Rains from a purely mathematical perspective  \cite{2003math......9252R}. 
This beautiful result has led to some interesting lines of research linking indices to more formal mathematical constructions---see {\it e.g.} \cite{Dolan:2008qi,Spiridonov:2008zr,Spiridonov:2009za,Spiridonov:2011hf, Kels:2017toi}.

The $A$-model of a given $\CN=1$ gauge theory is a topological field theory, and it is of course RG invariant. Therefore, Seiberg duality---or any $\CN=1$ infrared duality---implies an isomorphism between the $A$-models of the dual theories.
In particular,  the duality implies the existence of a one-to-one duality map:
\be
 \cD\; :\;  \cS_{\rm BE} \rightarrow \cS_{\rm BE}^D  \; : \; \h u \mapsto \h u^D
\ee
between Bethe vacua in the dual gauge theories. 
The most elementary observable of the $A$-model is the number of Bethe vacua, which can be identified with the $T^4$ partition function, or {\it regularized} Witten index.~\footnote{That is the Witten index regularized by turning on flavor fugacities on $T^4$, which lifts the moduli space of vacua. See \cite{Intriligator:2013lca} for a physical discussion of a related index in three dimensions. The results quoted here are for {\it generic} values of the flavor fugacities.}   We will compute that index in a few examples.  For example, for the $\cN=1$ $USp(2N_c)$ gauge theory with $2N_f$ flavors, we find:
\be\label{SQCD Sp}
 \cZ_{T^4} = |\cS_{BE}| = \binom{N_f-2}{N_c} \;\;\;\;\;\quad \text{for} \; USp(2N_c) \; \text{with} \; 2N_f \; \text{flavors}~,
\ee
Similarly, the Witten index of $\CN=1$ SQCD with special unitary gauge group is given by:
\be\label{SQCD conj}
 \cZ_{T^4} = |\cS_{BE}| = \binom{N_f-2}{N_c-1} \;\;\;\;\;\quad \text{for} \; SU(N_c) \; \text{with} \; N_f \; \text{flavors}~,
\ee
where each flavor consists of a pair of fundamental and anti-fundamental chiral multiplet. The formulas \eqref{SQCD Sp} and \eqref{SQCD conj} are nicely consistent with Seiberg duality \cite{Seiberg:1994pq, Intriligator:1995ne}.

The supersymmetric partitions functions \eqref{rel Mgp to S2T2 intro} are more complicated examples of $A$-model observables that should match across the duality.  It directly follows from the Bethe-vacua formula \eqref{Z BE intro} that the partition functions of dual theories will match, for all $\Mgp\times S^1$, provided that the $A$-model operators match on dual vacua:
\be\label{dual rels intro}
\CF(\h u, \nu; \tau)=\CF_D(\h u^D, \nu; \tau)~, \qquad\qquad
\CH(\h u, \nu; \tau)=\CH_D(\h u^D, \nu; \tau)~.
\ee
Similar relations must hold for the flavor flux operators---or for any insertions of mutually dual surface operators.
In this work, we check the duality relations \eqref{dual rels intro} for Seiberg duality with $Sp(2N_c)$ and $SU(N_c)$ gauge groups \cite{Intriligator:1995ne, Seiberg:1994pq}.  This will provide a proof of the equality of the twisted indices $\CZ_{\Sigma_g \times T^2}$ to all order, for these dualities; for $p \neq 0$, the equality of partition functions hinges on the first equality in \eqref{dual rels intro}, which can be checked to hold numerically or in some convenient limits, but would be interesting to prove analytically. For $S^3 \times S^1$, the equalities \eqref{dual rels intro}  provide us with a different perspective on the equality of the $S^3$ supersymmetric index, which might be conceptually simpler than Rains' integral identities.~\footnote{On the other hand, recall that we are restricted to the case ${\bf q}= {\bf p}=q$.}

\subsection*{Comments and outlook}
Let us briefly comment on the relation of our results to previous work.  In the case of $g=0$ and general $p$, we compute an index on the lens space $L(p, p-1)\cong S^3/\Z_p$.  However, the background we consider here differs from that considered in \cite{Benini:2011nc} for $p >2$ due to the presence of a non-trivial $R$-symmetry gauge field.~\footnote{See \cite{Closset:2017zgf} for further details in the context of the $3d$ lens space partition function.}  We also expect that our results in the case $g=0$ can be related to holomorphic blocks \cite{Beem:2012mb,Yoshida:2014qwa, Peelaers:2014ima, Nieri:2015yia}, where one also finds that the partition function is expressed as a sum over the supersymmetric Bethe vacua---the relation to our approach likely involves the $\Omega$-deformation---or ``quantization''---of the $4d$ $A$-model.  In the case of $\cN=2$ theories, one may also consider the fully topologically twisted partition function \cite{Witten:1988ze}, which is naively a different object than the partial twist considered here.  More relatedly, partial topological twists of $\cN=2$ theories along a Riemann surface have been considered, {\it e.g.}, in \cite{Kapustin:2006hi,Kapustin:2006pk,Gukov:2017zao}.

Surface operator expectation values play an important role in our story, since they are the basic observables of the $4d$ $A$-model.  These have appeared in the context of $4d$ partition functions in \cite{Gadde:2013ftv,Gaiotto:2014ina,Gomis:2014eya}.  We hope to return to a more systematic study of these operators in future investigation, and in particular to relate the $A$-model formalism to these earlier works.  It would  also be very interesting to study the generalized index in the large $N$ limit, and in particular to relate our results to the extremization principle of \cite{Hosseini:2017mds}.

Finally, we should note that the identification of the $A$-model partition function \eqref{Z BE intro} with the physical path integral over the supersymmetric $\Mgp\times S^1$ background is a subtle matter in the presence of 't Hooft anomalies. This is likely related to recent claims of an $\CN=1$ supercurrent anomaly \cite{Papadimitriou:2017kzw,An:2017ihs}. We will come back to this issue in future work.

\vskip0.2cm

\noindent This paper is organized as follows. In Section \ref{sec: 4dAtwist}, we study the theory on $\R^2 \times T^2$ through the $A$-model perspective.  In Section \ref{sec:Mgp}, we introduce the $\Mgp \times S^1$ partition function, explain how it is computed in terms of $A$-model operators, and discuss some of its properties.  In Section \ref{sec: free R}, we describe in detail cases, such as $\CM_{0,1}=S^3$, where the R-symmetry gauge field is topologically trivial, and one can relax the constraint that the R-charges be integer.  In Section \ref{sec: integral form and BE}, we sketch the derivation of the $\Mgp \times S^1$ partition function from a localization calculation, giving an alternative integral representation, and we relate our result to previous computations of the $S^3 \times S^1$ partition function that have appeared in the literature.  In Section \ref{sec:dualities}, we describe the computation of the $\Mgp \times S^1$ partition function for some SQCD-type theories that enjoy Seiberg duality, and we give strong evidence for the equality of the partition functions for dual theories.  Several appendices are included with further technical details.

\section{$\CN=1$ gauge theories on $\R^2 \times T^2$}\label{sec: 4dAtwist}
Consider a four-dimensional $\CN=1$ supersymmetric gauge theory on  $\R^2 \times T^2$, with a modular parameter $\tau$ for the torus. The $T^2$ compactification allows us to describe the four-dimensional theory in terms of a two-dimensional $\CN=(2,2)$ supersymmetric theory with an infinite number of fields corresponding to the Kaluza-Klein modes on $T^2$. In particular, the zero-modes of the $\CN=1$ vector multiplet give rise to an $\CN=(2,2)$ vector multiplet in $\R^2$. 
We will consider four-dimensional supersymmetric backgrounds that preserves the two supercharges $\CQ_-$ and $\t \CQ_+$ in the two-dimensional $\CN=(2,2)$ supersymmetry algebra. 
We may then write down an $A$-twisted effective field theory for two-dimensional vector multiplets spanning the classical Coulomb branch.

Given a gauge group $\GG$ with ${\rm Lie}(\GG)= \Fg$, the two-dimensional vector multiplet contains a complex scalar $u$,  which is also the lowest component of a $\Fg$-valued twisted chiral multiplet $U$---satisfying $[Q_-, U]= [\t Q_+, U]=0$. 
Let us consider the complex coordinate $w=x_1+\tau x_2$ on $T^2$, with
\be\label{def tau}
\tau= \tau_1 + i \tau_2~, \qquad \qquad \tau_2= {\beta_2\ov \beta_1}~.
\ee
Here  $\beta_1$, $\beta_2$ are the radii of $T^2 \cong S^1_{\beta_1}\times S^1_{\beta_2}$.
Let us denote by $a_{x^1}, a_{x^2}$ the holonomies along the torus:
\be \label{def abw}
a_{x^1} \equiv {1\ov 2 \pi} \int_{S^1_{\beta_1}}A_\mu dx^\mu~, \qquad  \qquad
a_{x^2} \equiv  {1\ov 2 \pi}  \int_{S^1_{\beta_2}} A_\mu dx^\mu~,
\ee
for the four-dimensional gauge field $A_\mu$,  and similarly for background gauge fields  $A_\mu^{(F)}$ for the flavor symmetries, and define:
\be
u \equiv \tau a_{x^1}- a_{x^2}~, \qquad \qquad\qquad \nu_F \equiv  \tau a_{x^1}^{(F)} - a_{x^2}^{(F)}~.
\ee
These fields are  dimensionless complex scalars in two dimensions. 
We pick a basis $e^a$ of the Cartan $\GH$ of $\GG$, and a basis $e_F^\alpha$ of the Cartan of $\GG_F$, such that:
\be
u = u_a e^a~,\qquad \qquad\qquad \nu_F= m_\alpha e_F^\alpha~.
\ee
We choose a basis $\{e^a\}$ that generates the coweight lattice $\Lambda_{\rm cw}$, so that $\rho(e^a)\equiv \rho^a \in \Z$ for all weights $\rho \in \Lambda_{\rm w}$. We similarly choose $\{e^\alpha_F\}$ such that $\omega(e^\alpha_F) \equiv \omega^\alpha \in \Z$, where $\omega$ denote the flavor weights.
We have the identifications:
\be
u_a\; \sim\; u_a+1  \:\sim \; u_a + \tau~, \qquad \qquad
\nu_\alpha\; \sim\; \nu_\alpha+1  \:\sim \; \nu_\alpha + \tau~,
\ee
under $U(1)_a$ and $U(1)_\alpha$ large gauge transformations on $T^2$, respectively.

The four-dimensional $A$-model is fully determined by the following effective action for the twisted chiral multiplets $U_a$, which governs the coupling of the theory to the curved space $\Sigma_g \times T^2$:
\bea\label{S TFT}
&S_{\rm TFT}&=&\; \int_{\Sigma_g} d^2 x \sqrt{g} \left(-2  {f_{1 \b1}}_a {\d \CW(u, \nu; \tau)\ov \d u_a} + \t\Lambda^a_{\b1}\Lambda_1^b {\d^2 \CW(u, \nu; \tau)\ov \d u_a\d u_b}   \right)\cr
&&&\; + \int_{\Sigma_g} d^2 x \sqrt{g} \left(-2  {f_{1 \b1}}_\alpha {\d \CW(u, \nu; \tau)\ov \d \nu_\alpha}   \right)  + {i\ov 2}  \int_{\Sigma_g} d^2 x \sqrt{g} \,  \Omega(u, \nu; \tau)\, R~.
\eea
Here $R$ is the Ricci scalar of $\Sigma_g$. The gauge fluxes $f_{1\b1 a}$ are to be summed over, while $f_{1\b1 \alpha}$ denote background flavor fluxes, such that:
\be\label{def nalpha}
{1\ov 2 \pi}  \int_{\Sigma_g} d^2 x \sqrt{g} \left(-2 i  {f_{1 \b1}}_\alpha\right) = \n_\alpha \in \Z~.
\ee
After the twist, there are also one-form fermions  $\t\Lambda_{\b1}$, $\Lambda_1$ that couple as indicated in \eqref{S TFT}, and provide $g$ pairs of fermionic zero-modes on $\Sigma_g$. The holomorphic function  $\CW$ is the {\it effective twisted superpotential}, and $\Omega$ is the {\it effective dilation} \cite{Closset:2017zgf}.  Both $\CW$ and $\Omega$ are locally holomorphic in $u$ and in the various ``mass'' parameters $\nu$ and $\tau$. In the following, we discuss these functions in detail for any ultraviolet-free four-dimensional gauge theory with a semi-simple gauge group $\GG$.  For simplicity, we also restrict ourselves to $\GG$ a simply-connected gauge group. (There can be interesting global issues for $\GG$ non simply-connected, which we leave for future work.)

\subsection{The twisted superpotential}
The twisted superpotential of a four-dimensional gauge theory only receives contributions from charged chiral multiplets, which obtain two-dimensional effective twisted masses proportional to $u_a$. For $\GG$ semi-simple, the $W$-bosons and their superpartners do not contribute.

\subsubsection{Chiral multiplet contribution}
Consider a four-dimensional chiral multiplet of charge $1$ under some $U(1)$ in the Cartan of $\GG$, with $u$ the $U(1)$ complexified flat connection. Integrating out all the KK modes, we obtain the formal twisted superpotential:
\be\label{W phi formal nm}
\CW_\Phi =-{1\ov 2 \pi i} \sum_{n, m\in \Z} (u+ n + \tau m)\left(\log{ (u+ n + \tau m)}-1\right)~.
\ee
where each KK mode contributes to the effective twisted superpotential as a $2d$ chiral multiplet \cite{Witten:1993yc}.  We first perform the sum over $n$, using the three-dimensional regularization discussed in \cite{Closset:2017zgf}. This gives:
\be\label{3d sum W}
\CW_\Phi ={1\ov (2\pi i )^2}\sum_{m\in \Z} \dilog(x q^m)~,
\ee
where we defined $x= e^{2\pi i u}$ and $q= e^{2\pi i \tau}$.
This formal sum can be further regularized to:
\be\label{full Wphi}
\CW_\Phi(u; \tau) =-{u^3\ov 6 \tau} + {u^2\ov 4}-{u\tau\ov 12} +{1\ov 24}+{1\ov (2\pi i )^2}  \sum_{k=0}^\infty \Big( \dilog\left(x q^k\right)- \dilog\left(x^{-1} q^{k+1}\right)\Big)~. 
\ee
The infinite sum in \eqref{full Wphi} converges absolutely for $\tau$ in the upper-half plane. The cubic polynomial encodes certain four-dimensional anomalies, which we will discuss momentarily. The derivation of \eqref{full Wphi}  is discussed in Appendix \ref{app: subsec W}.  

An equivalent definition of \eqref{full Wphi} can be given in terms of the following function, which we might call the ``elliptic dilogarithm'':
\be\label{def el dilog}
{\bm \psi}(u; \tau)\equiv -{1\ov 2\pi i} \int_{0}^u du' \log\theta(u'; \tau)~.
\ee
Here we introduced the theta-function:
\be\label{def theta}
\theta(u; \tau) = e^{-\pi i u} q^{1\ov 12}\,  \prod_{k=0}^{\infty}(1- x  q^k)(1- x^{-1}q^{k+1})~,
\ee
whose properties are discussed in Appendix \ref{app: el fct identities}.
The twisted superpotential \eqref{full Wphi} is equal to:
\be\label{Wphi bis}
\CW_\Phi(u; \tau) =-{u^3\ov 6 \tau} +{\bm \psi}(u; \tau)~.
\ee
It is easy to see that \eqref{full Wphi} and \eqref{Wphi bis} have the same derivative with respect to $u$; this establishes the identity of the two expressions up to an integration constant, which can be checked numerically.
The expression  \eqref{Wphi bis} is useful in order to study the analytic structure of the twisted superpotential. From its definition, one can see that ${\bm \psi}(u; \tau)$ has branch points at $u = n + m\tau$, $\forall m, n\in \Z$, with jumps by $u-n - m\tau$. This leads to the following branch  branch cut ambiguities of \eqref{Wphi bis}:
\be
\CW_\Phi \sim \CW_\Phi  + n' u + m \tau + n~, \qquad \qquad n', m, n \in \Z~.
\ee

\paragraph{A pair of massive chiral multiplets.} 
We can easily verify the identity:
\be \label{massivechiral}
 \cW_\Phi(u) + \cW_\Phi(-u) =- \half u~.
 \ee
up to a choice of branch. The left-hand-side of  \eqref{massivechiral} corresponds to the contribution of a pair of massive chiral multiplets $\Phi_1$, $\Phi_2$ with opposite charges $\pm 1$ under the background $U(1)$ symmetry. Such a pair can be integrated out with the four-dimensional superpotential $W= \Phi_1 \Phi_2$. We naively expect for such massive chiral multiplets to  decouple entirely. Instead, we find the linear term $-\half u$ in \eqref{massivechiral}, which leads to subtle signs in the partition function, in the presence of background $U(1)$ fluxes. Note that such signs only appear in the presence of an abelian background gauge field. Massive chiral multiplets that only couple to non-abelian gauge fields  will  decouple entirely.

\subsubsection{W-boson contribution}
The W-bosons enter the effective two-dimensional theory like chiral multiplets of gauge charge $\alpha^a$ and $R$-charge $2$, with $\alpha$ the roots of $\Fg$. Due to the pair-wise cancellation between roots $\alpha$ and $-\alpha$, the effect of the W-bosons on the twisted superpotential is extremely mild. Using the identity \eqref{massivechiral}, we find:
\be\label{Wvec triviial}
\CW_{\rm vec} = -\rho_W(u)~, \qquad \qquad  \rho_W\equiv \half \sum_{\alpha >0} \alpha~,
\ee
with $\rho_W$ the Weyl vector---the sum is over the positive roots. For $\GG$ semi-simple, all components $\rho_W^a$ are integers. Since the twisted superpotential is only defined modulo shifts by $n^a u_a$, $n_a \in \Z$, we can therefore ignore $\CW_{\rm vec}$ in the following.

\subsubsection{General gauge theory}
Consider an $\CN=1$ gauge theory for $\GG$ semi-simple, with chiral multiplets $\Phi_i$ in representations $\FR_i$ of $\Fg$. We also turn on generic background parameters $\nu_\alpha$ for any flavor symmetry $\GG_F$, with $\nu_i = \omega_i(\nu)$ and $\omega_i$ the flavor weight as defined above. We simply obtain:
\be\label{full W i}
\CW(u, \nu; \tau) = \sum_i \sum_{\rho\in \FR_i} \CW_\Phi(\rho_i(u) + \nu_i; \tau)~,
\ee
with $\CW_\Phi$ given by \eqref{full Wphi}. The sum is over all weights $\rho_i$ of the representations $\FR_i$. 
Importantly, as explained above, the twisted superpotential  generally has  branch cuts in $u_a$ and $\nu_\alpha$, where it jumps by:
\be \label{branchcuts} 
\CW \sim \CW +n^a u_a + n^\alpha \nu_\alpha + n + m \tau~, \qquad \qquad n^a, n^\alpha, n, m \in \Z~.
\ee
Therefore, $\CW$ is only defined up to such shifts. The linear ambiguity in $u_a$ reflects the fact that $U_a$ is a constrained twisted chiral fields, with the twisted $F$-term given by the two-dimensional gauge flux \cite{Nekrasov:2009uh, Beem:2012mb}. While $\CW$ itself is a multi-valued function of $(u_a, \nu_\alpha)$, the $A$-model observables will be well-defined, meromorphic functions of these chemical potentials.

\subsection{Large gauge transformations, modular transformations, and anomalies}
Let us consider $\prod_{\bf a}U(1)_{\bf a}$, the maximal torus of $\GG\times \GG_F$, where the index ${\bf a}= (a, \alpha)$ run over both the gauge and flavor group.
Classically, large gauge transformations of the background or dynamical gauge fields along the two torus cycles are symmetries of the action, which leads to the identifications:
\be 
{\bf u_a} \sim {\bf u_a} + n_{\bf a} + m_{\bf a} \tau~, \qquad \qquad \forall n_{\bf a}, m_{\bf a} \in \Z~,
\ee
where ${\bf u_a}\equiv (u_a,\nu_\alpha)$.  The same can be said about modular transformation---large diffeomorphisms of the torus---, with the two generators $S$ and $T$ acting as:
\be
 S\;:\;\; {\bf u_a} \rightarrow \frac{{\bf u_a}}{\tau}~, \qquad  \tau \rightarrow -\frac{1}{\tau}~, \;\;\;\;\;\;\;\; \qquad\qquad
T\;:\;\; {\bf u_a} \rightarrow{\bf u_a}~, \qquad \tau \rightarrow \tau + 1~.
\ee
Quantum anomalies can spoil these symmetries, however. In any anomaly-free gauge theory, we will see that the identifications $u_a \sim u_a+1\sim u_a+\tau$ for the gauge variables holds exactly, as required by consistency since the gauge fields must be integrated over. On the other hand, the A-model observables typically transform non-trivially under $\GG_F$ large gauge transformations, and under modular transformations, in a way which is governed by 't Hooft anomalies.

\subsubsection{Anomalies}
For convenience, and to set our notation, let us briefly review the various anomalies that can affect our four-dimensional gauge theory.

\paragraph{Gauge and flavor anomalies.}
Let $I=(\rho_i)$ run over all the chiral multiplets (that is, over all weights $\rho_i$  for each $i$). We define the anomaly coefficients:
\be\label{def anomalies}
\CA^{{\bf abc}}= \sum_I Q_I^{\bf a} Q_I^{\bf b} Q_I^{\bf c}~, \qquad \quad
\CA^{{\bf ab}}= \sum_I Q_I^{\bf a} Q_I^{\bf b}~, \qquad \quad
\CA^{{\bf a}}= \sum_I Q_I^{\bf a}~,
\ee
where $Q_I^{\bf a}$ is the integer-valued $U(1)_{\bf a}$ charge of $\Phi_I$---that is, $Q_{\rho_i}^a= \rho_i^a$ and $Q_{\rho_i}^\alpha= \omega_i^\alpha$ in terms of the weights of the gauge and flavor representations. The anomaly coefficients $\CA^{{\bf abc}}$ and $\CA^{{\bf a}}$ correspond to the perturbative cubic and mixed gauge-gravitational anomalies, respectively. For the dynamical  gauge symmetry  $\GG$, we must have:
\be\label{af cond 1}
\CA^{abc}= \CA^{a}=0~, \qquad \quad \CA^{ab\gamma}= \CA^{a\beta\gamma}=0~.
\ee
The first condition ensures that $\GG$ is non-anomalous, and the second condition ensures that the flavor symmetry is an actual symmetry in the quantum theory. On the other hand, we generally have non-vanishing 't Hooft anomaly coefficients $\CA^{\alpha\beta\gamma}$ and $\CA^\alpha$ for the flavor symmetry group $\GG_F$.

The coefficients $\CA^{\bf ab}$, on the other hand, will correspond to {\it non-perturbative} anomalies ---also known as global anomalies---for semi-simple gauge groups \cite{Witten:1982fp, Elitzur:1984kr, Zhang:1987pw}. Given \eqref{af cond 1} and the absence of perturbative anomalies, the absence of global anomalies for $\GG$ requires:
\be\label{global anomaly free cond}
\CA^{ab} \in 4\Z~.
\ee
The simplest example is $\GG=SU(2)$ with $n_f$ doublets, which has $\CA^{(2)}= 2  n_f$---we need $n_f$ to be even in order to satisfy \eqref{global anomaly free cond}, which is the condition for the absence of the well-known  $SU(2)$ global anomaly  \cite{Witten:1982fp}.~\footnote{More generally, a chiral multiplet in the spin ${j\ov 2}$ representation of $SU(2)$ contributes ${1\ov 3} j(j+1)(j+2)$ to $\CA^{(2)}$. This reproduces the global anomaly in that case \cite{Witten:1982fp, Bar:2002sa}.} 
More generally,  for any simple, simply-connected  group $\GG_s \subset \GG\times \GG_F$, the coefficients $\CA^{\bf ab}$ are given by the quadratic index:
\be
\CA^{\bf ab}\big|_{\GG_s}=  \sum_{\rho \in \FR} \rho^{\bf a} \rho^{\bf b}  \propto \Tr(T^{\bf a}_\FR T^{\bf b}_\FR)~,
\ee
with $\FR$ the (generally reducible) $\Fg_s$ representation for the chiral multiplets. 
For the flavor symmetry group $\GG_F$, the coefficients $\CA^{\alpha\beta}$ mod 4 may not vanish in general. 
In the presence of perturbative 't Hooft anomalies, there is no invariant meaning to the coefficients $\CA^{\alpha\beta}$, whether for abelian or non-abelian flavor symmetries, but it is useful to keep the same notation as a bookkeeping device. We may call $\CA^{\alpha\beta}$ the ``pseudo-anomaly'' coefficients.
Finally, note that we have  $\CA^{a\beta}=0$ for $\GG$ semi-simple.

\paragraph{Anomalies involving $U(1)_R$.} Last but not least, there are various anomalies involving the $R$-symmetry. The mixed gauge (or flavor)-$R$ anomalies coefficients read:
\be\label{def AR 1}
\CA^{{\bf a b R}}= \sum_I Q_I^{\bf a} Q_I^{\bf b}  (r_I-1)+\delta^{{\bf ab}, ab} \sum_{\alpha \in \Fg} \alpha^{a} \alpha^{b}~,
\qquad\qquad
 \CA^{{\bf a}RR}= \sum_I Q_I^{\bf a} (r_I-1)^2~,
\ee
while the cubic and gravitational $U(1)_R$ 't Hooft anomalies are given by:
\be\label{def AR 2}
\CA^{RRR}= \sum_I (r_i-1)^3 + {\rm dim}(\Fg)~, \qquad \qquad
\CA^{R}= \sum_I (r_i-1) + {\rm dim}(\Fg)~.
\ee
For future  reference, let us also define the quadratic ``pseudo-anomaly'' coefficients:
\be\label{def AR 3}
\CA^{{\bf a}R}= \sum_I Q_I^{\bf a}(r_i-1)~, \qquad \qquad \CA^{RR}= \sum_I (r_i-1)^2  + {\rm dim}(\Fg)~.
\ee
By assumption, our theory has a non-anomalous $U(1)_R$, therefore we must have:
\be\label{AR af}
\CA^{abR}= \CA^{a RR}=0~.
\ee
We also have $\CA^{a R}=0$ for $\GG$ semi-simple.

\subsubsection{Large gauge transformations of the twisted superpotential} 
The superpotential  $\CW(u, \nu; \tau)$ defined above is affected non-trivially by large gauge transformations of its parameters. This non-trivial behavior in turn determines the behavior of many $A$-model observables, including the supersymmetric partition function, under large gauge transformations. Let us define the finite variations:
\be
\Delta_n f(u) \equiv f(u+n) - f(u)~, \qquad \qquad \Delta_{m\tau} f(u) \equiv f(u+m\tau) - f(u)~,
\ee
for $n, m\in \Z$.
Similarly, for any function $F(\bf u)$ of multiple variables ${\bf u_a}$, we define
\be
\Delta_{ \delta_{\bf a} n} F({\bf u)} \equiv F({\bf u_a}+n)-F({\bf u_a})~, \qquad\quad
\Delta_{\delta_{\bf a} m \tau} F({\bf u}) \equiv F({\bf u_a}+m\tau)-F({\bf u_a})~,
\ee
where we shift $F({\bf u})$ along a single ${\bf u_a}$ at the time.

The behavior of \eqref{full W i} under shifts of ${\bf u_a}$ along the fundamental domain can be determined by direct computation on the building block $\CW_\Phi$ for a single chiral multiplet. Using the definition \eqref{full Wphi}, one can show that:
\bea\label{lgt W1 Wtau}
&\Delta_n \CW_\Phi(u; \tau) &=&\;
 -{1\ov \tau} \left({n^3\ov 6}+ {n^2 u\ov 2}+ {n u^2\ov 2}\right)+ {n^2\ov 4}+ {n u\ov 2} -{n\tau\ov 12}~, \cr
&\Delta_{m\tau} \CW_\Phi(u; \tau) 
& =& \;   {mu\ov 2}  +  {m(m-1) \tau\ov 4}  -{m\ov 12}~. 
\eea
for any $n, m\in \Z$. Note that the second line in \eqref{lgt W1 Wtau} is only determined up to a choice of branch.
The  anomalous transformations of the full superpotential directly follow:
\bea\label{anomalous shift W}
&\Delta_{\delta_{\bf a}}\CW=
 -{1\ov \tau} \left({\CA^{\bf aaa}\ov 6}+ {\CA^{\bf aab} {\bf u_b}\ov 2}+ {\CA^{\bf a bc}{\bf u_b u_c}\ov 2}\right) + {\CA^{\bf aa}\ov 4}+ {\CA^{\bf ab}{\bf u_b}\ov 2}-{\tau\ov 12} \CA^{\bf a} ~,\cr
&\Delta_{\delta_{\bf a}\tau}\CW=
 -\CA^{\bf a}\left({\tau\ov 4}-{1\ov 12}\right)\ + {\CA^{\bf aa} \tau \ov 4}+ {\CA^{\bf ab}{\bf u_b}\ov 2}~,
\eea
with the anomaly coefficients defined in \eqref{def anomalies}.
Note that the twisted superpotential would be well-defined in the absence of any anomalies. More precisely, imposing the conditions $\CA^{\bf abc}=\CA^{\bf a}=0$ and $\CA^{\bf ab}\in 4 \Z$ ensures that $\CW$ transforms as $\Delta\CW= n^{\bf a}{\bf u_a}+ n^\tau \tau + n^0$, with $n^{\bf a}, n^\tau, n^0 \in \Z$, which can be cancelled by a change of branch. In the presence of anomalies, however, the non-trivial transformation of $\CW$ is physically meaningful.

For non-anomalous four-dimensional theories satisfying eqref{af cond 1}-\eqref{global anomaly free cond}, the anomalous transformations \eqref{anomalous shift W} only depend on the flavor parameters $\nu_\alpha$, with coefficients precisely determined by the (pseudo-) 't Hooft anomalies $\CA^{\alpha}$, $\CA^{\alpha\beta}$ and  $\CA^{\alpha\beta\gamma}$ for $\GG_F$. 

\subsubsection{Modular transformations of the twisted superpotential} 
We should also consider the behavior of $\CW$ under modular transformations.  
For a single chiral multiplet, we find:
\bea\label{S and T on WPhi}
 &S\;:\;\;   &&\CW_\Phi\Big(\frac{u}{\tau}; -\frac{1}{\tau}\Big) \; \;=\;\;  \frac{1}{\tau} \CW_\Phi(u,\tau)+  \frac{u^3}{6 \tau^2} + \frac{u}{4\tau}~,\cr
& T\;:\;\; && \CW_\Phi\big(u; \tau+1\big)\; \;=\;\;     \CW_\Phi(u,\tau)+{u^3\ov 6 \tau(\tau+1)}  - \frac{u}{12}~,
 \eea
for the $S$ and $T$ generators acting on the superpotential. This is most easily derived using the expression \eqref{Wphi bis} and the modular properties of \eqref{def theta}. The transformations \eqref{S and T on WPhi} satisfy the $SL(2, \Z)$ relations $S^2 = C$ and $(ST)^3= C$,  where the center $C$ acts as  charge conjugation, with $C: (u, \tau) \mapsto (-u, \tau)$. The transformation:
\be
 C\;:\;\; \CW_\Phi(-u; \tau)\; \;=\;\;     -\CW_\Phi(u,\tau)   - {u\ov 2}
\ee
follows from the identity \eqref{massivechiral}.
Note that modular transformations and large gauge transformations are interrelated.  Namely, a large gauge transformation around a particular one-cycle in $T^2$, when composed with a modular transformation, should give the large gauge transformation around the modular-transformed one-cycle.  One can check that this is consistent with the above results. For instance, the $S$ transformation in \eqref{S and T on WPhi} implies:
\be
\Delta_{n\tau} \CW_\Phi(u; \tau)  = \tau S\big[ \Delta_{n} \CW_\Phi(u; \tau)  \big] - {1\ov 6 \tau}\left((u+n\tau)^3-u^3\right) - {n\tau\ov 4}~.
\ee
This  relation is satisfied by the large gauge transformations \eqref{lgt W1 Wtau}.
For a general four-dimensional gauge theory, we directly find:
\bea 
 &S\;:\;\;   && \CW\Big(\frac{{\bf u}}{\tau},-\frac{1}{\tau}\Big) \; \;=\;\;   \frac{1}{\tau}\CW({\bf u},\tau) + \frac{1}{6 \tau^2} \CA^{\bf abc} {\bf u_a}{\bf u_b}{\bf u_c} + \frac{1}{4\tau} \CA^{\bf a} {\bf u_a}~, \cr
& T\;:\;\; &&\CW\big({\bf u},\tau+1\big) \; \;=\;\;   \CW({\bf u},\tau)+ \frac{1}{6 \tau(\tau+1)} \CA^{\bf abc} {\bf u_a}{\bf u_b}{\bf u_c}- \frac{1}{12} \CA^{\bf a} {\bf u_a}~. 
\eea
Therefore, the modular properties of the twisted superpotential are fully determined by the perturbative anomalies.

\subsection{Flux operators and four-dimensional Bethe equations}
Given the twisted superpotential, we may define several $A$-model  operators  \cite{Closset:2017zgf}.
The {\it flux operator}  $\pif_{\bf a}$ is a local operator  that inserts one unit of $U(1)_{\bf a}$ flux on $\Sigma_g$. It is given by:
\be\label{def pif}
\pif_{\bf a} = \exp\left({2\pi i {\d\CW\ov \d {\bf u_a}}}\right)~.
\ee
For a single chiral multiplet with twisted superpotential \eqref{full Wphi}, we have the contribution:
\be\label{def pif Phi}
\pif^\Phi(u; \tau) \equiv \, e^{2\pi i \left(-{u^2\ov 2 \tau}+ {u\ov 2}-{\tau\ov 12}\right)}\,  {1\ov \theta_0(u; \tau)}~,
\ee
with $\theta_0(u; \tau)$ the reduced theta function:
\be
\theta_0(u; \tau) = \prod_{k=0}^{\infty}(1- x  q^k)(1- x^{-1}q^{k+1})~.
\ee
We thus obtain:
\bea\label{flux op a alpha}
&\pif_a(u, \nu; \tau) &= e^{2\pi i \d_{u_a} \CW} &= \prod_i \prod_{\rho_i \in \FR_i} \pif^\Phi(\rho_i(u) + \nu_i)^{\rho_i^a}~, \cr
&\pif_\alpha(u, \nu; \tau) &= e^{2\pi i \d_{\nu_\alpha} \CW} &= \prod_i \prod_{\rho_i \in \FR_i} \pif^\Phi(\rho_i(u) + \nu_i)^{\omega_i^\alpha}~, \cr
\eea
for the gauge and flavor flux operators, respectively.  

For future reference, it is useful to note that we may write the flux operators as:
\be\label{flux op gen ub}
\pif_{\bf a}({\bf u}; \tau) =   \, e^{-{\pi i \ov \tau}\CA^{\bf abc} {\bf u_b u_c}} \, \prod_{I}  \theta\big(Q_I({\bf u}); \tau\big)^{-Q_I^{\bf a}}~,
\ee
in terms of $\theta(u; \tau)$ defined in \eqref{def theta}, and of the cubic anomaly coefficients defined in \eqref{def anomalies}. In particular, we have:
\be\label{gauge flux op af full}
\pif_a(u, \nu; \tau) = \prod_i \prod_{\rho_i \in \FR_i} \theta\big(\rho_i(u)+\nu_i); \tau\big)^{-\rho_i^a}~,
\ee
for the gauge flux operators of an anomaly-free theory. 
The flux operators \eqref{flux op gen ub} satisfy the relations:
\bea\label{lgt pif}
&\pif_{\bf a}({\bf u_b}+1; \tau) &=&\;(-1)^{\CA^{\bf ab}} \, e^{-{\pi i\ov \tau}\left(\CA^{\bf abb}  +2 \CA^{\bf abc} {\bf u_c}\right)} \, \pif_{\bf a}({\bf u}; \tau)~,\cr
&\pif_{\bf a}({\bf u_b}+\tau; \tau) &=&\; (-1)^{\CA^{\bf ab}} \, \pif_{\bf a}({\bf u}; \tau)~,\cr
\eea
where we shift a single ${\bf u_b}$ in $\pif_{\bf a}({\bf u})$. This directly follows from \eqref{anomalous shift W} and from the definition of the flux operator. Note that $(-1)^{\CA^{\bf ab}}=(-1)^{\CA^{\bf aab}}=(-1)^{\CA^{\bf abb}}$. One can also show that, under a modular transformation $S$ of the torus, the flux operators transform non-trivially, with:
\be\label{mod prop of Pif a}
\pif_{\bf a}\left({{\bf u}\ov \tau}; -{1\ov \tau}\right)=e^{{\pi i \ov 2}\CA^{\bf a}} e^{ {\pi i\ov \tau}\CA^{\bf abc}{\bf u_b u_c}}\, \pif_{\bf a}({\bf u}; \tau)~.
\ee

The anomaly-free conditions \eqref{af cond 1}-\eqref{global anomaly free cond}  imply that the gauge flux operators \eqref{gauge flux op af full} are fully elliptic in all  of its parameters:
\be\label{pif elliptic}
\pif_a(u_b+ n + m\tau, \nu_\alpha+ n' + m' \tau; \tau)=\pif_a(u, \nu; \tau)~, \qquad \forall n, m, n', m' \in \Z~.
\ee
It also follows from \eqref{mod prop of Pif a} that the gauge flux operators are modular invariant.
On the other hand, the flavor flux operators $\pif_\alpha$ are elliptic in $u_a$, but transform non-trivially under shifts of $\nu_\alpha$ along the fundamental domain, as well as under modular transformations.

The set of Bethe-vacua for a four-dimensional $\CN=1$ supersymmetric gauge theory is given by:
\be\label{S BE 4d}
\cS_{BE} =\left\{ \;\h u_a \; \bigg| \; \Pi_a(\h u, \nu; \tau) = 1~, \;\; \forall  a~, \quad w \cdot \h u \neq \h u, \;\; \forall w \in W_\GG \; \right\} /W_\GG~,
\ee
with $u_a$ subject to the identifications $u_a \sim u_a +1 \sim u_a + \tau$. Here $W_\GG$ denotes the Weyl group of $\GG$, and $w\cdot u$ the Weyl group action on $\{u_a\}$. We are instructed to discard any solution that is not acted on freely by the Weyl group, as the corresponding would-be vacua are supersymmetry-breaking \cite{Hori:2006dk,Aharony:2016jki}. Note that the Bethe equations \eqref{S BE 4d} only make sense for a well-defined,  anomaly-free gauge theory, for which \eqref{pif elliptic} holds true.

Finally, we note in passing that the flux operators, which can be interpreted as surface operators supported on a $T^2$ fiber over $\Sigma_g$, have the form of the elliptic genus of a $2d$ $\cN=(0,2)$ theory of chiral and Fermi multiplets coupled to the $4d$ gauge and flavor symmetry, as computed in \cite{Gadde:2013ftv,Benini:2013nda}.

\subsection{The $T^2$ fibering operators}
Consider the torus $T^2 \cong S^1_{\beta_1}\times S^1_{\beta_2}$, with $\beta_1, \beta_2$ the radii of the two circles. The two-dimensional theory obtained by compactification on $T^2$ has a distinguished $U(1)_{{\rm KK}_1} \times U(1)_{{\rm KK}_2}$ global symmetry, whose conserved charges are the Kaluza-Klein (KK) momenta along both circles---corresponding to the integers $n$ and $m$ in \eqref{W phi formal nm}.

From the two-dimensional perspective, there must exist distinguished flux operators that insert background fluxes for the KK symmetries, the so-called {\it fibering operators} \cite{Closset:2017zgf}.
We denote them by $\CF_1$ and $\CF_2$ for $U(1)_{{\rm KK}_1}$ and $U(1)_{{\rm KK}_2}$, respectively.  Like any flux operator, they are fully determined in terms of the effective twisted superpotential. These operators introduce a non-trivial fibration of $T^2$ over the Riemann surface on which the $A$-model is defined. From the four-dimensional perspective, the fibering operator is a particular defect surface operator wrapped over $T^2$ in $\Sigma_g\times T^2$.

 With our definition \eqref{def tau} for the modular parameter $\tau$ and  reinstating dimensions, we have:
\be\label{mKK}
m_{{\rm KK}_1} = {\tau\ov \beta_2}~, \qquad m_{{\rm KK}_2} = {1\ov \beta_2}
\ee
for the two-dimensional twisted masses associated to $U(1)_{{\rm KK}_1} \times U(1)_{{\rm KK}_2}$.
One can then easily show that:
\be\label{def F1}
\CF_1(u, \nu; \tau)  =\exp\left({2\pi i {\d\CW\ov \d \tau}}\right)~.
\ee
for the first fibering operator, and 
\be\label{def F2}
\CF_2(u, \nu; \tau)  = \exp{\left(2\pi i \left(\CW- u_a {\d \CW\ov \d u_a} - \nu_\alpha {\d \CW\ov \d \nu_\alpha}- \tau  {\d \CW\ov \d \tau}  \right) \right)}~,
\ee
for the second fibering operator. 
Note that we have been using a ``three-dimensional'' regularization, wherein we first consider the theory as a three-dimensional theory on $\R^2 \times S^1_{\beta_2}$, and then regularize the remaining KK tower from $S^1_{\beta_1}$.~\footnote{A similar two-step regularization was used in \cite{Assel:2015nca}, where it was argued to be consistent with supersymmetry.} 
From the point of view of $\CM_3 \cong \R^2 \times S^1_{\beta_2}$, the modular parameter $\tau$ is the complexified fugacity associated to $U(1)_{{KK}_1}$ for $S^1_{\beta_1}$---an ordinary symmetry from the $\CM_3$ point of view---and all dimensions can be absorbed with the radius $\beta_2$, as in \eqref{mKK}. For this reason,   the formula \eqref{def F1} takes the same form as the formula \eqref{def pif} for an ``ordinary'' flux operator, and both \eqref{def F1} and   \eqref{def F2}  directly follow from the three-dimensional results of \cite{Closset:2017zgf}. 

\paragraph{Chiral multiplet contribution to $\CF_1$.}
The first fibering operator is given in terms of the function:
\be
\Gamma_0(u; \tau) \equiv  \prod_{n=0}^\infty  \left({1- x^{-1}q^{n+1}\ov 1- x q^{n+1}}  \right)^{n+1}~,
\ee
which is a specialization of the elliptic gamma function, $\Gamma_0(u; \tau) \equiv \Gamma_e(q x; q, q)$. (See Appendix \ref{app: el fct identities}.)  From the definition \eqref{def F1}, we find:
\be\label{CF1 Phi}
\CF_1^\Phi(u; \tau) = \exp\left(2 \pi i \left({u^3\ov 6 \tau^2}- {u\ov 12}\right) \right) \, \Gamma_0(u; \tau)
\ee
for the contribution of a single chiral multiplet of unit charge. 
A useful relation satisfied by \eqref{CF1 Phi} is:
\be\label{F1F1 id}
\CF_1^\Phi(u; \tau) \CF_1^\Phi(-u; \tau)= 1~.
\ee
This corresponds to a pair of massive four-dimensional chiral multiplets, which contribute trivially to the fibering operator.

\paragraph{Chiral multiplet contribution to $\CF_2$.}
To discuss the explicit form of the second fibering operator, it is useful to introduce the function:
\be\label{def fphi}
f_\Phi(u) \equiv  \exp{\left({1\ov 2 \pi i} \dilog\left(e^{2\pi i u}\right)+ u \log\left(1-e^{2\pi i u}\right)\right)}~,
\ee
which is meromorphic in $u$, with poles of order $n$ at $u=-n$, $n\in \Z_{>0}$. This is the fibering operator associated to the {\it three-dimensional} $\CN=2$ supersymmetric chiral multiplet   \cite{Closset:2017zgf}. It satisfies the identity:
\be\label{3d F id}
f_\Phi(u)f_\Phi(-u)= e^{\pi i \left(u^2- {1\ov 6}\right) }~.  
\ee 
For a four-dimensional chiral multiplet of unit $U(1)$ charge, plugging \eqref{full Wphi} into \eqref{def F2}, we directly obtain:
\be\label{CF2 Phi}
\CF_2^\Phi(u; \tau) = \exp\left(2 \pi i \left({u^3\ov 6 \tau} - {u^2\ov 4}+ {u\tau\ov 12}+{1\ov 24} \right)\right)\,   \prod_{k=0}^\infty{ f_\Phi(u+ k \tau) \ov f_\Phi(-u+ (k+1) \tau)}~.
\ee
The infinite product in \eqref{CF2 Phi} is convergent. 
Using  \eqref{3d F id}, we can also show that:
\be
\CF_2^\Phi(u; \tau) \CF_2^\Phi(-u; \tau)= 1~,
\ee
similarly to \eqref{F1F1 id}.

\paragraph{Large gauge transformations.}
The two fibering operators transform non-trivially under large gauge transformation, as follows from \eqref{lgt W1 Wtau}.  One can check that:
\bea\label{el pr F1}
&\CF^\Phi_1(u+n; \tau) &=&\;\; e^{-{\pi i n\ov 6}} e^{{2\pi i \ov \tau^2} \left({nu^2\ov 2} + {n^2 u\ov 2 }+ {n^3\ov 6 }\right)}  \CF^\Phi_1(u; \tau)~. \cr
&\CF^\Phi_1(u+m\tau; \tau) &=&\; \;  e^{-{\pi i \ov2} m^2} e^{-{\pi i \ov2} m}\, \pif^\Phi(u; \tau)^{-m}\, \CF^\Phi_1(u; \tau)~,
\eea
and
\bea\label{el pr F2}
&\CF^\Phi_2(u+n; \tau) &=&\;\; e^{-{\pi i\ov 2} n^2} e^{{2 \pi i \ov \tau} \left({n^2 u\ov 2}+ {n^3\ov 6}\right)} \, \pif^\Phi(u; \tau)^{-n}\, \CF^\Phi_2(u; \tau)~,\cr
&\CF^\Phi_2(u +m \tau; \tau) &=&\;\; e^{\pi i m\ov 6} \,\CF^\Phi_2(u; \tau)~,
\eea
for $n, m\in \Z$. Note the appearance of the chiral-multiplet flux operator $\pif^\Phi$ when we perform a large gauge transformation along the circle being fibered.
On general grounds, the two fibering operators are related by a modular transformation  $S$.
Indeed, one can prove that:
\be\label{F1 F2 chiral rel}
 \CF^\Phi_2\left({u\ov \tau}, -{1\ov \tau}\right) = \exp\left(-\pi i {u^3\ov 3 \tau^2}\right) \, \CF^\Phi_1(u, \tau)~.
\ee 
This directly follows from \eqref{S and T on WPhi}. It also follow from mathematical results about the elliptic gamma function \cite{1999math......7061F}.~\footnote{More precisely, one can check that \protect\eqref{F1 F2 chiral rel} is equivalent to Theorem 5.2 of \cite{1999math......7061F}. }

\paragraph{General gauge theory.}
In the full gauge theory, the fibering operators are simply given by:
\be
\CF_1({\bf u}; \tau)=\prod_I \CF_1^\Phi(Q_I({\bf u}); \tau)~, \qquad\qquad 
\CF_2({\bf u}; \tau)=\prod_I \CF_2^\Phi(Q_I({\bf u}); \tau)~.
\ee
From \eqref{F1 F2 chiral rel}, we have the modular transformation:
\be
 \CF_2\left({{\bf u}\ov \tau}; -{1\ov \tau}\right) = \exp\left(-{\pi i\ov 3 \tau^2} \CA^{\bf abc} {\bf u_a u_b u_c}\right) \, \CF_1({\bf u}; \tau)~,
\ee
It is clear that the two fibering operators must be related by an $S$ transformation. As we see, the exact relation involves  the cubic anomaly coefficients. In any anomaly-free theory, we simply have:
\be\label{F1 to F2 af theory}
 \CF_2\left({ u\ov \tau}, {\nu \ov \tau}; -{1\ov \tau}\right) = \exp\left(-{\pi i\ov 3 \tau^2} \CA^{\alpha\beta\gamma}  \nu_\alpha \nu_\beta \nu_\gamma \right) \, \CF_1(u, \nu; \tau)~,
\ee
where the exponential prefactor is given in terms of the flavor symmetry 't Hooft anomalies.
The fibering operators also transform non-trivially under large gauge transformations. We have:
\bea\label{ell F1}
&\CF_1({\bf u_a}+1, \tau) &=&\; \; e^{-{\pi i \ov 6}\CA^{\bf a}}\, e^{ {\pi i \ov \tau^2}\left(\CA^{\bf abc} {\bf u_b u_c} + \CA^{\bf aab}{\bf u_b}+{1\ov 3} \CA^{\bf aaa}\right)}\; \CF_1({\bf u}, \tau)~,\cr
&\CF_1({\bf u_a}+\tau, \tau)&=&\; \;  e^{- {\pi i\ov 2} \CA^{\bf a}} \, e^{ -{\pi i \ov 2}\CA^{\bf aa}}\,\pif_{\bf a}({\bf u};\tau)^{-1}\; \CF_1({\bf u}, \tau)~,
\eea
 and
\bea\label{ell F2}
&\CF_2({\bf u_a}+1; \tau) &=&\; \;  e^{ {\pi i \ov \tau}\left(\CA^{\bf aab} {\bf u_b} + {1\ov 3} \CA^{\bf aaa}\right)}\, e^{- {\pi i \ov 2}\CA^{\bf aa}} \, \pif_{\bf a}({\bf u};\tau)^{-1} \;\CF_2({\bf u}; \tau)~, \cr
&\CF_2({\bf u_a}+ \tau; \tau)&=&\; \; e^{{\pi i \ov 6} \CA^{\bf a}}\;\CF_2({\bf u}; \tau)~.
\eea
These important difference equations relate the flux and fibering operators. For any anomaly-free theory, the prefactors in \eqref{ell F1}-\eqref{ell F2} only involve the 't Hooft anomalies (as well as the ``pseudo 't Hooft anomaly'' coefficients), and therefore these prefactors only depend on $\nu_\alpha$, and are independent of the gauge variables $u_a$. This will be crucial later on.

\subsection{Effective dilaton and handle-gluing operator}
The second function  that determines the $A$-model \eqref{S TFT} is the effective dilaton $\Omega$.
For a four-dimensional gauge theory with matter fields in chiral multiplets $\Phi_i$ of $R$-charges $r_i \in \Z$, we have:
\be
 \Omega=  \Omega_{\rm mat}+  \Omega_{\rm vec}~,
\ee
such that:
\bea\label{Omega mat vec}
&\exp\Big(2\pi i \,\Omega_{\rm mat}(u, \nu; \tau) \Big)&=&\;\;  \prod_i \prod_{\rho_i \in \FR_i} \pif^\Phi(\rho_i(u)+ \nu_i; \tau)^{r_i-1}~, \cr
&\exp\Big(2\pi i \, \Omega_{\rm vec}(u, \nu; \tau) \Big)&=&\;\; \eta(\tau)^{-2\, \rk} \prod_{\alpha\in \Fg} \pif^\Phi(\alpha(u); \tau)~.
\eea
Here we have the contributions from the chiral and vector multiplets, respectively, which are given in terms of the chiral-multiplet flux operator \eqref{def pif Phi}.  With respect to the three-dimensional case \cite{Closset:2017zgf}, the only new ingredient in \eqref{Omega mat vec}  is the appearance of a contribution $\eta(\tau)^{-2}$ for each generator of the Cartan of $\Fg$, with $\eta(\tau)$ the Dedekind function---see Appendix \ref{app: el fct identities}.
Using the effective dilaton, one may define the  {\it handle-gluing operator } \cite{Nekrasov:2014xaa, Closset:2017zgf}, whose insertion on $\Sigma_g$ has the effect of adding a handle, changing the topology of the base to $\Sigma_{g+1}$.

Let us first define the Hessian determinant of the twisted superpotential:
\be\label{def H}
H(u, \nu; \tau)\, \equiv\, \det_{ab}{\d^2 \CW(u, \nu; \tau)\ov \d u_a \d u_b}\, =\,  \det_{ab}\left( {1\ov 2 \pi i} {\d \log \pif_a \ov \d u_b}\right)~.
\ee
Note that $H$ is fully elliptic in all parameters $u$ and $\nu$, as follows from \eqref{pif elliptic}.
On the other hand, it has a non-trivial  modular transformation:
\be\label{lgt H}
H\left({u\ov \tau}, {\nu\ov \tau}; -{1\ov \tau}\right)= \tau^\rk\, H(u, \nu; \tau)~,
\ee
as evident from \eqref{def H}. 
Combining \eqref{Omega mat vec} and \eqref{def H}, we construct the handle-gluing operator:
\be\label{hgo def}
\CH(u, \nu; \tau) =e^{2\pi i \,\Omega(u, \nu; \tau)} \, H(u, \nu; \tau)~.
\ee
It is clear from the effective action \eqref{S TFT} that the insertion of $e^{2\pi i \Omega}$ corresponds to concentrating the curvature of a single handle to a point. The appearance of the Hessian in \eqref{hgo def} is due to the gaugino coupling in \eqref{S TFT}, because each handle comes together with two gaugino zero modes $\t\Lambda$, $\Lambda$, to be integrated over.

The behavior of the handle-gluing operator under large gauge transformations follows simply from the properties of $\pif^\Phi$. It is easy to see that:
\bea\label{lgt CH}
&\CH({\bf u_a}+ 1; \tau) &=&\;\; (-1)^{\CA^{{\bf a}RR}}  e^{-{\pi i \ov \tau}\left(\CA^{{\bf a a R}}+2 \CA^{{\bf a b R}}{\bf u_b}\right)}\, \CH({\bf u_a}; \tau)~, \cr
&\CH({\bf u_a}+ \tau; \tau) &=&\;\; (-1)^{\CA^{{\bf a}RR}} \CH({\bf u_a}; \tau)~.
\eea
Here the anomaly coefficients are the ones defined in \eqref{def AR 1}.
By assumption, our four-dimensional gauge theory has an anomaly-free R-symmetry, so that \eqref{AR af} holds true.
This implies that the handle-gluing operator is elliptic in $u_a$, while it retains non-trivial transformation properties under shifts of $\nu_\alpha$ in the presence of mixed $U(1)_R$-$\GG_F$ 't Hooft anomalies.
We  can similarly show that $\CH$ is invariant under modular transformations up to 't Hooft anomalies. One can see that:
\be
\CH\left({{\bf u}\ov \tau}; -{1\ov \tau}\right) = e^{{\pi i \ov 2}\CA^{R}} e^{ {\pi i\ov \tau}\CA^{{\bf ab} R}{\bf u_a u_b}}\,   \CH({\bf u}; \tau)~,
\ee
with $\CA^R$ defined in \eqref{def AR 2}. This crucially relies on the contribution of the Cartan vector multiplet contribution in \eqref{Omega mat vec}, whose modular transformation cancels the one from \eqref{lgt H}.
Finally, let us note that, upon imposing the anomaly-free constraint \eqref{AR af}, we may write the handle-gluing operator \eqref{hgo def} explicitly as:
\bea\label{CH explicit}
&\CH(u, \nu; \tau)=\;   e^{-{\pi i \ov \tau}\CA^{\alpha\beta R}\nu_\alpha\nu_\beta } \,  H(u, \nu; \tau)\cr 
&\qquad\quad\quad\times \prod_i\prod_{\rho_i \in \FR_i} \theta(\rho_i(u)+ \nu_i; \tau)^{1-r_i}\,
\;{1\ov \eta(\tau)^{2\, \rk} } \, \prod_{\alpha\in \Fg} {1\ov \theta(\alpha(u); \tau)}~.
\eea
This will be useful below.

\section{The $\Mgp$ index and its properties}\label{sec: 3}
\label{sec:Mgp}
In the previous section, we defined and computed the flux, fibering, and handle-gluing operators:
\be\label{all ops}
\pif_a(u, \nu; \tau)~, \qquad \pif_\alpha(u, \nu; \tau)~, \qquad \CF_1(u, \nu; \tau)~, \qquad \CF_2(u, \nu; \tau)~, \qquad \CH(u, \nu; \tau)~,
\ee
 for any well-defined four-dimensional $\CN=1$ supersymmetric gauge theory with a semi-simple, simply-connected gauge group $\GG$ and matter fields in chiral multiplets.  The operators \eqref{all ops} are local operators in {\it the four-dimensional A-model}, which is a two-dimensional effective field theory with $\CN=(2,2)$ supersymmetry which describes the light degrees of freedom of the four-dimensional gauge theory compactified on $T^2$, in the presence of arbitrary ``complexified chemical potentials''  $\nu_\alpha$ for the flavor symmetry. (More precisely, $\nu_\alpha$ are $T^2$ flat connections for the Cartan of the flavor group $\GG_F$.) We can also think of  the operators \eqref{all ops} as particular half-BPS four-dimensional surface operators wrapped on $T^2$.

For generic-enough flavor parameters $\nu_\alpha$,  the four-dimensional A-model is a massive theory with discrete vacua, called the Bethe vacua. These vacua are in one-to-one correspondence with the solutions to the Bethe equations \eqref{S BE 4d}, as explained above. The Bethe equations are given explicitly by:
\be\label{S BE 4d bis}
\Pi_a(u, \nu; \tau) = \prod_i \prod_{\rho_i \in \FR_i} \theta\big(\rho_i(u)+\nu_i; \tau\big)^{-\rho_i^a}= 1~, \;\; \forall  a~, 
\ee
as equations for the elliptic variables $u_a$ ($a=1, \cdots, \rk$), with the additional constraint that a valid solution $\{\h u_a\}_{a=1}^{\rk}$ must be acted freely on by the Weyl group.

\subsection{Fibering the four-dimensional A-model}\label{subsec: fibering T2}
Consider a complex four-manifold $\CM_4$ given as a $T^2$ fibration over $\Sigma_g$:
\be\label{M4 fibration}
T^2 \longrightarrow \CM_4 \longrightarrow \Sigma_g~,
\ee
with $T^2 \cong S^1_{\beta_1}\times S^1_{\beta_2}$.
From the A-model point of view, the non-trivial fibration is realized by inserting the fibering operators $\CF_1$ and $\CF_2$ at a point on $\Sigma_g$, thus introducing a non-trivial fibration of $S^1_{\beta_1}$ and $S^1_{\beta_2}$, respectively, over $\Sigma_g$. 
To realize the $\CM_4$ background \eqref{M4 fibration}, we must turn on fluxes:
\be\label{p1 p2 fluxes}
{1\ov 2 \pi } \int_{\Sigma_g} dA_{{\rm KK}_{1}}= p_{1}~,\qquad \qquad
{1\ov 2 \pi } \int_{\Sigma_g} dA_{{\rm KK}_{2}}= p_{2}~, \qquad \qquad p_1,p_2 \in \Z~,
\ee
 for the $U(1)_{{\rm KK}_1}\times U(1)_{{\rm KK}_2}$ symmetry on $\Sigma_g$. In addition, we may turn on arbitrary background fluxes \eqref{def nalpha}, denoted by  $\n_\alpha\in \Z$, for the flavor symmetry $\GG_F$.
The partition function of the four-dimensional $A$-model on $\Sigma_g$, with those insertions, is then given by:
\be\label{Zgp1p2}
\CZ_{g, \, p_1,\, p_2}(\nu; \tau) =\sum_{\h u \in \CS_{\rm BE}}  \CF_1(\h u,\nu; \tau)^{p_1}\,  \CF_2(\h u,\nu; \tau)^{p_2}\, \CH(\h u,\nu; \tau)^{g-1} \,\prod_{\alpha} \pif_\alpha(\h u,\nu; \tau)^{\n_\alpha}~,
\ee 
as a sum over the Bethe vacua \eqref{S BE 4d}.  It is important to note that this formula only makes sense if the gauge theory is anomaly-free, with vanishing mixed gauge-flavor and gauge-$R$ anomalies. If and only if those condition are satisfied, the operators \eqref{all ops} are truly elliptic in the parameters $u_a$, and therefore the Bethe equations \eqref{S BE 4d bis} and the partition function \eqref{Zgp1p2} are well-defined.

Importantly, we may always perform a modular transformation of $T^2$ such that:
\be\label{p1 p2 to p}
(p_1, p_2)= (p,0)~.
\ee
Indeed, the graviphoton fluxes \eqref{p1 p2 fluxes} have the effect of shifting the twisted spins of the KK modes by $\delta s=m p_1 + n p_2$ in two dimensions, which can be mapped to $\delta s= m'p$ by a modular transformation. There always exists an $SL(2,\Z)$ matrix:
\be
A= \mat{l_1 & -{p_2\ov p} \\ l_2 &{ p_1\ov p}}~, \qquad  p_1 l_1+ p_2 l_2 = p~, \qquad p=\gcd(p_1, p_2)~,
\ee
such that:
\be\label{A of Op}
A[\CF_1^{p_1} \CF_2^{p_2}]\propto  \CF_2^p~, \qquad A[\CH]\propto \CH~, \qquad A[\pif_\alpha]\propto \pif_\alpha~, 
\ee
where $A[\CO]$ denotes the corresponding $SL(2,\Z)$ action on the operator $\CO$. The proportionality factors in \eqref{A of Op} are given entirely in terms of 't Hooft anomalies, as we saw explicitly in the previous section in the case $A=S$.  Therefore, in the following and without loss of generality, we mostly consider \eqref{p1 p2 to p}. We will further discuss the modular properties of the four-dimensional $A$-model in subsection \ref{sec: modularity}.

This general discussion is compatible with the known classification of supersymmetric backgrounds in four dimensions. Four-dimensional $\CN=1$ supersymmetric backgrounds coupling to the $\CR$-multiplet~\footnote{That is, a supercurrent  multiplet including the conserved $R$-symmetry current. See \protect\cite{Dumitrescu:2011iu} and references therein.} are only possible for $\CM_4$ a complex manifold \cite{Dumitrescu:2012ha, Klare:2012gn}. Moreover, our two-supercharge  background is the pull-back of the two-dimensional $A$-twist through the fibration $\Sigma_g$ in \eqref{M4 fibration}. This implies that $T^2$ must be elliptically fibered, and any such elliptic fibration has the topology:
\be
\CM_4 \cong \CM_3 \times S^1~.
\ee
More specifically, we are considering principal elliptic bundles, which implies that $\CM_3\cong \Mgp$.
The three-manifold $\Mgp$ is itself a principal circle bundle over $\Sigma_g$ \cite{Closset:2017zgf}. This family of four-dimensional supersymmetric backgrounds was also discussed thoroughly in \cite{Nishioka:2014zpa}.

An important special case is  $g=0$ and $p=1$,  in which case we have $\CM_{0,1}\cong S^3$ and $\CM_4$ is a primary Hopf surface $\CM_4^{q,q}=S^3\times S^1$ defined by the quotient:
\be
(z_1, z_2) \sim (q z_1, q z_2)~, \qquad \qquad q= e^{2 \pi i \tau}~.
\ee
Recall that, in general, the primary Hopf surface $\CM_4^{{\bf p},{\bf q}}$ is determined by two complex structure parameters ${\bf p}$ and  ${\bf q}$, and the partition function on  $\CM_4^{{\bf p},{\bf q}}$ computes the index \eqref{index pq} \cite{Closset:2013vra, Assel:2014paa}. In this paper, we have set ${\bf p}={\bf q}=q$ in order to use the $A$-model point of view, as we explained in the introduction.
We are therefore computing the corresponding specialization of the $S^3$ supersymmetric index.

Incidentally, we should also note that the ``$S$-transformed'' $S^3\times S^1$ background, corresponding to $(p_1, p_2)= (0,1)$, has appeared before in the literature in the form of the so-called ``modified $S^3$ index'' \cite{Brunner:2016nyk}.  In our language, this simply corresponds to using the fibering operator $\CF_2$, instead of $\CF_1$, in the construction of the three-sphere index. 
In the case of conformal theories, related modular properties of partition functions have been discussed in \cite{Shaghoulian:2016gol}.

\subsection{The $\Mgp\times S^1$ partition function from the $A$-model}

Specializing \eqref{Zgp1p2} to \eqref{p1 p2 to p}, we obtain:
\be\label{ZMgp}
\CZ_{\Mgp \times S^1}(\nu; \tau) =\sum_{\h u \in \CS_{\rm BE}}  \CF_1(\h u,\nu; \tau)^{p}\, \CH(\h u,\nu; \tau)^{g-1} \,\prod_{\alpha} \pif_\alpha(\h u,\nu; \tau)^{\n_\alpha}~.
\ee 
  By construction, the $A$-model partition function \eqref{ZMgp} is locally holomorphic in the flavor parameters $\nu_\alpha$, and in the complex structure parameter $\tau$. This is in agreement with general constraints on supersymmetric partition functions  \cite{Closset:2013vra, Closset:2014uda}. The precise identification between the $A$-model partition function and the ``physical'' $\Mgp\times S^1$ partition function is a non-trivial matter, due in particular to possible anomalous corrections to the naive supersymmetric Ward identities \cite{Papadimitriou:2017kzw}. 

Nonetheless,  the four-dimensional $A$-model is well-defined by itself for any four-dimensional gauge theory. We claim that \eqref{ZMgp} computes the expected ``supersymmetric partition function'' \eqref{ZM3S1 gen intro}, possibly up to simple anomalous prefactors. In section \ref{sec: free R} and \ref{sec: integral form and BE}, we will  explain the exact relation between the $A$-model partition function and the standard supersymmetric index on $S^3$. We will also give an alternative derivation of \eqref{ZMgp}, for any $\Mgp$, by supersymmetric localization.

Note that the background fluxes $\n_\alpha$ in \eqref{ZMgp} are actually valued in $\Z_p$, a torsion subgroup of $H^2(\Mgp, \Z)$. Under large gauge transformations for the flavor group  along the $T^2$ fiber, we have:
\be
(\nu_\alpha, \n_\alpha)\;\sim\; (\nu_\alpha+1, \n_\alpha)\;\sim\;  (\nu_\alpha+\tau, \n_\alpha+p)~,
\ee
for any Cartan element $U(1)_\alpha\subset \GG_F$.
It follows from the general properties of the operators \eqref{all ops} that the partition function \eqref{ZMgp} transforms non-trivially under large gauge transformations for the background gauge fields. These transformations properties are determined by the 't Hooft anomalies:
\bea
&\CZ_{g, p, \n}(\nu_\alpha+1)=(-1)^{\left[(g-1)\CA^{\alpha R}+\n_\beta\CA^{\alpha\beta}\right]} \,
e^{-{\pi i p \ov 6} \CA^\alpha} \, e^{{\pi i  p\ov \tau^2}\left(\CA^{\alpha\beta\gamma}\nu_\beta\nu_\gamma + \CA^{\alpha\alpha\beta}\nu_\beta + {1\ov 3}\CA^{\alpha\alpha\alpha} \right)}\cr
& \qquad\qquad\qquad \times e^{-{\pi i \ov \tau}\left((g-1)\CA^{\alpha\alpha R}+\n_\beta\CA^{\alpha\alpha\beta}+ 2\left((g-1)\CA^{\alpha\gamma R}+\n_\beta\CA^{\alpha\beta\gamma} \right)\nu_\gamma\right)}\, \CZ_{g, p, \n}(\nu)~,\cr
&\CZ_{g, p, \n_\alpha+p}(\nu_\alpha+\tau)=   (-1)^{\left[(g-1)\CA^{\alpha R}+ \n_\beta \CA^{\alpha\beta}+ {p\ov 2}( \CA^\alpha+ \CA^{\alpha\alpha})\right]} \, \CZ_{g, p, \n}(\nu)~,
\eea
as follows from \eqref{lgt pif}, \eqref{ell F1} and \eqref{lgt CH}.  Note that the prefactor in the last line is a pure sign. 

\subsection{Supersymmetric Casimir energy and the $\Mgp$ index}
As explained in the introduction, while the supersymmetric partition function should exactly compute the index \eqref{def gen index}, it is sometimes convenient to isolate the vacuum contribution like in \eqref{ZM3S1 gen intro}.
Consider the $\Mgp\times S^1$ partition function \eqref{ZMgp} for a non-anomalous gauge theory. We define the ``normalized''  $\Mgp$ index by:
\be\label{fac I E}
\CZ_{\Mgp \times S^1}(\nu; \tau) = e^{2 \pi i \tau\, \CE_{\Mgp}(\nu; \tau)}  \; \CI_{\Mgp}(\nu; \tau)~,
\ee
where $\CE_{\Mgp}$ is defined to be:
\bea\label{CE def}
&\CE_{\Mgp}(\nu; \tau) =\; p \left({\CA^{\alpha\beta\gamma}\ov 6 \tau^3} \nu_\alpha\nu_\beta\nu_\gamma- {\CA^\alpha\ov 12 \tau}\nu_\alpha\right)\cr
&\; \qquad\qquad\quad\;\; - (g-1)\left({\CA^{\alpha\beta R}\ov 2 \tau^2}\nu_\alpha \nu_\beta+{\CA^R\ov 12} \right) 
- \n_\alpha\left({\CA^{\alpha\beta\gamma}\ov  2\tau^2} \nu_\beta \nu_\gamma+{\CA^\alpha\ov 12}\right)~.
\eea
It is natural to conjecture that \eqref{CE def} gives the vacuum expectation value of the Hamiltonian generating the $S^1$ translation on $\Mgp\times S^1$---on this particular $A$-twist background (see Appendix \ref{app:  SUSY backg}). To verify this conjecture, one should perform the Hamiltonian quantization on $\Mgp$,  similarly to the $S^3$ computation of \cite{Lorenzen:2014pna, Assel:2015nca}. We leave this for future investigation.

Note that the would-be supersymmetric Casimir energy \eqref{CE def} is given entirely in terms of perturbative `t Hooft anomalies. Therefore, it is scheme-independent. In particular, dual field theories on $\Mgp$ will have the same value of $\CE_{\Mgp}$.
The normalized $ \CI_{\Mgp}$ is given by:
\be\label{IMgp0}
 \CI_{\Mgp}(\nu; \tau)=   \sum_{\h u \in \CS_{\rm BE}}  \CJ_\CF(\h u, \nu; \tau)^p \,   \CJ_\CH(\h u, \nu; \tau)^{g-1}\, \prod_{\alpha} \CJ_{\pif_\alpha}(\h u, \nu; \tau)^{\n_\alpha}~,
\ee
as a sum over Bethe vacua. Here we defined:
\bea\label{def J index}
& \CJ_\CF(u, \nu; \tau) &=&\; \;\prod_i\prod_{\rho_i}  \Gamma_0\big(\rho_i(u)+ \nu_i; \tau\big)~,\cr
& \CJ_\CH(u, \nu; \tau) &=&\;\; e^{\pi i \CA^{\alpha R}\nu_\alpha} H(u, \nu; \tau) \prod_i\prod_{\rho_i} \theta_0\big(\rho_i(u)+ \nu_i; \tau\big)^{1-r_i}\cr
&&& \;\;  \times (q; q)_\infty^{-2\rk}  \,\prod_{\alpha\in \Fg}   \theta_0\big(\alpha(u); \tau\big)^{-1}~,\cr
& \CJ_{\pif_\alpha}(u, \nu; \tau)&=&\; \; e^{\pi i \CA^{\alpha \beta}\nu_\beta}\prod_i\prod_{\rho_i} \theta_0\big(\rho_i(u)+ \nu_i; \tau\big)^{-\omega^a_i}~. 
\eea
By construction, the index \eqref{IMgp0} has an expansion in integer powers of $q=e^{2\pi i \tau}$ and $y=e^{2\pi i \nu}$.  Indeed, the summand is a power series in $q$, $y$, and $x=e^{2\pi i u}$, and the solutions of the Bethe equations $\h x=  \h x(q, y)$, can be computed order-by-order in $q$ and $y$, in principle, since the gauge flux operators themselves have a natural expansion in integer powers of $q$, $x$ and $y$. Note, however, that \eqref{def J index} also contains overall factors of $y^{\half}$, determined by the quadratic (pseudo-)anomaly coefficients $\CA^{\alpha R}$ and $\CA^{\alpha \beta}$, whose physical signification is more mysterious.~\footnote{These terms originate from the linear term in the exponential in \eqref{def pif Phi}.  }

\subsection{Small circle limit and Cardy-like formula}
Consider any $\CN=1$ supersymmetric background preserving two supercharges with topology $\CM_3\times S^1$, with $S^1$ a circle of radius $\beta$.
It has been argued, on general grounds,   that the $\CM_3\times S^1$ supersymmetric partition function has a small-$\beta$ limit governed by the mixed flavor- and $R$-gravitational anomalies \cite{DiPietro:2014bca}. In particular, in the absence of flavor  background vector multiplets, we must have the universal contribution:
\be\label{gen res dPK}
\log \CZ_{\CM_3\times S^1} = - { \pi \Tr(R)\ov 24 \beta} L_{\CM_3}+ O(\beta^0)~,
\ee
with $\Tr(R)= \CA^R$.
Here $L_{\CM_3}$ is a constant depending only on the three-dimensional supersymmetric background $\CM_3$.~\footnote{We adopted the normalization of \cite{DiPietro:2016ond} for $L_{\CM_3}$. We should also note that our $\beta$ corresponds to $\beta/2 \pi$ in \cite{DiPietro:2014bca, DiPietro:2016ond}. Similarly, we identify $\t\beta$, the radius of the fibered circle in $\Mgp$, with the $\CM_3$ radius  $r_3$  in those papers.} 
In the case of our three-manifold $\Mgp$, one can compute:
\be\label{LM3 gp}
L_{\Mgp} = {1\ov \pi^2} \int_{\Mgp} d^3 x \sqrt{g}\left({1\ov 4} R - \half H^2 \right) = 4 \t\beta \, (1-g)~,
\ee
This can be easily obtained from the three-dimensional supersymmetric background studied in \cite{Closset:2017zgf}. The quantity \eqref{LM3 gp} is essentially an ``FI term'' for the $R$-symmetry in three dimensions.
The $A$-model partition function \eqref{ZMgp} corresponds to $\beta=\beta_2$ and $\t \beta=\beta_1$.

This general result is elegantly reproduced by the Bethe-vacua formula \eqref{ZMgp}.  In our notation, the correct three-dimensional limit is:
\be\label{limit 3d}
-{1\ov \tau} \rightarrow i \infty~, \qquad\qquad {u\ov \tau}~,  {\nu\ov \tau}  \quad {\rm fixed}~.
\ee
The second condition ensures that the three-dimensional parameters---the complexified real masses on $\Mgp$---are kept finite as we send $\beta$ to zero. 
Using the modular properties of the flux, fibering, and handle-gluing operators, it is easy to prove that:
\be\label{small beta limit Z}
\log \CZ_{\Mgp\times S^1} = -{2\pi i \ov \tau} \left((1-g){ \CA^R\ov 12} +{\CA^{\alpha} \ov 12}\left(p{\nu_\alpha \ov \tau}-\n_\alpha\right)\right)+ O(\beta_2^0)~,
\ee
in the limit \eqref{limit 3d}. In addition to \eqref{gen res dPK}, we also have terms involving the gravitational-flavor 't Hooft anomalies, which appear in the presence of background $\GG_F$ vector multiplets. This fully agrees with  \cite{DiPietro:2014bca}. Note that \eqref{small beta limit Z} is invariant under the large gauge transformations $(\nu_\alpha, \n_\alpha)\sim (\nu_\alpha + \tau, \n_\alpha +p)$ for the background $U(1)_\alpha$ gauge field on $\Mgp$.  This is expected from three-dimensional gauge invariance.

\subsection{Dimensional reduction}
\label{sec:dimred}
Having discussed the leading, divergent contribution in the $\beta \rightarrow 0$ limit, we may also consider the finite piece, which can be obtained by subtracting off the contribution \eqref{small beta limit Z}.  In the limit \eqref{limit 3d},  one finds:
\be\label{3d chiral1}
 \pif^\Phi(u; \tau)\Big|_{\text{finite},\, \beta \rightarrow 0} =  e^{-{\pi i\ov2}+\pi i \t u}\, {1\ov 1-\t x} + O(\beta)~,
 \ee
 and:
 \be\label{3d chiral2}
 \CF_1^\Phi(u; \tau)\Big|_{\text{finite},\, \beta \rightarrow 0} =   e^{-{\pi i\ov 2} \t u^2+{\pi i \ov 12}}\,  f_\Phi\left(\t u\right) + O(\beta)~, 
 \ee
 for the four-dimensional chiral multiplet.
Here $f_\Phi$ is the function defined in \eqref{def fphi}, and we defined the variables $\t u \equiv {u\ov \tau}$ and $\t x\equiv e^{2\pi i \t u}$, which is the $S$-transformed $u$ variable kept finite in this limit.~\footnote{The variable $\t u$ in  \protect\eqref{3d chiral1}-\protect\eqref{3d chiral2} should be identified with the three-dimensional $u$ in \protect\cite{Closset:2017zgf}. This is simply because we constructed $\Mgp$ with the first fibering operator $\CF_1$, so that $\beta= \beta_2$. An identical limit can be obtained in term of $u$ if we consider the limit $\beta_1\rightarrow 0$ on $\pif^\Phi$ and $\CF_2^\Phi$, respectively.}
  The finite terms  \eqref{3d chiral1}-\eqref{3d chiral2}  precisely correspond to the contributions of a three-dimensional chiral multiplet to the $\cM_{g,p}$ partition function, in a regularization that preserves three-dimensional parity at the expense of gauge invariance \cite{Closset:2017zgf}.

From these building blocks, one can deduce that, from  a general $4d$ theory with gauge group $\GG$ and chiral multiplets $\Phi_i$ in representations $\FR_i$ of $\Fg$, one obtains in the $\beta \rightarrow 0$ limit the flux, fibering, and handle-gluing operators for the $3d$ theory with gauge group $\GG$ and the same matter content.  Since the divergent factors depend only on anomalies, we find that, for the non-anomalous gauge flux operators:
\be
\pif_a(u, \nu; \tau) = \prod_i \prod_{\rho_i \in \FR_i} \theta\left(\rho_i(u)+\nu_i); \tau\right)^{-\rho_i^a} \;\;\; \underset{\beta \rightarrow 0}{\rightarrow} \;\;\; \prod_i \prod_{\rho_i \in \FR_i}\bigg( \frac{e^{\pi i(\rho_i(\t u)+\nu_i)}}{1- e^{2 \pi i(\rho_i(\t u)+\nu_i)}}\bigg)^{-\rho_i^a}
\ee
with no divergent factor.  Thus the $4d$ Bethe equations descend directly to the $3d$ Bethe equations.  More precisely, the above relation holds when we keep the parameters $\t u$ finite as $\beta \rightarrow 0$. There may also be solutions to the $4d$ Bethe equations which do not stay finite in this limit, and so are not captured by the above three-dimensional limit.  In a favorable case in which  all the four-dimensional Bethe vacua survive in the three-dimensional limit, we directly obtain (suppressing parameter dependence for simplicity):
\bea \label{3dlim} \lim_{\beta \rightarrow 0} e^{-\frac{\epsilon}{\beta}} \cZ_{\cM_{g,p} \times S^1} &= \lim_{\beta \rightarrow 0} e^{-\frac{\epsilon}{\beta}} \sum_{\hat{u}\in\cS_{BE}} \cF^p \, \cH^{g-1}\,  {\Pi_\alpha}^{\n_\alpha} \\
& =  \sum_{\hat{u}\in\cS^{(3d)}_{\rm BE}} {\cF_{(3d)}}^p \, {\cH_{(3d)}}^{g-1} \, {\Pi_{(3d) \alpha}}^{\n_\alpha}  
&= \cZ_{\cM_{g,p}}^{(3d)} \eea
where $\frac{\epsilon}{\beta}$ is the divergent factor \eqref{small beta limit Z}, which we subtract off, and on the last line we obtain the $\cM_{g,p}$ partition function of the $3d$ $\cN=2$ theory with the same gauge group and matter content as the $4d$ theory we started with.  
On the other hand, if some of the four-dimensional Bethe vacua are not captured in the above limit, we may find additional terms in the $\beta \rightarrow 0$ limit of the $\cM_{g,p} \times S^1$ partition function, in addition to the above $\cM_{g,p}$ partition function of the naive $3d$ theory.
The three-dimensional limit of the supersymmetric index has been discussed in detail by  \cite{Ardehali:2015bla, DiPietro:2016ond}, where potentially related issues arose. It would be interesting to compare the two approaches further.

Note that the flavor symmetry parameters we obtain in $3d$ descend from those in $4d$, which were constrained to be non-anomalous---$\cA^{a\alpha \beta} =\cA^{ab \alpha} =0$ for all flavor indices $\alpha,\beta$ and gauge indices $a,b$.  In $3d$ there are no such anomalies, so no {\it a priori} reason to make this restriction.  However, as discussed in \cite{Aharony:2013dha},  $3d$ theories  obtained from $4d$ by dimensional reduction  generally contain  non-perturbative superpotential terms depending on three-dimensional chiral monopole operators, generated by twisted instantons in the $4d$ theory on a circle,  which have the effect of breaking precisely those $3d$ symmetries which are anomalous in $4d$. The more precise statement, therefore, is that the $3d$ theory whose $\cM_{g,p}$ partition function we obtain in \eqref{3dlim} is the theory with these monopole superpotential terms included.  

In section \ref{sec:dualities},  we will discuss pairs of $4d$ dual theories and show that their $\cM_{g,p} \times S^1$ partition functions agree; then the argument above shows that, by taking the $\beta \rightarrow 0$ limit on both sides, the $\cM_{g,p}$ partition functions of their reductions agree, providing strong evidence for their duality in $3d$, as discussed in the case $\cM_{g,p}=S^3$ in  \cite{Aharony:2013dha}.  This statement is however subject to the caveat mentioned above.

\subsection{Modular transformations of the $A$-model}\label{sec: modularity}
Since the four-dimensional $A$-model is defined by compactification on a torus, one expects that is behaves naturally under modular transformations. Let us denote by $S$ and $T$ the generators of the modular group, corresponding to the $SL(2,\Z)$ matrices:
\be
S= \mat{0& 1 \cr -1 & 0}~, \qquad \qquad T= \mat{1 & 1 \cr 0 & 1}~. 
\ee
We have:
\be
S^2 = C~, \qquad (ST)^3= C~,
\ee
with  $C= -{\bf 1}$ the central element.  The modular group acts on  the $A$-model fields and parameters according to:
\be
S\, : \, ( {\bf u}, \tau) \mapsto \left({{\bf u}\ov \tau}, -{1\ov \tau}\right)~, \qquad  \qquad
T\, : \, ({\bf u}, \tau)   \mapsto ( {\bf u}, \tau+1) ~, 
\ee
while $C$ inverts the sign of all chemical potentials, $C\, : \, ({\bf u}, \tau)   \mapsto ( -{\bf u}, \tau)$. In order to give an explicit presentation of $SL(2, \Z)$ on the $A$-model, it is more convenient to consider the generator:
\be\label{def t T}
\t T \equiv CSTS= \mat{1& 0 \cr -1 & 1}~,
\ee
instead of $T$. This gives an equivalent presentation of $SL(2, \Z)$, with 
\be
(S \t T)^3 =C~.
\ee
The action of $S$ and $\t T$ on the $A$-model operators is given by:
\bea\label{STt on operators}
&S[\CF_1]  &=&\;\; e^{{\pi i\ov 3 \tau} \CA^{\bf abc} {\bf u_a u_b u_c}} \;  {\CF_2}^{-1}~,\qquad \qquad
&& \t T[\CF_1] &=&\;\;  { \CF_1}\,{\CF_2}~,\cr
&S[\CF_2]  &=&\;\; e^{-{\pi i\ov 3 \tau^2} \CA^{\bf abc} {\bf u_a u_b u_c}} \; \CF_1~,  \qquad \qquad
&& \t T[\CF_2]  &=&\;  {\CF_2}~,\cr
&S[\pif_{\bf a}] &=&\;\; e^{{\pi i \ov 2}\CA^{\bf a}} e^{ {\pi i\ov \tau}\CA^{\bf abc}{\bf u_b u_c}}\; \pif_{\bf a}~, \qquad \qquad 
&& \t T[\pif_{\bf a}] &=&\;\;  e^{-{ \pi i \ov 6}\CA^{\bf a}}\; \pif_{\bf a}~,\cr
&S[\CH] &=&\;\;  e^{{\pi i \ov 2}\CA^R} e^{ {\pi i\ov \tau}\CA^{R{\bf bc}}{\bf u_b u_c}}\; \CH~,\qquad \qquad 
&& \t T[\CH] &=&\;\;  e^{-{\pi i \ov 6}\CA^R}\; \CH~,
\eea
where all the operators are evaluated at $({\bf u}, \tau)$. The $S$ transformations were already discussed in section \ref{sec: 4dAtwist}. The $\t T$ transformations can also be proven by direct computation, using the results of that section. Given  \eqref{STt on operators}, we can construct the action of any $A\in SL(2, \Z)$ on the $A$-model. For instance, we easily check that:
\be
C[\CF_1]={ \CF_1}^{-1}~, \quad C[\CF_2]={ \CF_2}^{-1}~,\quad
C[\pif_{\bf a}]=(-1)^{\CA^{\bf a}} \, \pif_{\bf a}~,\quad
C[\CH]=(-1)^{\CA^{R}} \, \CH~,
\ee 
for the central element $C$.  
Consider the $A$-model partition function \eqref{Zgp1p2}, which takes the form: 
\be\label{Zgp1p2 bis}
\CZ_{g, \, p_1,\, p_2} = \sum_{\h u \in \CS_{\rm BE}}  {\CF_1}^{p_1}\,  {\CF_2}^{p_2}\, \CH^{g-1} \,{\pif_\alpha}^{\n_\alpha}~,
\ee 
where we suppressed the arguments to avoid clutter.
As we mentioned in section \ref{subsec: fibering T2}, we can use the $SL(2, \Z)$ action on the $A$-model  to set $(p_1, p_2)= (p,0)$. This gives a particular construction of the $\Mgp\times S^1$ partition function, with $\tau$ a complex structure parameter of the four-manifold. Note that the four-dimensional supersymmetric background breaks $SL(2, \Z)$ explicitly unless $p= 0$. By performing an $SL(2, \Z)$ transformation, one simply obtains a different realization of the $\Mgp\times S^1$ partition function.
Related modular properties of the $S^3$ index have been discussed in the literature, in connection with 't Hooft anomalies \cite{Spiridonov:2012ww, Brunner:2016nyk}. In our formalism, these modular properties are simply explained in terms of the $T^2$ compactification necessary to define the four-dimensional $A$-model.~\footnote{We are only explaining a certain $SL(2, \Z)$ action in this way. We have nothing new to say about the more general $SL(3,\Z)$ action that one may define on the index \protect\eqref{index pq} when ${\bf p}\neq {\bf q}$ \cite{1999math......7061F, Spiridonov:2012ww}.}

Finally, let us note that the $\Sigma_g\times T^2$ partition function---that is, the special case $p=0$---enjoys natural modular properties:
\bea
&S[\CZ_{g, \, 0,\, 0}] &=&\; e^{{\pi i \ov 2} \left(\n_\alpha \CA^\alpha + (g-1)\CA^R\right)}\, e^{{\pi i \ov \tau} \left(\n_\alpha  \CA^{\alpha\beta\gamma}\nu_\beta\nu_\gamma + (g-1) \CA^{R\beta\gamma}\nu_\beta\nu_\gamma\right)}\; \CZ_{g, \, 0,\, 0}~,\cr
&\t T[\CZ_{g, \, 0,\, 0}] &=&\;  e^{-{ \pi i \ov 6} \left(\n_\alpha \CA^\alpha + (g-1)\CA^R\right)} \;\CZ_{g, \, 0,\, 0}~,
\eea
which are closely related to the modular transformation properties of the $\CN=(0,2)$ elliptic genus \cite{Benini:2013nda, Benini:2013xpa}. Indeed, $\CN=1$ theories on $\Sigma_g\times T^2$ can be related to $\CN=(0,2)$ theories on the torus, by dimensional reduction on the $\Sigma_g$ factor  \cite{Closset:2013sxa, Gadde:2015wta}.

\section{Freeing the $R$-charge on a trivial $U(1)_R$ bundle}\label{sec: free R}
Our discussion so far described the $A$-model perspective on supersymmetric partition functions. In that approach, it is important that the $R$-charges be integer-quantized so that the A-twisted theory can be defined on any closed Riemann surface $\Sigma_g$. More generally, we  should consider $R$-charges such that $r_i (g-1)\in \Z$, at fixed genus $g$. This Dirac quantization arises because the $U(1)_R$ background gauge field has non-trivial flux:
\be\label{R flux sec4}
{1\ov 2\pi} \int_{\Sigma_g} dA^{(R)}= g-1~,
\ee
across the Riemann surface $\Sigma_g$ on which the A-model is defined.
Here, $A^{(R)}_\mu$ is a connection on $L^{(R)}\cong \b\CK^{-\half}$, with $\b\CK$ the anti-canonical line bundle over $\Sigma_g$. In four dimensions, our supersymmetric background is a pull-back of the two-dimensional background. Our choice of supersymmetry imposes a choice of complex structure on the four-dimensional manifold $\CM_4\cong \Mgp\times S^1$. (We briefly review this in Appendix \ref{app:  SUSY backg}.) In particular, the anti-canonical line bundle of $\CM_4$, denoted $\b\CK$ as well, is the pull-back of the two-dimensional anti-canonical line bundle over $\Sigma_g$, and similarly for the $R$-symmetry line bundles. 

When  $T^2$  is non-trivially fibered over $\Sigma_g$---that is, for $p\neq 0$---, the  $R$-symmetry line bundle over $\CM_4$, denoted $L^{(R)}$, is a torsion line bundle, with first Chern class:
\be
c_1\left(L^{(R)}\right) = g-1\in \Z_p~.
\ee
In particular, whenever:
\be\label{trivial LR condition}
g-1= 0 \mod p~, 
\ee
the $U(1)_R$ line bundle is topologically trivial, and the $R$-charges need {\it not} be quantized. This is the case we will discuss further in this section. The most important instance  of the trivial-bundle condition \eqref{trivial LR condition} is $p=1$, $g=0$, corresponding to the Hopf surface $\CM_4^{q,q}\cong S^3\times S^1$.

\subsection{A tale of two gauges}
Even as we consider a background with $L^{(R)}$ topologically trivial, it need not be trivial as a {\it holomorphic} line bundle. A holomorphic line bundle $L$ over the complex manifold $\CM_4 \cong \Mgp\times S^1$ is determined by the parameters $(\nu, \n)$, with $\nu$ its complex modulus, which is valued in the first Dolbeault cohomology $H^{0,1}(\CM_4, \C)$, and $\n$ its first Chern class valued in $H^2(\CM_4, \Z)$.  The modulus $\nu$ corresponds to a choice of flat connection on the $T^2$ fiber.~\footnote{In general, $L$ is also determined by other complex moduli, but no physical observable depend on them  due to supersymmetry \cite{Closset:2013vra}. Similarly, the line bundle $L$ must be torsion only (for $p\neq 0$) by supersymmetry.} Let us choose the $SL(2, \Z)$ frame in which $\Mgp\times S^1$ is constructed using the first fibering operator, $\CF_1$, so that $S^1_{\beta_1}$ is the circle being non-trivially fibered. A large gauge transformations along $S^1_{\beta_2}$ identifies: $(\nu, \n) \sim (\nu+1, \n)$, while a large gauge transformation along  $S^1_{\beta_1}$ identifies:
\be
(\nu, \n) \sim (\nu+ \tau, \n+p)~.
\ee
A topologically trivial line bundle has $\n= 0$ (mod $p$).

When we specify our supersymmetric background, the choice of gauge for the $U(1)_R$ line bundle must be specified. In most of this paper, we are choosing the so-called ``A-twist gauge'', in which:
\be\label{Atwist gauge}
\nu_R= 0~, \qquad \n_R =g-1~,\qquad  \qquad\qquad \text{ (``A-twist'')}~.
\ee
However, if the condition \eqref{trivial LR condition} is satisfied, we can also take:
\be\label{physical gauge}
\nu_R= \left({1-g\ov p}\right)\tau~, \qquad \n_R =0~,\qquad  \qquad\qquad \text{ (``physical'')}~,
\ee
which is related to \eqref{Atwist gauge} by a large gauge transformation. For lack of a better term, we call \eqref{physical gauge} the ``physical gauge''. On the round $S^3$, this corresponds to a $U(1)_R$ gauge that is trivial along the $S^1$ and has a fixed holonomy along the $S^1$---see Appendix \ref{app:  SUSY backg}.

The choice of gauge is important because it determines the $R$-charge dependence of the supersymmetric partition function \cite{Closset:2014uda}. If we choose $r_i \in \Z$ for all the $R$-charges, either gauge leads to the same partition function (up to  some relative sign). On the other hand, the  physical gauge allows us to easily obtain the correct result for {\it real} $R$-charges. Consider varying the $R$-charge by mixing it with some flavor symmetry $U(1)_F$:
\be\label{R to F mix}
R\rightarrow R+t \, F~,
\ee
with $t$ a mixing parameter. In general, we should take $t\in \Z$, to preserve the Dirac quantization condition on the $R$-charge, but in the case that we are considering now, we may take $t\in \R$. 
The mixing \eqref{R to F mix} corresponds to a tensor product of the $U(1)_F$ line bundle with the $R$-symmetry line bundle,  $L^{(F)} \rightarrow L^{(F)} \otimes (L^{(R)})^t$.
Accordingly, the flavor parameters are shifted according to: 
\be\label{mixing nuR}
\nu_F\rightarrow  \nu_F+ t\, \nu_R~, \qquad \qquad
\n_F\rightarrow  \n_F+ t\, \n_R~.
\ee
In the A-twist gauge \eqref{Atwist gauge}, the flavor modulus $\nu_F$ stays invariant, and the $R$-charge only appears through the background fluxes, including \eqref{R flux sec4}. The shift \eqref{mixing nuR} only makes sense for $t\in \Z$, because $\n_F$ is integer-quantized. In the physical gauge \eqref{physical gauge}, on the other hand, the flavor fluxes stay constant, and we can take $t\in \R$. In this case, the $R$-charge dependence of the partition function appears  through the holomorphic moduli, through the particular combination:
\be\label{nuR in phys gauge}
\nu_i + \nu_R (r_i-1)~, \qquad \qquad \nu_R = \left({1-g\ov p}\right)\tau~,
\ee
for each chiral multiplet $\Phi_i$.
This is the case considered in the literature, in the case of the $S^3$ supersymmetry index, where the $R$-charges appear through the combination $\nu_i + \tau (r_i-1)$ for each chiral multiplet. Let us also define the integer:
\be\label{def lR}
l_R \equiv {1-g\ov p} \in \Z~,
\ee
for future reference.

\subsection{Physical-gauge twisted superpotential and flux operators}
While we lose the straightforward $A$-model interpretation in the physical gauge, we may still define an effective two-dimensional theory for the light degrees of freedom. The effective twisted superpotential takes the form:
\bea\label{W with R}
&\CW^{\rm phys}(u, \nu, \nu_R; \tau) &=&\;\; \sum_i \sum_{\rho\in \FR_i} \CW_\Phi\big(\rho_i(u) + \nu_i+ \nu_R (r_i-1); \tau\big) \cr
&&&\;\;  + \sum_{\alpha\in \Fg} \CW_\Phi\big(\alpha(u) + \nu_R; \tau\big) \cr
&&&\;\;  +  \CW_{\Fh}\big(\nu_R; \tau\big)~.
\eea
Here  the function $\CW_\Phi(u; \tau)$ is the one defined in \eqref{full Wphi}, with its argument shifted according to \eqref{nuR in phys gauge}. In addition to the chiral multiplet contribution, we have the contribution from the W-bosons on the second line. Finally, the $u$-independent term $\CW_{\Fh}(\nu_R; \tau)$ in \eqref{W with R} is the contribution from the vector multiplets along the Cartan $\Fh$ of $\Fg$. We have:
\be\label{CW cartan}
 \CW_{\Fh}(\nu_R; \tau) = \rk \;  \CW_{U(1)}(\nu_R; \tau)~,
\ee
with the $\CW_{U(1)}(\nu_R; \tau)$ the contribution of a $U(1)$ vector multiplet after removing all its zero-modes. We will come back to this term momentarily.
If we set $\nu_R=0$ in \eqref{W with R}, we recover \eqref{full W i} plus the trivial contribution  \eqref{Wvec triviial}.

Let us use the notation ${\bf u_a}= (u_a, \nu_\alpha)$, as in previous sections. The physical-gauge flux operators are naturally defined in terms of \eqref{W with R}, by the formula:
\be
\pif^{\rm phys}_{\bf a}({\bf u}, \nu_R; \tau) = \exp\left({2\pi i {\d\CW^{\rm phys}\ov \d {\bf u_a}}}\right)~.
\ee
One directly obtains:
\be\label{pif phys expl}
\pif^{\rm phys}_{\bf a}({\bf u}, \nu_R; \tau) = \prod_I \pif^\Phi\big(Q_I({\bf u}) + \nu_R (r_I-1); \tau\big)^{Q_I^{\bf a}}~.
\ee
Note that the $W$-boson terms in \eqref{W with R} do not contribute to the gauge flux operators, because of the simple identity:
\be
\prod_{\alpha\in \Fg} \pif^\Phi\big(\alpha(u)+ \nu_R; \tau \big)^{\alpha^a}= (-1)^{2\rho_W^a}= 1~, \qquad \quad \forall a~, \qquad \forall \nu_R \in \tau \Z~, 
\ee
where $\rho_W$ is the Weyl vector, and the last equality is for $\GG$ semi-simple.
The elliptic properties of \eqref{pif phys expl} are similar to \eqref{lgt pif}. We find:
\bea\label{lgt pif phys}
&\pif^{\rm phys}_{\bf a}({\bf u_b}+1, \nu_R; \tau) &=&\;(-1)^{\CA^{\bf ab}} \, e^{-{\pi i\ov \tau}\left(\CA^{\bf abb}  +2 \CA^{\bf abc} {\bf u_c}+2 \CA^{{\bf ab} R} \nu_R \right)} \, \pif^{\rm phys}_{\bf a}({\bf u}, \nu_R; \tau)~,\cr
&\pif^{\rm phys}_{\bf a}({\bf u_b}+\tau, \nu_R; \tau) &=&\; (-1)^{\CA^{\bf ab}} \, \pif^{\rm phys}_{\bf a}({\bf u}, \nu_R; \tau)~.\cr
\eea
Note the appearance of the anomaly coefficient $\CA^{\bf ab R}$, as defined in \eqref{def AR 1}.~\footnote{Importantly,  the anomaly coefficients involving the $R$-symmetry are not integers because $r_I \in \R$. In the special case when all $R$-charges are integer, then $e^{-{2\pi i\ov \tau}  \CA^{{\bf a b}R}\nu_R}= 1$, since $\nu_R \in \tau \Z$.} In particular, in any anomaly-free theory, the gauge flux operators $\pif^{\rm phys}_a$ are elliptic in all their parameters $u_a$, $\nu_\alpha$. It is also modular invariant.

In the special case when the $R$-charges are all integer, the physical-gauge flux operators \eqref{pif phys expl} and $A$-twist-gauge flux operators \eqref{flux op a alpha} are identical  up to a sign:
\be\label{P to P phys sign}
\pif^{\rm phys}_{\bf a}({\bf u}, \nu_R; \tau) = (-1)^{l_R \CA^{{\bf a} R}} \, \pif_{\bf a}({\bf u}, \tau)~,\qquad \quad {\rm if}\; r_I \in \Z~, \; \; \forall I~,
\ee
with $l_R$ defined in \eqref{def lR}. In particular, the two gauge flux operators are exactly identical in that case.

\subsection{The physical-gauge fibering and handle-gluing operators}
We can similarly introduce a generalization of the fibering and handle-gluing operators.  The following results can be derived by considering various one-loop determinants in the physical gauge, as we explain in Appendix \ref{app: subsec phys g}. In the following, we also restrict ourselves to the modular frame \eqref{p1 p2 to p}, for simplicity. This is the frame in which we can most easily compare our results to standard results for the supersymmetric index. 
The physical-gauge fibering operator is given by:
\bea\label{physical fibering op G}
&\CF_1^{\rm phys}({\bf u}, \nu_R; \tau)  &=&\; \;    \exp\left({2\pi i {\d\CW^{\rm phys}\ov \d\tau}}\right) \cr
&&=& \; \;   \prod_I \CF_1^\Phi\big(Q_I({\bf u}) + \nu_R (r_I-1); \tau\big) \cr
&  && \times \prod_{\alpha\in \Fg} \CF_1^\Phi\big(\alpha({\bf u}) + \nu_R; \tau\big)\;\CF_{U(1)}(\nu_R; \tau)^{\rk}~.
\eea
The contribution from \eqref{CW cartan}, for the Cartan of the gauge group,
is given explicitly by:
\be\label{CF U1}
\CF_{U(1)}(\nu_R; \tau) = (-1)^{l_R(l_R+1)\ov 2}  \; \eta(\tau)^{2 l_R}~.
\ee
Let us also define the Hessian determinant of the twisted superpotential \eqref{W with R}:
\be\label{def H phys}
H^{\rm phys}(u, \nu, \nu_R; \tau)\, \equiv\, \det_{ab}{\d^2 \CW^{\rm phys}(u, \nu; \tau)\ov \d u_a \d u_b}\, =\,  \det_{ab}\left( {1\ov 2 \pi i} {\d \log \pif^{\rm phys}_a \ov \d u_b}\right)~,
\ee
which generalizes \eqref{def H}.  On the other hand, there is no effective dilaton in the physical-gauge picture, because $\n_R=0$ in \eqref{physical gauge}.
 
It will be convenient to consider the combination:
\be\label{def G2}
\CG^{\rm phys}({\bf u}, \nu_R; \tau)\,  \equiv\, \CF_1^{\rm phys}({\bf u}, \nu_R; \tau)  \;  H^{\rm phys}(u, \nu, \nu_R; \tau)^{- l_R}~,
\ee
with the Hessian \eqref{def H phys}.
For integer $R$-charges, the operator \eqref{def G2} is equivalent to the product of the fibering and handle-gluing operators in the $A$-twist gauge. More precisely, one can check that:
\be\label{G to H sign}
\CG^{\rm phys}({\bf u}, \nu_R; \tau)  = (-1)^{\half \left[l_R^2 \CA^{RR}+ l_R \CA^R\right]}\, \CF_1({\bf u}; \tau) \, \CH({\bf u}; \tau)^{-l_R}~,\qquad   {\rm if}\; r_I \in \Z~, \; \; \forall I~,
\ee
Here the relative sign is given in terms of the (pseudo-)anomaly coefficients \eqref{def AR 2}-\eqref{def AR 3}.

\subsection{Supersymmetric partition function, Casimir energy and  index}
Given the ingredients introduced above, we can easily construct the supersymmetric partition function. Consider the supersymmetric Bethe vacua, defined by:
\be\label{S BE 4d phys}
\cS_{BE}^{\rm phys} =\left\{ \;\h u_a \; \bigg| \; \pif^{\rm phys}_a(\h u, \nu; \tau) = 1~, \;\; \forall  a~, \quad w \cdot \h u \neq \h u, \;\; \forall w \in W_\GG \; \right\} /W_\GG~.
\ee
These vacua are completely equivalent to the $A$-model vacua \eqref{S BE 4d}. Given a solution $\{\h u_a\}$ to \eqref{S BE 4d}, we obtain a solution to \eqref{S BE 4d phys} by the substitution:
\be
\h u_a(\nu_i) \rightarrow \h u_a(\nu_i + \nu_R (r_i-1))~.
\ee
The ``physical'' partition function is given by:
\be\label{ZMgp phys}
\CZ_{\Mgp \times S^1}^{\rm phys}(\nu, \nu_R; \tau) =\sum_{\h u \in \CS_{\rm BE}^{\rm phys}}  \CG^{\rm phys}(\h u,\nu, \nu_R; \tau)^{p}\, \prod_{\alpha} \pif^{\rm phys}_\alpha(\h u,\nu, \nu_R; \tau)^{\n_\alpha}~.
\ee
For  integer $R$-charges and whenever \eqref{trivial LR condition} holds true,  \eqref{ZMgp phys} is equal to \eqref{ZMgp} up to a sign, as follows from \eqref{P to P phys sign}, \eqref{G to H sign}, and from the total ellipticity of the gauge flux operators.

\paragraph{Supersymmetric Casimir energy.} One can again consider a factorization of \eqref{ZMgp phys} into a ``normalized index'' and a supersymmetric Casimir energy contribution:
\be\label{index E fact phys}
\CZ_{\Mgp \times S^1}^{\rm phys}(\nu, \nu_R; \tau) = e^{2\pi i \tau E_{\Mgp}(\nu, \nu_R; \tau)}\; \CI_{\Mgp }^{\rm phys}(\nu, \nu_R; \tau)~.
\ee
The supersymmetric Casimir energy is given by:
\be
 E_{\Mgp}(\nu, \nu_R; \tau) = p  E_{\Mgp}^{(\CG)}(\nu, \nu_R; \tau) +\sum_\alpha \n_\alpha E_{\Mgp}^{(\alpha)}(\nu, \nu_R; \tau)~, 
\ee
with
\bea\label{Casimir E phys}
&E_{\Mgp}^{(\CG)}   &=& \;\;{1\ov 6 \tau^3}\CA^{\alpha\beta\gamma}\nu_\alpha\nu_\beta \nu_\gamma
- {1\ov 12\tau} \CA^\alpha \nu_\alpha 
+{1\ov 2 \tau^3}\CA^{\alpha\beta R}\nu_\alpha\nu_\beta\nu_R+{1\ov 2 \tau}\CA^{\alpha RR}\nu_\alpha\nu_R^2  \cr
&&&\; \; + {1\ov 6 \tau^3}\CA^{RRR} \nu_R^3 -{1\ov 12 \tau}\CA^R \nu_R~,\cr
& E_{\Mgp}^{(\alpha)}   &=& \;\; -{\CA^{\alpha\beta\gamma}\ov  2\tau^2} \nu_\beta \nu_\gamma -{\CA^{\alpha\beta R}\ov  \tau^2} \nu_\beta \nu_R  -{\CA^{\alpha RR}\ov  2\tau^2} \nu_R^2 -{\CA^\alpha\ov 12}~.
\eea
The expression $E_{\Mgp}^{(\CG)}$ in \eqref{Casimir E phys} reproduces exactly the three-sphere supersymmetric Casimir energy \cite{Assel:2015nca, Bobev:2015kza}, as one would expect. In particular, if we set the flavor chemical potential to zero and choose the $R$-charge to the superconformal $R$-charge in the infrared, we get:
\be
E_{\Mgp}^{(\CG)}(0, \nu_R; \tau)= {8\ov 27} \big(5 a- 3 c\big)l_R^3 - {4\ov 3} \big(a-c\big)l_R~,
\ee
with $a$, $c$ the four-dimensional conformal anomalies. Setting $l_R=1$, we reproduce the correct result for the round $S^3$ \cite{Assel:2015nca}.

\paragraph{The generalized index.}  Comparing \eqref{index E fact phys} and  \eqref{ZMgp phys}, we read off the complete expression for the generalized $\Mgp$ index in the physical gauge. It is given by:
\be\label{Index phys}
\CI_{\Mgp }^{\rm phys}(\nu, \nu_R; \tau) =\sum_{\h u \in \CS_{\rm BE}^{\rm phys}}  \CJ_{\CG}^{\rm phys}(\h u,\nu, \nu_R; \tau)^{p}\, \prod_{\alpha} \CJ_{\pif_\alpha}^{\rm phys}(\h u,\nu, \nu_R; \tau)^{\n_\alpha}~,
\ee
with:
\bea\label{def J index phys}
& \CJ^{\rm phys}_\CG(u, \nu, \nu_R)  =\;  \prod_i\prod_{\rho_i}  \Gamma_0\big(\rho_i(u)+ \nu_i+\nu_R(r_i-1)\big)  \, \prod_{\alpha\in \Fg} \Gamma_0\big(\alpha(u)+\nu_R\big)\cr
&\qquad\qquad\qquad\;\; \times \left[(-1)^{l_R(l_R-1)\ov 2} e^{-{\pi i \tau \ov 3} l_R (l_R^2-1)}\, (q; q)_\infty^{2 l_R} \right]^\rk \,  H^{\rm phys}(u, \nu, \nu_R)^{- l_R}~,\cr
& \CJ^{\rm phys}_{\pif_\alpha}(u, \nu, \nu_R) =\; e^{\pi i \left(\CA^{\alpha \beta}\nu_\beta +\CA^{\alpha R}\nu_R \right)} \prod_i\prod_{\rho_i} \theta_0\big(\rho_i(u)+ \nu_i+ \nu_R(r_i-1)\big)^{-\omega^a_i}~. 
\eea
We suppressed some of the $\tau$ dependence in \eqref{def J index phys} to avoid clutter. Note the appearance of a subtle phase for each Cartan element, on the second line of \eqref{def J index phys}. This follows from the computation of Appendix \ref{app: subsec phys g}.

\subsection{A new evaluation formula for the $S^3$ index.}
Let us consider the important special case $p=l_R=1$, $\nu_R=\tau$, which corresponds to the round three-sphere background.  According to  \eqref{Index phys}, the $S^3$ supersymmetric index can be written as a sum over Bethe vacua:
\be\label{Bethe vacua formula index}
\CI_{S^3}^{\rm phys}(\nu; \tau) = \sum_{\h u \in \CS_{\rm BE}^{\rm phys}}   \CJ_{S^3}(\h u, \nu; \tau)~.
\ee
The summand $ \CJ_{S^3}$ can be written suggestively as:
\be\label{JS3}
 \CJ_{S^3}(u, \nu; \tau) =\;  (q; q)_\infty^{2\rk}  {\prod_i\prod_{\rho_i}  \Gamma_0\big(\rho_i(u)+ \nu_i+\tau(r_i-1); \tau\big)\ov \prod_{\alpha\in \Fg} \Gamma_0\big(\alpha(u)-\tau; \tau\big)}\; {1\ov H^{\rm phys}(u, \nu, \tau; \tau)}~,
\ee
where we used the inversion formula $\Gamma_0(u; \tau)= \Gamma_0(-u; \tau)^{-1}$ to put the $W$-boson contribution in the denominator. The summand \eqref{JS3} is  precisely the integrand of the Romelsberger integral for the $S^3$ index \cite{Romelsberger:2007ec, Dolan:2008qi}, multiplied by  $(H^{\rm phys})^{-1}$.

In the next section, we will prove that Bethe-vacua formula \eqref{Bethe vacua formula index} for the round three-sphere index precisely agrees with the Romelsberger index (specialized to ${\bf p}={\bf q}=q$). We will also present a generalization of the integral formula for any $\Mgp \times S^1$ partition function.


\section{Integral formulas and Bethe equations}\label{sec: integral form and BE}

In this section, we present explicit integral formulas for the $\cM_{g,p} \times S^1$ partition function. They can be derived by supersymmetric localization, similarly to the three-dimensional case studied in \cite{Closset:2017zgf}---see also \cite{Benini:2013nda, Benini:2013xpa, Benini:2015noa, Benini:2016hjo, Closset:2016arn, Closset:2017zgf}.  We will only sketch this derivation, insisting on the specificities of the four-dimensional case.  We will also relate this result to the standard integral formula for the $S^3$ index.

These integral formulas can argued to be equivalent to the sum-over-Bethe vacua formula of the previous sections, at least {\it formally}. The important caveat is that, for genus $g$ larger than zero, the $W$-bosons introduce additional subtle contributions, which we will not address systematically here. The net effect of these contributions is to restrict the sum over Bethe vacua to the physical abelian vacua, discarding any potential contribution from supersymmetry-breaking non-abelian vacua (that is, the would-be solutions of the Bethe equations that are not acted on freely by the Weyl group). Similar discussions appeared in \cite{Benini:2016hjo, Closset:2016arn}.

\subsection{Localization to a contour integral}

The $\cM_{g,p} \times S^1$ partition function of an $\CN=1$ gauge theory can be computed directly by localization with the UV action, in principle.  Let us consider the case $p \neq 0$, unless otherwise stated. The case $p=0$ was studied in \cite{Benini:2016hjo}.  In this section, we consider the $\Mgp\times S^1$ background corresponding to the first fibering operator, $\CF_1$.

To deal with the vector multiplet, we follow the abelianization method of Blau and Thompson \cite{Blau:1993tv, Blau:1994rk, Blau:2006gh}. The gauge field can be localized to flat connections along the $T^2$ fiber, which can  be conjugated to the Cartan torus $\GH \subset \GG$:
\be
u_a = \tau a^a_1- a^a_2~, \qquad a^a_i\equiv \frac{1}{2 \pi} \int_{S^1_{\beta_i}} A^a_\mu dx^\mu~, \quad\qquad  i=1, 2~, \quad a=1, \cdots, {\rm rk}(\GG)~.
\ee
In addition to these, we also have flat connections along the  base of the Riemann surface:
\be\label{aSigmag}
a_{\Sigma_g} = \sum_{I=1}^g \alpha^a_I w_z^I dz + \t \alpha^a_I w_{\bar z}^I d\bar z~, \qquad [w^I]\in H^1(\CM_{g,p},\mathbb{R})~, \qquad I=1, \cdots,g~.
\ee
The one-form $w^I$ are pulled-back from the $g$ closed one-forms on $\Sigma_g$.~\footnote{Moreover, the holomorphic one-form $w_z^I dz$ also pull-back to representatives of the four-dimensional first Dolbeault cohomology of $\CM_4 \cong \Mgp \times S^1$.}
The parameters $\alpha^a, \t\alpha^a$ live on a compact domain. The supersymmetric equations also allows for torsion bundles over $\CM_{g,p}$. For $p \neq 0$, we should sum over  $\frak{m}$ valued in:
\be\label{top sectors}
\Gamma_{\mathbf{G^\vee}}^{(p)} = \{ \frak{m}  \in \frak{g}: \rho(\frak{m})\in \mathbb{Z}\ , \forall\rho\in \Gamma_{\GG} \}/\{ \frak{m} \in \frak{g} : \rho(\frak{m})\in p \mathbb{Z} \} \cong \mathbb{Z}_p^\rk~.
\ee
This sum over topological sectors is the price to pay for abelianization over $\Mgp$ \cite{Blau:2006gh}.

For $p=0$, $u_a= \tau a^a_1 - a^a_2$ take values in the torus $T^{2\rk}$, due to the identifications $a^a_1 \sim a^a_1+1$ and $a^a_2 \sim a^a_2 \sim 1$. For $p \neq 0$, on the other hand, $S^1_{\beta_1}$ is a torsion one-cycle, so that $a_1^a$ takes values in $\mathbb{R}$. More precisely, 
as we explained in \cite{Closset:2017zgf}, $a_1^a$ receives a topologically non-trivial contribution if $A_\mu$ is the connection of a torsion bundle $L$. For any $U(1) \subset \GH$, have:
\be
a_1 =  \h a_1 + a_1^{\rm (flat)}~, \qquad \h a_2 \in \R~, \qquad e^{-i \int_{S^1_{\beta_1}} A^{(\text{flat})} }= e^{2\pi i {\m \ov p}}~,
\ee
where $\m \in \Z_p$ the first Chern class of $L$. This implies  the identifications:
\be\label{lgt sec 5}
(u_a,\frak{m}_a) \sim (u_a+\tau,\frak{m}_a+p) \sim (u_a+1,\frak{m}_a)~,
\ee
under large gauge transformations along $S^1_{\beta_1}\times S^1_{\beta_2}$, with the complexified flat connections $u_a$ valued in  $\mathbb{C}^{\rk}$, and the integer-valued torsion fluxes $\m_a$.

These flat connections have fermionic superpartners, which consist of scalar and one-form zero-modes for the gauginos on the $A$-twist background. 
In order to regulate the singularities of the one-loop determinant, we also turn on some constant modes $D_0$ for the auxiliary field $D$ in the abelianized vector multiplets.   
After careful integration over the fermionic zero modes and over $D_0$, we find the expression:
\be \label{intform} 
\cZ_{\cM_{g,p} \times S^1}(\nu) = \frac{1}{|W_\GG|}\sum_{\m \in \Z_p^\rk}\int_{\cC^\eta} \prod_{a=1}^\rk du_a  \; 
I_{\frak m} (u,\nu)~.
\ee
Here the sum is over the topological sectors \eqref{top sectors}, indexed by $\m$. For each $\m$, the  integrand reads:
\be\label{general integrand}
I_{\frak m}(u,\nu)=\CF_1(u, \nu)^p\,  \CH(u,\nu)^{g-1}\,  H(u,\nu)\,  \prod_{a=1}^\rk \pif_a(u,\mu)^{\m_a}\,  \prod_{\alpha=1}^{{\rm rk}(\Fg_F)} \pif_\alpha(u,\nu)^{\n_\alpha}~, 
\ee
in terms of the fibering, handle-gluing, gauge flux and flavor flux operators, and with $H$ the Hessian determinant \eqref{def H}. This integrand can also be written as:
\be
I_{\frak m}(u,\nu) = \CZ^\oneloop(u,\nu)\, H(u,\nu)^g~,
\ee
where $\CZ^\oneloop$ is a one-loop determinant around the supersymmetric background corresponding to $(u, \m)$ and $(\nu, \n)$---as discussed in Appendix \ref{App: oneloop dets}---, while the Hessian to the power $g$ comes from integrating out the $g$ gaugino one-form zero-modes on $\Mgp\times S^1$.
The prefactor $|W_\GG|$ in \eqref{intform} is the order of the Weyl group of $\GG$.

\subsection{The Jeffrey-Kirwan contour integral}
The remaining ingredient necessary to properly define \eqref{intform} is the integration contour $\CC^\eta$, which is a certain middle-dimensional contour inside $\{u_a\} \cong (\C^*)^\rk$. Its precise determination is highly non-trivial. For simplicity, let us focus on the case of a rank-one gauge group, although we expect that a similar story holds more generally.

\begin{figure}[t]
\begin{center}
\includegraphics[width=13cm]{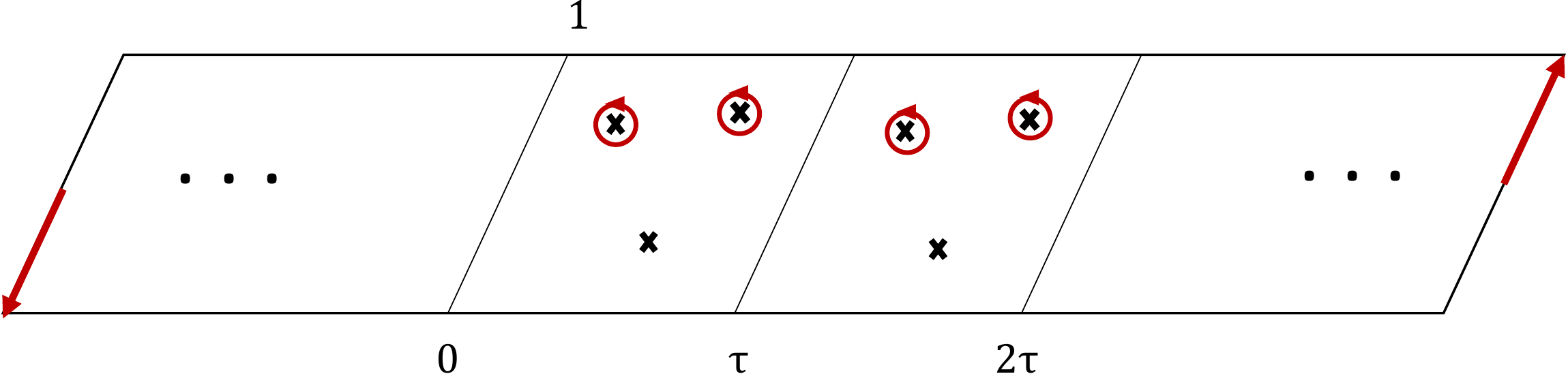}
\caption{Integration contour $\CC_R^\eta$ for $p>0$ in the rank-one case. The contour in the bulk is determined by the JK residue. Two vertical segments represent the potential singularities at $u = \rightarrow \pm \tau \infty$. } 
\label{domain1}
\end{center}
\end{figure}

\begin{figure}[t]
\begin{center}
\includegraphics[width=14cm]{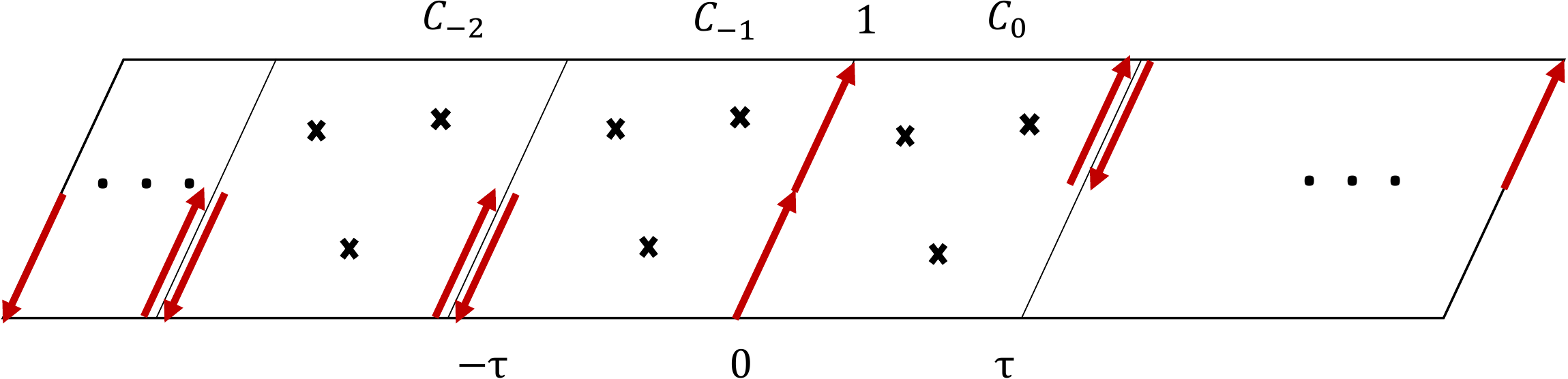}
\caption{Decomposition of the boundary integral $\CC^\eta_R = \sum_{k\in \mathbb{Z}}\CC^\eta_k$. By an appropriate choice of the $R$-charges and of $\eta$, the contour in the bulk cancel each other and only the unit circle contour $a_1=0$ remains.} 
\label{decomposition1}
\end{center}
\end{figure}

Following  \cite{Benini:2015noa, Benini:2016hjo, Closset:2016arn}, the contour $\CC^\eta$ in \eqref{intform} 
can be related to the Jeffrey-Kirwan (JK) residue integral \cite{JK1995, 1999math......3178B}.
More precisely, the contribution of the  ``bulk'' singularities---that is, poles of the integrand at finite values of $u$---, corresponding to the matter chiral multiplets, are captured by JK residues at those singularities, with auxiliary parameter $\eta \in i \Fg^\ast$.
 There  are also potential singularities at the ``boundary'' $u_a \rightarrow \pm \tau \infty$.  We have shown the schematic picture for the rank-one case in Figure \ref{domain1}.  We may understand these boundary contributions by cutting off the integral at $a_1 = R$ for some large $R \in \R_{>0}$, which we take to infinity at the end of the calculation.  Following the discussion of the three-dimensional case \cite{Closset:2017zgf}, one can show that the regulated contour $\CC_R^\eta$ is given by:
\be\label{CR def}
\CC_R^\eta = \big\{u \in \partial \hat{\frak{M}}_R ~\big|~ \text{sign(Im}(\partial_u\CW)) = -\text{sign}(\eta)\big\}\ ,
\ee
where $\partial \hat{\frak{M}}_R$ is the contour that encircles all the bulk and boundary singularities. 
The integrand  \eqref{general integrand} transforms as:
\be
I_{\frak m}(u+ k \tau,\nu)  = \pif(u, \nu)^{-k p} \, I_{\frak m}(u,\nu)= I_{\frak m- k p}(u,\nu)~, \qquad \qquad \forall k\in \Z~,
\ee
 under large gauge transformations \eqref{lgt sec 5} along $S^1_{\beta_1}$, with $\pif$ the gauge flux operator. This directly follows from the properties of the fibering operator $\CF_1$ for a non-anomalous gauge theory. Using this relation, we can write:~\footnote{Here the sum over $k \in \Z$ is understood to be cut off at $|k|=R$, which we take to infinity at the end of the calculation.}
\bea \label{boundary integral}
\sum_{\m\in\mathbb{Z}_p} \oint_{\CC_R^\eta}\frac{dx}{2\pi i x} I_{\frak m}(u, \nu)
&=\frac{1}{|W|}\sum_{k\in \mathbb{Z}}\sum_{\m \in \Z_p}\int_{\cC^\eta_k}  du \; I_{\frak m}(u, \nu)\cr
&= \frac{1}{|W|}\sum_{k\in \mathbb{Z}}\sum_{\m \in \Z_p}\int_{\cC^\eta_0} du  \; I_{\frak m- k p}(u,\nu)\cr
&=\frac{1}{|W|}\sum_{\m \in \Z} \int_{\cC^\eta_0} du \; I_{\m}(u,\nu)~.
\eea
Here, we defined $\CC_k^\eta$ the contour $\CC^\eta$ restricted to the region $a_1 \in [k, (k+1)]$,  such that:
\be\label{decomposition}
\CC^\eta_R = \sum_{k\in \mathbb{Z}}\CC^\eta_k~,
\ee
as depicted in Figure~\ref{decomposition1}.
The contour $\CC_0^\eta$ in the last expression is the ``JK-contour" restricted to the fundamental domain of the torus.  Generalizing to higher rank, we expect similar boundary contributions  \cite{Closset:2017zgf}, and we conjecture a formula:
\be \label{zmgpmsum}
\cZ_{\cM_{g,p}\times S^1}(\mu_i,s_i) =\frac{1}{|W|}\sum_{\m \in \Z^\rk} \int_{\cC^\eta_0} \prod_{a=1}^\rk du_a  \; 
I_{\frak m}(u,\nu)~,
\ee
In the case $p=0$, the boundary contributions are trivial, and  \eqref{zmgpmsum} agrees with the formula for the $\Sigma_g \times T^2$ partition function in \cite{Benini:2016hjo}. 

More generally, the manipulation \eqref{boundary integral} brings the $\Mgp\times S^1$ partition function \eqref{intform} to a form analogous to the $p=0$ case, including the sum over all gauge fluxes on $\Sigma_g$ in  \eqref{zmgpmsum}.  At the level of these integral formulas, this simply corresponds to the relation \eqref{rel Mgp to S2T2 intro} between spaces of different topologies.

\subsubsection{The  $W$-boson contribution.}  The above analysis holds with one very important caveat. So far, we did not mention the potential singularities of the integrand \eqref{general integrand} originating from poles in the one-loop determinant for the $W$-bosons, for any $\GG$ non-abelian. Such $W$-bosons singularities exist if and only if $g>1$.

A deeper study of this important issue, which would lead to a more rigorous supersymmetric localization argument in the UV, is beyond the scope of this paper. For our purposes, it will be enough to  note   that the Bethe-vacua result can be obtained by naively summing over topological sectors with a JK residue that includes the matter chiral multiplets singularities only (plus contributions from the boundary), thus obtaining a naive sum over Bethe vacua of the abelianized theory, and then excluding by hand the would-be Bethe vacua not acted on freely by the Weyl group, as in \cite{Benini:2016hjo}.

\subsection{The unit-circle contour integral}

Starting from the formula \eqref{intform} for $p \neq0$, we can also deform the contour $\CC^\eta$ to a simpler one, which can be related to previous results in the literature \cite{Romelsberger:2007ec, Assel:2014paa}.  Under a certain assumption on the flavor and R-symmetry backgrounds, which we will specify below, and for a suitable choice of $\eta$ for each $k$, we claim that:  
\be\label{unit circle}
 \CC^\eta \cong \sum_{k \in \Z^{\rk}} \cC^{\eta(k)}_k \cong \prod_a \mathbb{T}_{x_a}~,
\ee
where $\mathbb{T}_{x_a}$ is the unit circle in the $x_a$-plane, where $x_a=e^{2 \pi i u_a}$.
This correspond to an integration over flat connections $a_2$ over the $S^1$ in $\Mgp \times S^1$. Interestingly, this is also the ``naive'' localization result one would obtain by imposing the standard reality conditions for the fields, so that we localize on four-dimensional flat connections, $F_{\mu\nu}=0$---see in particular \cite{Assel:2014paa}. For $p\neq 0$, such flat connections include the connections of non-trivial flat torsion bundles \eqref{top sectors}, which we sum over.

  For the rank-one case, the equivalence \eqref{unit circle} can be shown by taking:
\be
\eta(k)>0\text{ for }k\geq 0~,\quad  \text{ and }\quad \eta(k)<0\text{ for }k<0.
\ee
Then, as illustrated in Figure \ref{decomposition1}, all the vertical segments in the bulk cancel each other except for the contour at $a_1 =0$. The two boundary segments at either end do not contribute, because the integrand vanishes as we take them to infinity.  There remain potential bulk contributions, but we claim that these do not contribute when a certain condition---see \eqref{scassum} below---holds.  The remaining contour is precisely the unit circle contour \eqref{unit circle}. We claim that the equivalence holds for the higher-rank theories as well. We then have:
\be \label{zmgpuc}
 \cZ_{\cM_{g,p \neq 0}\times S^1} (y) = \frac{1}{|W_\GG|} \sum_{\m \in \Z_p^\rk}\oint_{\prod_a \mathbb{T}_{x_a}} \prod_{a=1}^{\rk} \frac{dx_a}{2\pi i x_a}  \, I_{\frak m}(x,y)~, 
 \ee
where we wrote the integrand \eqref{general integrand} as a function of $x_a$ and $y_\alpha= e^{2\pi i \nu_\alpha}$.  

We can also argue for the formula \eqref{zmgpuc}  by directly relating it to the Bethe-vacua formula \eqref{ZMgp}.  This argument is analogous to the one in \cite{Closset:2017zgf} for the three-dimensional $\cM_{g,p}$ partition function. Let us again consider the rank-one case, for simplicity.  
Starting from \eqref{zmgpuc}, we may perform the sum over $\m$ explicitly:
\be
\sum_{\m=0}^{p-1}  I_{\frak m}(x,y) =  {1- \pif(x,y)^p\ov 1- \pif(x,y)} \, \CF_1(x,y)^p\,  \CH(x,y)^{g-1}\,  H(x,y)~.
\ee
Here and in the following,  $\pif$ is the gauge flux operator, while we omitted the flavor flux operators $\pif_\alpha$ to avoid clutter.
Using the difference equation $\pif_u(x, y) \CF_1(x, y)= \CF(q^{-1}x, y)$, we can rewrite \eqref{zmgpuc} as:
\bea\label{contrw}
 &\cZ_{\cM_{g,p \neq 0}\times S^1} & =&\, \frac{1}{|W_\GG|}  \int_{|x|=1} {dx\ov 2 \pi i x}  \, { \CF_1(x,y)^p - \CF_1(q^{-1} x, y)^p  \ov 1- \pif(x, y) }\CH(x,y)^{g-1}\,  H(x,y)\cr
&& =&\,  \frac{1}{|W_\GG|}  \int_{\Delta} {dx\ov 2 \pi  x}  \, {I_0(x,y)\ov 1- \pif(x,y)}~,
 \eea
 where $I_0(x, y)$ is the integrand \eqref{general integrand} at $\m=0$. Here, $\Delta$ is the difference of the contour at $|x|=1$ and the contour at $|x|=|q|^{-1}$.  Here we have used the fact that all factors in the integrand, apart from $\CF(x, y)$, are invariant under $x \rightarrow qx$.  As noted above, this relies on the absence of any anomalies for the gauge symmetry.

This contour integral is equal to the sum of the residues of all poles contained in the region $1<|x|<|q|^{-1}$.  These poles may come from the numerator or denominator.  However, we claim that, if we assume the relation \eqref{scassum}  below, there are no poles in this region coming from the numerator.   We will argue for this momentarily. Assuming it is true, the only poles in \eqref{contrw} that lie inside this region are those at $\pif(x, y)=1$.  These are precisely the solutions to the ``naive'' Bethe equation (including, potentially, non-abelian vacua that one should discard).  We thus find:
\be\label{Z res to BE inter}
 \cZ_{\cM_{g,p} \times S^1}(y) =- \frac{2 \pi i}{|W_\GG|}  \sum_{\hat{x} \; | \; \pif(\hat{x}, y)=1}  \text{Res}_{x = \hat{x}} \frac{\CF(x,y)^p \CH(x,y)^{g-1} H(x,y)}{1-\Pi(x,y)}~. 
 \ee
Using \eqref{def H}, one finds the residue of the $\frac{1}{1-\Pi}$ factor cancels the factor of $H$, and therefore:
\be
-2\pi i \text{Res}_{x = \hat{x}} \frac{\CF(x,y)^p \CH(x,y)^{g-1} H(x,y)}{1-\Pi(x,y)}  
 =\CF(\h x,y)^p \CH(\h x,y)^{g-1}~.
\ee
A similar argument holds for higher-rank theories. 
The acceptable Bethe  solutions $\{\h x_a\}$ to $\pif_a(\h x, y)=1$ come in Weyl-group orbits of maximal size $|W_\GG|$, canceling the Weyl-symmetry factor in \eqref{Z res to BE inter}. The contribution from the non-acceptable solutions, which are not acted on freely by the Weyl group, should be discarded by hand. Reinstating the flavor flux operators, this leaves us with:
\be \label{ans}
 \cZ_{\cM_{g,p} \times S^1}(y) = \sum_{\hat{x} \in \cS_{BE}} \cF(\hat{x}, y)^p \, \cH(\hat{x}, y)^{g-1}\, \pif_\alpha(\h x, y)^{\n_\alpha}~,
 \ee
which is precisely the sum-over-Bethe-vacua formula \eqref{ZMgp}.

In this sense, the relation between the integral formula and the Bethe-vacua formula is only formal, as we already anticipated. For $g=0$, the contribution from the non-acceptable Bethe vacua vanishes, because $\CH^{-1}$ vanishes on such solutions---for $g>1$, on the other hand, it would diverge. Since the Bethe-vacua formula has a rather simple derivation and itself passes a number of physical consistency checks, the open challenge would be to specify the exact contour $\CC^\eta$ in \eqref{intform}, valid for any $g>0$, that reproduces the Bethe-vacua answer without further assumption. We leave this for future work.

This important remark notwithstanding, it remains to derive the conditions under which there are no poles coming from the numerator in \eqref{contrw}.  Apart from the Hessian factor, a single chiral multiplet of gauge charge $Q$ and $R$-charge $r$ contributes to this numerator as:
\bea
&Z^\Phi_{\m=0}(x)&=&\; \cF^\Phi(x^Q y)^p\,  \Pi^\Phi(x^Q y)^{ \n + (r-1)(g-1)}\cr
& \, &\sim\,&\;
 \prod_{m \in \Z} \left[1\ov 1 - q^m x^Q y\right]^{p m  + \n + (r-1)(g-1)}~, 
\eea
in the $A$-twist gauge---see Appendix \ref{App: oneloop dets}. Here, $y$ and $\n$ are the net flavor parameters the chiral multiplet couples to.
  This expression can have a pole when $x^Q = y^{-1} q^{-m}$ for some $m \in \Z$.  In addition, the chiral multiplet contributes a simple pole at the same points to the Hessian determinant, $H(x)$.  Generically, this is the only source of a pole at this point, so the behavior of the numerator $I_0(x)$  here is given by:
\be 
I_0(x) \underset{x^Q \sim y^{-1} q^{-m} } \sim Z^\Phi_{\m=0}(x)\, H(x)^g \sim  (1 - q^n x^Q y)^{-(p  m+ \n + (r-1)(g-1) + g)}~.
\ee
Let us fix $p>0$, for definiteness.
Picking $m$ so that the exponent is negative, one finds that a pole can only arise when:
\be\label{xQqrel}
 |x|^Q \geq |y|^{-1} |q|^{{\n+r(g-1) \ov p}}~.
 \ee
Now, we want to impose that there are no poles of the integrand \eqref{contrw} coming from $I_0$, in the region $1<|x|<|q|^{-1}$, for this chiral multiplet. For $Q<0$, we should pick the parameters so that the RHS of \eqref{xQqrel} is bounded below by $1$.  For $Q>0$, we would naively bound the RHS below by $|q|^{-Q}$; however, since the denominator of \eqref{contrw} also blows up at these points, we may actually allow sufficiently mild poles in the numerator, and one finds that RHS need only be bounded below by $1$.  Summarizing, in either case, we find the condition:
\be \label{scassum}
 |y| \; |q|^{-{\n+r(g-1)\ov p}}  <  1~.
\ee
If we impose this relation for all the chiral multiplets in the theory, then the numerator contributes no poles at all to  \eqref{contrw}, and this completes the proof. 
The  restriction \eqref{scassum} implies that we may only perform this calculation in some subset of parameter space, and this may not be compatible with the physical constraints imposed by the superpotential and anomaly cancellation.  However, having performed the computation here, we may then extend it to the rest of parameter space by analytic continuation.  Moreover, by starting in such a region of parameter space and continuously varying parameters, one may deform the contour such that it is not crossed by any poles, and one may, in principle, obtain the correct integration contour for any point in parameter space in this way.  The argument for higher rank is a straightforward extension \cite{Closset:2017zgf}.

An interesting special case of \eqref{scassum} is when we consider real chemical potentials only, so that $|y|=1$. In that case, the condition is simply that $\n + r(g-1)<0$. In particular, on the three-sphere with $\n=0$, we simply need all the chiral multiplet $R$-charges to be positive, $r>0$.

\subsection{The three-sphere supersymmetric index}
In the case $\cM_{g,p} = S^3$, we may also work in the physical gauge and at  real $R$-charges, $r\in \R$, as explained in section \ref{sec: free R}. Consider the physical gauge for any $\Mgp$ such that \eqref{trivial LR condition} holds. The chiral multiplet one-loop determinant can have poles at $x^Q= y^{-1} q^{-m - l_R(r-1)}$, with $l_R$ defined in \eqref{def lR}, for $m$ such that $p m + \n >0$. By the same reasoning as above for $p>0$, this leads us to the same condition as in \eqref{scassum},  with the difference that $r$ can be real. In particular, for $S^3$, we should have:
\be\label{S3 phys cond}
|y_i|\; |q|^{r_i}  <1~,
\ee
for every chiral multiplet $\Phi_i$.
The integral formula \eqref{zmgpuc} then reads:
\be
\CZ^{\rm phys}_{S^3 \times S^1}(y; q) = \frac{1}{|W_\GG|}  \oint_{\prod_a \mathbb{T}_{x_a}} \prod_{a=1}^{\rk} \frac{dx_a}{2\pi i x_a}  \, \CF_1^{\rm phys}(x,y; q)~, 
\ee
where we have a single topological sector because $p=1$. The integrand is the ``physical'' fibering operator \eqref{physical fibering op G} for $\nu_R= \tau$. Once we strip away the supersymmetric Casimir energy term,
\be
\CZ^{\rm phys}_{S^3 \times S^1}(y; q)  = q^{E_{S^3}} \, \CI_{S^3}^{\rm phys}(y; q)~,
\ee
with $E_{S^3}$ given by the first equality in \eqref{Casimir E phys} with $\nu_R= \tau$, we are left with the index formula:
\be\label{S3 index integral expl}
 \CI_{S^3}^{\rm phys}(y; q) = \frac{(q; q)^{2\rk}_\infty}{|W_\GG|}  \oint_{\prod_a \mathbb{T}_{x_a}}  \prod_{a=1}^{\rk} \frac{dx_a}{2\pi i  x_a} \,   {\prod_i\prod_{\rho_i}  \Gamma_0\big(\rho_i(u)+ \nu_i+\tau(r_i-1); \tau\big)\ov \prod_{\alpha\in \Fg} \Gamma_0\big(\alpha(u)-\tau; \tau\big)}~,
\ee
with $y= e^{2 \pi i}$, $q= e^{2\pi i \tau}$. This is the standard expression for the $S^3$ supersymmetric index \cite{Romelsberger:2007ec, Dolan:2008qi} with ${\bf p}={\bf q}=q$. From \eqref{S3 phys cond}, we see that this formula is valid, in particular, for real chemical potentials $\nu_i$ and positive $R$-charges, $r_i >0$. The contour integral in \eqref{S3 index integral expl} has a standard interpretation as a projection onto gauge-invariant states. For more general parameters, we can deform the contour appropriately, as mentioned above.

Finally, we note that our simple proof of the equality between the integral formula  and the Bethe-vacua formula goes through without problem, so that \eqref{S3 index integral expl}  is exactly equal to \eqref{Bethe vacua formula index}.

\section{Supersymmetric dualities}
\label{sec:dualities}

The exact computation of supersymmetric partition functions can provide detailed evidence for supersymmetric dualities. In this section, we compute the $\cM_{g,p} \times S^1$ partition function for dual pairs of four-dimensional gauge theories, using the Bethe-equation approach.   We focus on Seiberg dualities between $\cN=1$ gauge theories with $USp(2N_c)$ and $SU(N_c)$ gauge groups. We verify that the dual partition functions agree. This new test of Seiberg duality can also be viewed as a strong   consistency check of our general results.

\subsection{Generalities: Mapping Bethe vacua and surface operators}
Before turning to the examples, let us make a few general comments, closely following the three-dimensional discussion of  \cite{Closset:2017zgf}.  Consider some gauge theory $\CT$. We know from \eqref{ZMgp} that we may write the $\cM_{g,p} \times S^1$ partition function as:
\be
\CZ_{\Mgp \times S^1}(\nu,n) =\sum_{\h u \in \CS_{\rm BE}}  \CF_1(\h u,\nu)^{p}\, \CH(\h u,\nu)^{g-1} \, \pif_\alpha(\h u,\nu)^{\n_\alpha}~,
\ee 
where $\nu_\alpha$ and $\n_\alpha$ are the flavor fugacities and background fluxes, respectively. (The geometric parameter $\tau$ is kept implicit.)  If the theory has an infrared-dual description as another gauge theory, $\CT^D$, one may write the dual partition function similarly:
\be
\CZ_{\Mgp \times S^1}^D(\nu,n) =\sum_{\h u^D \in \CS^D_{\rm BE}}  \CF_1^D(\h u^D,\nu)^{p}\, \CH^D(\h u^D,\nu)^{g-1} \, \pif^D_\alpha(\h u^D,\nu)^{\n_\alpha}~.
\ee 
Here we assumed that the dual gauge theories have isomorphic flavor symmetries,~\footnote{This is not necessarily the case, since one or both theories could have accidental symmetries in the infrared. In that case, one should still be able to map fugacities for the flavor group common to $\CT$ and $\CT^D$ in the UV.}
 and we identified the flavor parameters appropriately.  To prove the equality of supersymmetric partition functions,   on any four-manifold $\Mgp\times S^1$ and for any set of flavor symmetry background fluxes, it suffices to exhibit a ``duality map'' between the two-dimensional Bethe vacua, such that dual operators are equal when evaluated on dual vacua. More precisely,  we must find a bijection:
\be \label{dualitymap}
 \cD\;: \;\; \cS_{BE} \rightarrow \cS_{BE}^D\;\; : \; \; \{\h u\} \mapsto \{\h u^D\}~,
\ee
such that:
\be\label{dual relation F CH}
\CF_1(\h u, \nu) =  \CF_1(\h u^D, \nu)~,  \qquad
\CH(\h u, \nu) = (-1)^{s(\Delta \CH)}\, \CH(\h u^D, \nu)~,
\ee
for the fibering and handle-gluing operators, and
\be\label{dual relation pif}
\pif_\alpha(\h u, \nu) = (-1)^{s(\pif_\alpha)}\, \pif_\alpha(\h u^D, \nu)
\ee
for the flavor flux operators. The matching of these operators in every supersymmetric vacuum immediately implies the equality
 \be\label{duality ZZD}
 \CZ_{\Mgp \times S^1}(\nu,n) =(-1)^{(g-1)s(\CH)+ \n_\alpha s(\pif_\alpha)}\; \CZ_{\Mgp \times S^1}^D(\nu,n)
 \ee
for the partition function.
Note that the duality relation \eqref{dual relation F CH} for the fibering operator implies  \eqref{dual relation pif}, due to the difference equations \eqref{ell F1}, and the fact that 't Hooft anomalies must match between dual theories. Similarly, the matching of $\CF_1$ also implies the matching of the other fibering operator, $\CF_1(\h u) = \CF^D_1(\h u^D)$, since the two fibering operators are related by \eqref{F1 to F2 af theory}.

Note the appearance of subtle signs $(-1)^{s(\CH)}$ and $(-1)^{s(\pif_\alpha)}$ in the duality relations. They are given explicitly by:
\bea\label{def sCH spif}
& s(\CH) \equiv {1\ov 2}\left( \CA^{RR}- \CA^{RR}_D\right) -  \dim(\GG) + \dim(\GG^D)  \;\qquad ({\rm mod}\; 2)~, \cr
& s(\pif_\alpha) \equiv {1\ov 2}\left( \CA^{\alpha\alpha}- \CA^{\alpha\alpha}_D\right) \;\qquad ({\rm mod}\; 2)~, 
\eea
with  $\GG$ and $\GG^D$ the gauge groups of the dual theories $\CT$ and $\CT^D$, respectively. Here $\CA^{RR}$ and $\CA^{\alpha\beta}$ are the quadratic ``pseudo-anomalies'':
\be
\CA^{RR}= \sum_I (r_I-1)^2 + \dim(\GG)~, \qquad \qquad
\CA^{\alpha\alpha} = \sum_I Q_I^\alpha Q_I^\alpha~,
\ee
and similarly for $\CA^{RR}_D$ and $\CA^{\alpha\alpha}_D$ in the dual theory.
One can show that $s(\CH)$ and $s(\pif_\alpha)$ are integers. The sign $(-1)^{s(\pif_\alpha)}$ follows from the equality of the dual fibering operators together with the second line in  \eqref{ell F1}, and the sign $(-1)^{s(\CH)}$ can be similarly determined by consistency. 
We find that the relative sign in \eqref{duality ZZD} is given in terms of the quadratic  ``pseudo-anomalies'', but the physical meaning of this observation is unclear.~\footnote{Note that the sign $(-1)^{s(\CH)}$ disappear if we match the partition functions in the physical gauge, for instance on $S^3\times S^1$, because of the relative sign in \eqref{G to H sign}.  
}
 In particular, we should note that trivial modifications of the UV completion of a theory, such as the inclusion of very massive fields charged under the flavor symmetry, can modify the sign of the $A$-model partition function. We leave a more thorough understanding of this point for future work.

To conclude with our general remarks, we would like to emphasize that the duality map \eqref{dualitymap} allows us to map any pairs of operators in the dual four-dimensional A-models---in principle. That is,  we may consider the expectation values of any half-BPS surface operators $\cS$ wrapping the torus fiber, which are computed by insertions the corresponding $A$-model operator $\cS(u,\nu)$ in the sum over Bethe vacua. We have:
\be
 \langle \cS \rangle_{\Mgp\times S^1} =\sum_{\h u \in \CS_{\rm BE}}  \cS(\h u,\nu)\,   \CF_1(\h u,\nu)^{p}\, \CH(\h u,\nu)^{g-1} \, \pif_\alpha(\h u,\nu)^{\n_\alpha}~,
 \ee
 for the unnormalized expectation value---or correlation functions, since these correlators are independent of the insertion points---, and similarly for surface operators in the dual theory.  Thus, if we can find a pair of dual surface operators, which must satisfy:
\be
\cS(\h u,\nu)= \cS^D(\h u^D,\nu)
\ee
in all Bethe vacua, we may also infer the equality of their expectation values on $\Mgp\times S^1$ in the dual theories.

\subsection{Example: $\cN=1$ $SU(2)$ gauge theory with $N_f$ flavors}

Before discussing some larger families of dualities, we pause here to describe the computation of the $\Mgp \times S^1$ partition function in a simple example.  This serves to make some of the abstract considerations above more concrete. 
The example we consider will be the $\cN=1$ $SU(2)$ gauge theory with $N_f$ flavors, {\it i.e.} $2N_f$ chiral multiplets in the fundamental representation of $SU(2)$.~\footnote{Recall there must be an even number of doublets to cancel the global anomaly.}  Let us introduce the parameters $u$ for the Cartan of the $SU(2)$ gauge symmetry, and $\nu_i$, $i=1,\cdots,2N_f$, for that of the $SU(2N_f)$ flavor symmetry, which satisfy $\sum_{i=1}^{2N_f} \nu_i= 0$.  The twisted superpotential of this theory reads:
\be 
\cW(u, \nu) = \sum_{i=1}^{2 N_f} \big( \cW_\Phi(u+\nu_i; \tau) + \cW_\Phi(-u+\nu_i; \tau)  \big)~.
 \ee
As usual, there is no contribution from the W-bosons.

\paragraph{$A$-model operators and Bethe vacua.}
From the twisted superpotential, we may construct the gauge flux operators:
\be \label{expiu} \Pi_u = e^{2 \pi i \partial_u \cW} = \prod_{i=1}^{2N_f} \frac{\Pi^\Phi(u+\nu_i)}{\Pi^\Phi(-u+\nu_i)} = \prod_{i=1}^{2N_f}  \frac{\theta(-u+\nu_i;\tau)}{\theta(u+\nu_i;\tau)}~,
 \ee
 the flavor flux operators:
\be \label{expij} \Pi_i = e^{2 \pi i \partial_{\nu_i} \cW} =  \frac{e^{\frac{\pi i}{\tau} (u^2 + {\nu_i}^2)}}{\theta(u+\nu_i;\tau) \theta(-u+\nu_i;\tau)}~,
 \ee
the fibering operator:
\be \label{exf2}
 \CF_1 = \prod_{i=1}^{2N_f} \cF^\Phi_1(u + \nu_i) \cF^\Phi_1(-u + \nu_i)  = e^{\frac{2\pi i}{3 \tau^2} \sum_{i=1}^{2N_f} {\nu_i}^3 } \prod_{i=1}^{2N_f} \Gamma_0(u+\nu_i;\tau) \Gamma_0(-u+\nu_i;\tau)~, 
\ee
and the handle-gluing operator:
\be \label{exh} \cH =H \prod_{i=1}^{2N_f} \Big[ e^{\frac{2\pi i}{\tau} {\nu_i}^2 }  \theta(u+\nu_i;\tau)\theta(-u+\nu_i;\tau) \Big]^{1-r_j}\;  \frac{1}{\eta(\tau)^2} \, \frac{1}{\theta(2u;\tau) \theta(-2u;\tau)}~, 
\ee
where:
\be
 H = \frac{\partial^2 \cW}{\partial u^2} = -{1\ov 2 \pi i}\sum_{i=1}^{2N_f} \bigg( \frac{\theta'(u+\nu_i;\tau)}{\theta(u+\nu_i;\tau)}+ \frac{\theta'(-u+\nu_i;\tau)}{\theta(-u+\nu_i;\tau)} \bigg)~.
\ee
More precisely, each flux operator $\Pi_i$ by itself corresponds to an anomalous symmetry.
The non-anomalous $SU(2N_f)$ flux insertions are:
\be\label{expijna}
 \pif_{SU(2N_f)} = \prod_{i=1}^{2N_f} {\pif_i}^{\n_i}, \;\;\;\;\quad \text{such that} \qquad \sum_{i=1}^{2N_f} \n_i = 0~.
\ee
Then, $ \pif_{SU(2N_f)}$ is an elliptic function of $u$.  For the handle-gluing operator, we must pick a non-anomalous R-symmetry by assigning the chiral multipets R-charges $r_j \in \Z$ satisfying the anomaly-free condition:
\be 
\CA^{uuR}=\sum_{i=1}^{2N_f} (r_i-1) + 4 = 0~.
\ee
The vacua of the theory are determined by the Bethe equation:
\be \label{exbe}
 \pif_u =\prod_{i=1}^{2N_f}  \frac{\theta(-u+\nu_i;\tau)}{\theta(u+\nu_i;\tau)}= 1~.
  \ee
As described in more detail in the next subsection, for generic $\nu_i$, this has $2N_f$ solutions in a given fundamental domain of the torus:
\be \label{exusol}
 \hat{u} \in  \{  0,-\frac{1}{2}, -\frac{\tau}{2}, \frac{1+\tau}{2}\} \cup \{ \hat{u}_l, -\hat{u}_l \} , \;\;\; l=1,\cdots,N_f-2~,
\ee
for some $\hat{u}_l$ depending on the $\nu_i$'s.  A solution corresponding to a valid vacuum must be acted freely by the Weyl symmetry, $\hat{u} \rightarrow -\hat{u}$, which excludes the first four solutions in \eqref{exusol}, and solutions are considered up to this symmetry, so the set of Bethe vacua is:

\be \label{exsbe} \cS_{BE} = \{ \hat{u}_l, \;\;\; l=1,\cdots,N_f-2 \} \ee
In particular, this theory has $N_f-2$ massive vacua when quantized on a torus with generic flavor fugacities $\nu$. That is, we find the Witten index:
\be
\CZ_{T^4} = N_f-2~.
\ee

\paragraph{Bethe sum formula for the partition function.}
The $\cM_{g,p} \times S^1$ partition function is given by:

\be\label{Z SU2 bv exp} 
\cZ_{\cM_{g,p} \times S^1}(\nu) = \sum_{l=1}^{N_f-2} \CF_1(\hat{u}_l,\nu)^p\,  \cH(\hat{u}_l,\nu)^{g-1} \, \pif_i(\hat{u}_l,\nu)^{\n_i}~,
 \ee
where the sum runs over the supersymmetric vacua in \eqref{exsbe}, and the dependence on $\tau$ is  implicit.  In practice, it may be difficult to compute \eqref{Z SU2 bv exp}  explicitly since we only know the $\hat{u}_l$ implicitly through \eqref{exbe}.  One way to proceed is by working perturbatively in $q=e^{2\pi i \tau}$.  We use the exponentiated variables, $x=e^{2 \pi i u}$ and $y_i = e^{2 \pi i \nu_i}$, and expand:
\be \label{expiuexp} 
\pif_u = \pif_u^{(0)} + q \pif_u^{(1)} + \cdots, \;\;\;\;\; \hat{x}_l = {\hat{x}_l}^{(0)} + q  {\hat{x}_l}^{(1)} + \cdots
 \ee
Then the ${\hat{x}_l}^{(0)}$ are given by solutions to the polynomial equation:
\be 
 \pif_u^{(0)} = \prod_{i=1}^{2N_f} \frac{x-y_i}{1-x y_i}  = 1 \qquad \Leftrightarrow \qquad \prod_{i=1}^{2N_f}(x-y_i)-  \prod_{i=1}^{2N_f}(1-x y_i)=0~.
\ee
One can then correct these solutions order by order in $q$.  For example, one finds that the leading correction is given by:~\footnote{Here and below, we consider the basic Weyl-invariant function of the solutions, $\hat{z}_l = \hat{x}_l + {\hat{x}_l}^{-1}$, as any gauge-invariant observables can be expressed as a function of these.}
\be \label{exloc} \hat{z}_l \equiv \hat{x}_l + {\hat{x}_l}^{-1} =  \hat{z}_l^{(0)} +  q \; \frac{\big({\mbox{$\hat{z}_l$}^{(0)}}^2-4\big)\displaystyle \sum_{i=1}^{2N_f} (y_i - {y_i}^{-1})}{\displaystyle \sum_{i=1}^{2N_f} \frac{y_i-{y_i}^{-1}}{ y_i+{y_i}^{-1}- {\mbox{$\hat{z}_l$}^{(0)}}} } + O(q^2)~.
 \ee
One can systematically compute higher order corrections in a similar way; however, this quickly becomes quite cumbersome to do analytically.

\paragraph{The $N_f=3$ case.} For concreteness, let us consider in more detail the case $N_f=3$.  In this case, there is a single Bethe vacuum, up to the Weyl symmetry, and one computes the leading order solution:
\be
 \hat{z}^{(0)} = \frac{\sum_{i<j} (y_i y_j - {y_i}^{-1} {y_j}^{-1})}{\sum_i (y_i - {y_i}^{-1})}~, \qquad \text{if} \quad N_f=3~.
 \ee
One can use \eqref{exloc} to compute the solution to the next order in $q$.

We can use this solution to compute the $\cM_{g,p} \times S^1$ partition function to leading order in $q$.  Let us assume, for simplicity, that the flavor symmetry fluxes vanish, $\n_i=0$.  Then, we simply need evaluate the fibering and handle-gluing operators at the Bethe solution.  For general $N_f$ (writing $\hat{z}=\hat{x} + {\hat{x}}^{-1}$ as above, and stripping off the Casimir energy factors for simplicity), we find:
\bea \label{exfhexp}
& \CF_1 = 1 + q  \hat{z} \sum_{i=1}^{2N_f} \big( y_i - {y_i}^{-1} \big) + O(q^2)~,\cr
&  \cH =H \frac{\prod_{i=1}^{2N_f} ( y_i+{y_i}^{-1} - \hat{z} )^{1-r_i}}{\hat{y}^2-4}\cr
&\qquad \qquad \times \bigg( 1 + q  \big(2\hat{z}^2-2 + \sum_{i=1}^{2N_f} (r_i-1) \hat{z} (y_i + {y_i}^{-1}) \big) \bigg) + O(q^2)~, 
\eea
with
\be
H =  \sum_{j=1}^{2N_f} \bigg( \frac{v_j-{v_j}^{-1}}{v_j+{v_j}^{-1} - \hat{y}} - q \hat{y} (v_j - {v_j}^{-1}) \bigg) + O(q^2)~.
\ee
After plugging in the form of $\hat{z}$ above, one eventually finds a relatively simple results for $N_f=3$ at first order in $q$:
\bea  \label{exos}
& \CF_1 = 1 + q \sum_{i<j} (y_i y_j - {y_i}^{-1} {y_j}^{-1}) + O(q^2)~, \cr
&  \cH = \prod_{i=1}^{6} {y_i}^{2 r_i}  \prod_{i<j} (1- y_i y_j)^{1-r_i-r_j} \bigg( 1 + q \sum_{i<j} (r_i+r_j-1)( y_i y_j + {y_i}^{-1} {y_j}^{-1}) \bigg) + O(q^2)~.
\eea
One can verify that these precisely agree with the leading order $q$-expansions of the fibering and handle-gluing operator for a theory of $15$ free chiral multiplets, which can be identified with the mesons, $Q_i Q_j$, of the $SU(2)$ $N_f=3$ theory.  This gives a simple example of the duality of \cite{Intriligator:1995ne}, with R-charges mapped appropriately across the duality. We will discuss this duality in more generality in the next subsection.  The $\cM_{g,p} \times S^1$ partition function is simply given by:
\be \label{exbseval}
 \cZ_{\cM_{g,p} \times S^1} = {\CF_1}^p \cH^{g-1}
  \ee
with $\CF_1$ and $\cH$ given, to leading order in $q$, by \eqref{exos}.

\paragraph{Integral formula} 
We may alternatively use the integral formulas in Section \ref{sec: integral form and BE}.  Let us consider the case $p \neq 0$, so that we may use the formula \eqref{zmgpuc}, which gives:
\be
  \cZ_{\cM_{g,p} \times S^1}(y) = \frac{1}{2} \sum_{\frak{m} \in {\Z_p}}\oint_{\mathbb{T} } \frac{dx}{x}  \cF(x,y)^p\,  \cH(x,y)^{g-1} \, H(x,y)\, \pif_u(x,y)^{\frak{m}}~.
\ee
We have argued in the previous section that this agrees with the known formula for the $S^3 \times S^1$ index  in the case $g=0$, $p=1$, but let us check this explicitly here.  In this case, the above becomes:
\bea\nn
& \cZ_{S^3 \times S^1} (y) \;=\; \frac{1}{2} \oint_{\mathbb{T} } \frac{dx}{x} \cH(x,y)^{-1} H(x,y) \cF_1(x,y)\cr
& = \frac{1}{2} \oint_{\mathbb{T} } \frac{dx}{x} (x-x^{-1})^2 \prod_{i=1}^{2N_f} ( y_i+{y_i}^{-1} - x-x^{-1} )^{r_j-1} \;  \Bigg[  1+ q \; \bigg[  (x+x^{-1}) \sum_{j=1}^{2N_f} \big( y_j - {y_j}^{-1} \big)  \cr
&\; \qquad\qquad\qquad\qquad - 2(x+x^{-1})^2+ 2 + \sum_{j=1}^{2N_f} (1-r_j) (x+x^{-1}) (y_j + {y_j}^{-1}) \bigg] \Bigg] + O(q^2)~.
\eea
Here, one must take care because the terms in the $q$-expansion are typically rational functions of $x$, rather than polynomials as in the superconformal $S^3 \times S^1$ index.~\footnote{Specifically, in a unitary theory the superconformal R-charges of all chiral multiplets are positive, which lifts any zero modes on $S^3$, leading to a finite number of states at each order in $q$.  In our case with integer $R$-charges, such zero modes may arise, which leads to an infinite number of states, and so such rational functions can appear in the $q$-expansion.}  To deal with these, we formally make the assumption \eqref{scassum}, which imposes that the $|y_i|$ are small, which determines which poles are enclosed by the unit circle contour.  Taking as an explicit example the anomaly-free R-charge assignments $\{r_i\}=\{0,0,0,0,1,1\}$, we find poles at $x={y_i}^{\pm}$, $i=1,...,4$, and taking those poles with the top sign, which lie inside the unit circle, one computes:
\bea
& \cZ_{S^3 \times S^1} (y) =  \frac{{y_5}^2 {y_6}^2 (1-y_5 y_6)}{\prod_{1\leq i<j \leq 4} (1- y_i y_j)} \bigg( 1 + q \bigg( -y_5 y_6 - {y_5}^{-1} {y_6}^{-1} + \sum_{1\leq i<j \leq 4} ( y_i y_j + {y_i}^{-1} {y_j}^{-1})  \cr
&\qquad\qquad\qquad + \sum_{i<j} (v_i v_j - {v_i}^{-1} {v_j}^{-1}) \bigg) \bigg)+ O(q^2)~,
\eea
which one can check agrees with \eqref{exbseval} in this case.

\subsection{$USp(2N_c)$ duality}
\label{subsec:uspduality}

Now let us consider the general duality of \cite{Intriligator:1995ne}, which relates the following four-dimensional $\cN=1$ gauge theories:~\footnote{Here, $USp(2N_c)$ is the compact symplectic group of rank $N_c$, which has dimension $N_c(2N_c+1)$. }

\begin{itemize}
\item A gauge group $USp(2N_c)$, with the vector multiplet coupled to $2N_f$ fundamental chiral multiplets $Q_i$.
\item A gauge group $USp(2N_f-2N_c-4)$, with the vector multiplet coupled to  $2N_f$ fundamental chiral multiplets $q^i$. In addition, the theory contains $N_f(2N_f-1)$ gauge-singlet chiral multiplets, denoted $M_{ij}$, $\scriptsize{1 \leq i<j \leq 2N_f}$, which interact with the charged multiplets through the superpotential $W=M_{ij} q^i q^j$.
\end{itemize}
Both of these theories have an $SU(2N_f)$ flavor symmetry. Note that the number of flavors is even in order to cancel the $USp(2N_c)$ global anomaly. The charges of the chiral multiplets are summarized in Table \ref{tab:sp}. The gauge-singlet ``mesons'' $M_{ij}$ of the second theory, in the antisymmetric representation of $SU(2N_f)$, are identified with the gauge-invariant mesons $Q_i Q_j$ in the first theory. The $R$-charges $r_i$ must satisfy:
\be\label{Rs Sp AF cond}
\sum_{i=1}^{2 N_f}  (r_i-1) + 2N_c +2=0~, 
\ee
in order to cancel the $U(1)_R$-(gauge)$^2$ anomaly.
We may choose any $r_i\in \Z$ satisfying \eqref{Rs Sp AF cond}.
Let us also introduce $\nu_i$ and $\n_i$ the $SU(2N_f)$ flavor chemical potential and fluxes, subject to the traceless condition:
\be\label{Sp traceless}
\sum_{i=1}^{2 N_f} \nu_i = 0 ~, \qquad\qquad \sum_{i=1}^{2 N_f} \n_i = 0~.
\ee

\subsubsection{'t Hooft anomaly matching and relative signs}
One of the earliest, non-trivial test of Seiberg duality was the matching of 't Hooft anomalies \cite{Seiberg:1994pq, Intriligator:1995ne}.
For $\GG_F = SU(2N_f)$, we have:
\be\label{t hooft an sp match}
\CA^{\alpha\beta\gamma} \nu_\alpha\nu_\beta\nu_\gamma = 2N_c \sum_{i=1}^{2N_f} \nu_i^3~, \qquad\quad \CA^{R\, \alpha\beta} \nu_\alpha\nu_\beta = 2N_c \sum_{i=1}^{2N_f}(r_i-1)\, \nu_i^2 ~,
\ee
in the $USp(2N_c)$ theory. It is easy to check that these anomalies are reproduced by the dual $USp(2N_f-2N_c-4)$ theory, given the relations \eqref{Rs Sp AF cond} and \eqref{Sp traceless}. We can similarly check the matching of  $\CA^R$ and $\CA^{RRR}$ across the duality. One can also check that the quadratic $SU(2N_f)$ pseudo-anomaly vanishes (mod 4) in both theories. On the other hand, we have a non-trivial sign $(-1)^{s(\CH)}$ as defined by \eqref{def sCH spif}. One finds:
\be
(-1)^{s(\CH)}= (-1)^{N_c+N_f+1}~,\qquad \qquad (-1)^{s(\pif_i)}=1~,
\ee
with $\pif_i$ the $SU(2N_f)$ flux operators.

\begin{table}
\begin{center}
\begin{tabular}{|c|c|c|c|c|}
\hline
 & $USp(2N_c)$ & $USp(2N_f-2N_c-4)$ & $SU(2N_f)$ & $U(1)_R$ \\
\hline
$Q_i$ & ${\bf 2 N_c}$ & $-$ & ${\bf 2N_f}$ & $r_i$ \\
\hline
$q^i$ & $-$ & ${\bf 2(N_f-N_c-2)}$ & $\overset{\vspace{1mm}}{\overline{\bf 2N_f}}$ & $1-r_i$ \\
$M_{ij}$ & $-$ & ${\bf 1}$ & ${\bf N_f(2N_f-1)}$ & $r_i+r_j$ \\
\hline
\end{tabular}
\end{center}
\caption{Charges for the $USp(2N_c)$ theory its $USp(2N_f-2N_c-4)$ dual.}
\label{tab:sp}
\end{table}

\subsubsection{ $USp(2N_c)$ Bethe equations and duality map}
To check the duality at the level of the A-model, we must first study the set of Bethe vacua in the two theories.  Starting with the first theory, let $u_a, \;a=1,\cdots,N_c$, denote the parameters for the Cartan of the gauge symmetry. The twisted superpotential is given by:
\be
\CW_\Phi(u, \nu) = \sum_{a=1}^{N_c} \sum_{i=1}^{2N_f}  \big( \CW_\Phi(u_a + \nu_i) + \CW_\Phi(-u_a + \nu_i)  \big)~,
\ee
with $\CW_\Phi(u)$ defined in \eqref{full Wphi}. The corresponding Bethe equations are:
\be \label{spvac}
\exp\left( 2 \pi i {\d \cW \ov \d u_a}\right) =\pif_0(u_a)=1 ~, \;\;\qquad a=1,\cdots,N_c ~,
\ee
where we defined:
\be\label{pif0 of u}
\pif_0(u) \equiv \prod_{i=1}^{2N_f} \frac{\theta(-u + \nu_i)}{\theta(u + \nu_i)}~.
\ee
Note that $\pif_0(u)$  is an elliptic function of $u$. (It is also elliptic in the parameters $\nu_i$ modulo the traceless constraint.)  Since $\pif_0(u)-1$ has $2N_f$ poles in $u$, in a given fundamental domain, it must also have  $2N_f$ distinct zeros, for generic values of the parameters. Let us denote these ``Bethe roots'' by $\tilde{u}_k$, $k=1,\cdots,2N_f$.

A Bethe vacuum is determined by an assignment of the $N_c$ eigenvalues $u_a$ to these $2N_f$ solutions.  However, recall that we exclude solutions which are fixed by any Weyl symmetry generators,  which act by either permuting the eigenvalues or exchanging $u_a$ with $-u_a$.  One can check that $ u=0, \half, {\tau\ov 2}, {\tau+1\ov 2}$,  are always solutions to $\pif_0(u)=1$.
Since these four solutions are fixed by  the Weyl symmetry, they are not allowable Bethe roots.  The remaining $2N_f-4$ Bethe roots come in pairs, $\pm \h u_k$. We then have:
\be\label{bethe roots all sp}
\Big\{\t u_k\Big\}_{k=1}^{2N_f}  \,=\,  \Big\{ 0~,  -\half~,\; -{\tau\ov 2}~,\; {1+\tau\ov 2}\Big\} \, \cup\, \Big\{\h u_l,  -\h u_l\Big\}_{l=1}^{N_f-2}~.
\ee
A Bethe vacuum is therefore a choice of $N_c$ of the $N_f-2$ Bethe roots $\h u_l$, up to the Weyl symmetry. Therefore, the number of vacua is given by:
\be \label{spcount} 
|\cS_{BE}| = \binom{N_f-2}{N_c}~.
\ee
In the dual theory, the vacuum equations are given by in terms of the same elliptic function \eqref{pif0 of u}. Denoting by $u^D_{\b a}$, $\b a = 1, \cdots, N_f-N_c-2$,  the eigenvalues for the dual gauge group, we simply find:
\be \label{spvac2} 
\pif_0(u^D_{\b a})=1 \;\;\; \qquad{\bar a}=1,\cdots,N_f-N_c-2 
 \ee
The solutions are again given in terms of  \eqref{bethe roots all sp}. By the same argument as above, one finds:
\be 
|\cS_{BE}^D| = \binom{N_f-2}{N_f-N_c-2}
 \ee
This is equal to the number of Bethe vacua \eqref{spcount} in the first theory, which provides a simple new test of the duality. Indeed, the number \eqref{spcount} should be understood as a Witten index for the gauge theory with flavors.~\footnote{As usual, the $T^4$ Witten index is not well-defined for theories with moduli spaces. However, one can regularize the theory by turning on generic fugacities for the flavor symmetry, which is what we are doing here.}

To perform more refined tests, we must construct the duality map.  Let $\cP$ be the set of $N_f-2$ pairs of non-trivial solutions, $\{\h u_l,- \h u_l\}$.  A Bethe vacuum of the first theory corresponds to a subset $A\subseteq\cP$ of size $N_c$, while a vacuum of the second theory corresponds to a subset $A^D\subseteq\cP$ of size $N_f-N_c-2$.  The natural guess for the duality map---which turns out to be correct---is:
\be\label{dual map sp}
\cD : A \rightarrow A^D = A^c 
\ee
\ie, the subset $A$ is mapped to its complement $A^c$ in  $\cP$.  Given the duality map, we can check the matching of the operators involved in constructing supersymmetric partition functions.

\subsubsection{Matching the flux and handle-gluing operators}
Let us start with the $SU(2N_f)$ flux operators. The flux operators for the first theory are given by:~\footnote{Note that this flux operators is related to the general definition \protect\eqref{flux op gen ub} by a prefactor involving the 't Hooft anomalies \protect\eqref{t hooft an sp match}. Since 't Hooft anomalies match independently across the duality, we ignore all such prefactors in the following to avoid clutter.}
\be
\pif_i(u, \nu) =\prod_{a=1}^{N_c} {1\ov \theta(u_a+ \nu_i) \theta(-u_a+ \nu_i)}~.
\ee 
For the second theory, we have:
\be
\pif_i^D(u, \nu) =\prod_{{\bar a}=1}^{N_f-N_c-2}  \theta(u^D_{\ba}- \nu_i) \theta(-u^D_\ba- \nu_i)\; \prod_{\substack{j=1\\ j\neq i}}^{2 N_f} {1\ov \theta(\nu_i + \nu_j)}
\ee
where the second term is the contribution from the gauge singlets $M_{ij}$. Duality predicts the relation:
\be\label{dual flux 1 sp}
\prod_{i=1}^{2N_f}\pif_i(\h u, \nu)^{\n_i}  = \prod_{i=1}^{2N_f}\pif_i^D(\h u^D, \nu)^{\n_i}~,
\ee
for any integers $\n_i$ such that $\sum_i\n_i=0$, and for any pair  of dual Bethe vacua $\{\h u_a\}$ and $\{\h u_\ba^D\}$, respectively. It directly follows, using the duality map \eqref{dual map sp}, that \eqref{dual flux 1 sp} can be massaged into:
\be\label{dual flux 2 sp}
\prod_{i=1}^{2N_f}\prod_{l=1}^{N_f-2} \Big[ \theta(\h u_l+ \nu_i) \theta(-\h u_l+ \nu_i)\Big]^{\n_i} = \prod_{\substack{i, j=1\\ i>j}}^{2 N_f} \theta(\nu_i+ \nu_j)^{\n_i+\n_j}~.
\ee
To prove this relation, consider the identity:
\be\label{sp id 00}
 \prod_{i=1}^{2N_f} {\theta(-u + \nu_i)\ov \theta(u + \nu_i) } -1 = \t C(\nu, \tau)\, {\prod_{k=1}^{2N_f} \theta(u-\t u_k)\ov \prod_{i=1}^{2N_f}\theta(u + \nu_i)}
\ee
for a single variable $u$ with the identifications $u\sim u+1 \sim u+\tau$. The relation holds on general grounds, for some $u$-independent function $\t C(\nu, \tau)$,  because both sides are elliptic functions of $u$ with the same poles and zeros. The zeros are at $u=\t u_k$, with  $\t u_k$  the Bethe roots defined above. Using some simple $\theta$-function identities, the identity \eqref{sp id 00} is equivalent to:
\be\label{sp id 01}
 \prod_{i=1}^{2N_f}\theta(-u + \nu_i) -  \prod_{i=1}^{2N_f}\theta(u + \nu_i)  = C(\nu, \tau) \,\theta(2 u) \, \prod_{l=1}^{N_f-2} \theta(u+ \h u_l) \theta(-u+ \h u_l)~,
\ee
with $C= q^{-{1\ov 4}}\t C$, and the $\h u_l$ as in \eqref{bethe roots all sp}. Plugging $u= \nu_i$ in \eqref{sp id 01}, it is easy to prove \eqref{dual flux 2 sp}. Note that the dependence on the unknown function $C(\nu, \tau)$ drops out  in \eqref{dual flux 2 sp} because of the traceless condition $\sum_i \n_i=0$.

By a similar computation involving the identity \eqref{sp id 01} and its first derivative, we can prove the matching of the handle-gluing operators. We find:
\be
\CH(\h u, \nu)=(-1)^{N_c+N_f+1} \, \CH^D(\h u^D, \nu)~,
\ee
for any pair of dual vacua, in agreement with expectations.
We explain this computation more thoroughly in Appendix \ref{app: sp dual}.

\subsubsection{Matching the fibering operators}
The fibering operator of the first theory reads:
\be
\CF_1(u, \nu) = \prod_{a=1}^{N_c} \prod_{i=1}^{2 N_f} \CF_1^\Phi(u_a +\nu_i) \CF_1^\Phi(-u_a +\nu_i)~,
\ee
and that of the second theory is given by:
\be
\CF^D_1(u, \nu) =\prod_{\ba=1}^{N_f-N_c-2}\, \prod_{i=1}^{2 N_f} \CF_1^\Phi(u_\ba^D -\nu_i) \CF_1^\Phi(-u_\ba^D -\nu_i) \;  \prod_{\substack{i, j=1\\ i>j}}^{2 N_f} \CF_1^\Phi(\nu_i+ \nu_j)~. 
\ee
The matching of the fibering operator, 
\be\label{match F Sp}
\CF_1(\h u, \nu) =\CF^D_1(\h u^D, \nu)~, 
\ee
in any pair of dual vacua,  is equivalent to the following simple-looking identities for the reduced elliptic gamma-function $\Gamma_0(u)$. Given the $N_f-2$ Bethe roots $\{\h u_l\}$ defined by \eqref{bethe roots all sp}, we must have: 
\be\label{match sp ell gamma fct id}
\prod_{l=1}^{N_f-2} \, \prod_{i=1}^{2 N_f} \Gamma_0(\h u_l + \nu_i) \Gamma_0(-\h u_l + \nu_i) =  \prod_{\substack{i, j=1\\ i>j}}^{2 N_f} \Gamma_0(\nu_i+ \nu_j)~.
\ee
We leave a direct analytic proof of this identity for future work.  As a consistency check, we can also verify directly that \eqref{match sp ell gamma fct id} implies \eqref{dual flux 2 sp}, using the elliptic properties of $\Gamma_0(u)$; conversely, \eqref{dual flux 2 sp} implies that the ratio of the two sides of \eqref{match sp ell gamma fct id} is at most an elliptic function of the $\nu_i$.  Some indirect evidence also follows from the identity of the dual $S^3$ supersymmetric indices \cite{Dolan:2008qi} and from our general relation between the index and the Bethe-equation formula.  However, the identity above implies the identity of partition functions on the infinite class of manifolds, $\cM_{g,p} \times S^1$, and so is a more powerful statement.

In principle, one can also check this, and other $\cM_{g,p} \times S^1$ partition function identities, perturbatively in $q$.  Specifically, as we saw in the previous subsection, we first expand the Bethe equations \eqref{pif0 of u} as a series in $q$:
\be \label{bep}
1 \;\; = \;\;  \prod_{i=1}^{2N_f} \frac{\theta(-u_a + \nu_i)}{\theta(u_a + \nu_i)} \;\; = \;\; \prod_{i=1}^{2N_f} \frac{x_a - v_i}{1- x_a v_i }  + \sum_{n=1}^\infty q^n \Pi^{(n)}(u_a,\nu_i) 
\ee
where we defined $x_a=e^{2 \pi i u_a}, v_i=e^{2 \pi i \nu_i}$.  The leading piece gives a polynomial equation for $x_a$, which is precisely the Bethe equation for the dimensionally reduced $3d$ $USp(2N_c)$ theory, as discussed in Section \ref{sec:dimred}.  This has $N_f-2$ non-trivial pairs of solutions, ${\h u}_l^{(0)}$.  Then we may correct these solutions order by order in $q$ so that they solve \eqref{bep} at each order, generating perturbative Bethe solutions:

\be \h u_l = {\h u}_l^{(0)} + q {\h u}_l^{(1)} + q^2 {\h u}_l^{(2)} + \cdots \ee
Finally, we expand \eqref{match sp ell gamma fct id} perturbatively in $q$ and substitute these solutions to check that the identity holds.  In practice this procedure can be quite cumbersome to perform analytically, as even the leading solutions, ${\h u}_l^{(0)}$, are complicated algebraic functions of the flavor parameters.  However one may also substitute some generic numerical values and check this identity numerically.  We have performed such checks and found that the identity \eqref{match sp ell gamma fct id} holds for the first several orders in $q$.  

\subsection{$SU(N_c)$ duality}\label{subsec:suduality}

As our second example, we consider the original Seiberg duality for $\CN=1$ SQCD \cite{Seiberg:1994pq}. It relates the following two theories:
\begin{itemize}
\item A gauge group $SU(N_c)$, with the vector multiplet coupled to  $N_f$ fundamental and $N_f$ anti-fundamental chiral multiplets, $Q_i$ and $\tilde{Q}_j$.
\item A gauge group $SU(N_f-N_c)$, with the vector multiplet coupled to with $N_f$ fundamental and $N_f$ anti-fundamental chiral multiplets, $q^i$ and $\tilde{q}^j$. In addition, the theory contains ${N_f}^2$ gauge-singlet chiral multiplets, $M_{ij}$, coupled through the superpotential $W= M_{ij} q^i \t q^j$. 
\end{itemize}
The flavor group is $\GG_F \cong SU(N_f) \times SU(N_f) \times U(1)_B$, with the charges shown in the Table \ref{tab:su}.  Let us denote the gauge symmetry parameters as $u_a$, $a=1,..,N_c$ and $u_{\ba}^D$, $\ba =1, \cdots, N_f-N_c$ for the first and second theory, respectively, and the  flavor symmetry parameters as $\nu_i$, $\t\nu_j$ and $\mu_B$ for $SU(N_f) \times SU(N_f) \times U(1)_B$, such that:
\be
\label{sumasscon} \sum_{a=1}^{N_c} u_a = 0~, \qquad \qquad  \sum_{\ba=1}^{N_f-N_c} u_\ba^D = 0~, \qquad \qquad  \sum_{i=1}^{N_f}\nu_i = \sum_{j=1}^{N_f}\t \nu_j=0~, 
\ee
We similarly have $\sum_i \n_i =\sum_j \t\n_j=0$ for the $SU(N_f)\times SU(N_f)$ background fluxes. Note that we normalized $U(1)_B$ in the standard way, to give charge $\pm 1$ to the ``baryons'' of either theory. 

The integer-valued $R$-charges for the chiral multiplets of the first theory are denoted by $r_i$, $\t r_j$. They must satisfy the anomaly-free condition for $U(1)_R$:
\be
\sum_{i=1}^{N_f} (r_i-1) + \sum_{j=1}^{N_f} (\t r_j-1) +2 N_c= 0~.
\ee
In the second theory, the gauge singlets $M_{ij}$ are identified with the mesons $Q_i \t Q_j$ of the first theory, which fixes the $R$-charge of $M_{ij}$. On the other hand, the $R$-charges of the dual quarks, 
\be
r_i^D = 1+ \Delta- r_i~, \qquad \t r_j^D= 1- \Delta - \t r_j~, \qquad \quad \Delta \equiv-1+ {1\ov N_f-N_c}\sum_{i=1}^{N_f} r_i~,
\ee
are fixed by the superpotential and by matching baryons across the duality \cite{Seiberg:1994pq}. We should restrict $r_i \in \Z$ to be such that $\Delta \in \Z$, so that all elementary fields have integer $R$-charges in the second theory.\footnote{Alternatively, for choices of $r_i \in \Z$ such that $\Delta \notin \Z$, it is still possible to assign the dual quarks integer charges if we mix the R-symmetry with the $U(1)^{N_f-N_c-1}$ maximal torus of the gauge symmetry.  This follows because all gauge-invariant chiral operators have integer R-charge.  For simplicity, we will restrict to the case of integer $\Delta$ below.}

\begin{table}
\begin{center}
\begin{tabular}{|c|c|c|c|c|c|c|}
\hline
 & $SU(N_c)$ &  $SU(N_f-N_c)$ & $SU(N_f)$& $SU(N_f)$ & $U(1)_B$ & $U(1)_R$ \\
\hline
$Q_i$ & ${\bf N_c}$ & $-$ & ${\bf N_f}$ & ${\bf 1}$ & $\frac{1}{N_c}$ & $r_i$ \\
$\tilde{Q}_j$ & $\overline{\bf N_c}$ & $-$ & ${\bf 1}$ & ${\bf N_f}$ & $-\frac{1}{N_c}$ & $\underset{\vspace{1mm}}{\tilde{r}_j}$ \\
\hline
$q^i$ & $-$ & $\overline{\bf N_f-N_c}$ & $\overline{\bf N_f}$ & ${\bf 1}$ & $\frac{1}{N_f-N_c}$ & $1+\Delta-r_i$ \\
$\tilde{q}^j$ & $-$ & ${\bf N_f-N_c}$ & ${\bf 1}$ & $\overline{\bf N_f}$ & $-\frac{1}{N_f-N_c}$ & $1-\Delta -\tilde{r}_j$ \\
$M_{ij}$ & $-$ & ${\bf 1}$ & ${\bf N_f}$ & ${\bf N_f}$ & $0$ & $r_i+\tilde{r}_j$ \\ 
\hline
\end{tabular}
\end{center}
\caption{Charges for the $SU(N_c)$ theory and the dual $SU(N_f-N_c)$ theory. } 
\label{tab:su}
\end{table}

\subsubsection{'t Hooft anomaly matching and relative signs}
The cubic 't Hooft anomalies for the flavor group are encoded in:
\be
\CA^{\alpha\beta\gamma} \nu_\alpha\nu_\beta\nu_\gamma = N_c \left(\sum_i \nu_i^3 + \sum_j \t\nu_j^3\right) + 3 \mu_B\left( \sum_i \nu_i^2-\sum_j \t\nu_j^2\right)~.
\ee
One can easily check that this is matched by the dual theory, and similarly for all 't Hooft anomalies involving $U(1)_R$.  For future reference, we also compute the handle-gluing operator relative sign:
\be
(-1)^{s(\CH)}= (-1)^{N_f+ N_c N_f+(N_f+1)\sum_i r_i}~.
\ee
The relative signs for the flavor flux operators are trivial.

\subsubsection{ $SU(N_c)$ Bethe equations and duality map}
The twisted superpotential of the first theory is given by:
\bea
&\CW_\Phi(u, \nu, \lambda) &=&\; \sum_{a=1}^{N_c} \sum_{i=1}^{2N_f}  \Big( \CW_\Phi\big(u_a + \nu_i+ {1\ov N_c} \mu_B\big) + \CW_\Phi\big(-u_a +\t \nu_i-{1\ov N_c} \mu_B\big)  \Big)\cr
&&&\; + \lambda \sum_{a=1}^{N_c}  u_a~.
\eea
Here, following \cite{Hori:2006dk}, we have introduced a Lagrange multiplier, $\lambda$, which imposes the traceless condition
\be
 \sum_{a=1}^{N_c}  u_a = 0 
\ee
Physically, $\lambda$ can be thought of as a complexified Fayet-Iliopoulos parameter in the $U(N_c)$ theory, which is taken to be dynamical.
For simplicity of notation, it is useful to introduce the parameters:
\be
v_a \equiv u_a +{1\ov N_c}\mu_B~, \qquad\qquad v^D_\ba = u^D_\ba - {1\ov N_f-N_c}\mu_B~.
\ee
The Bethe equations of the first theory are:
\be\label{su BE SU 1}
\pif_0(v_a, \lambda)=1~, \quad a=1, \cdots, N_c~, \qquad\quad
\qquad\sum_{a=1}^{N_c} v_a = \mu_B
\ee
where we defined the following elliptic function of $v$:
\be\label{pif0 def SU}
\pif_0(v, \lambda)\equiv e^{2 \pi i \lambda} \prod_{i=1}^{N_f} {\theta(-v + \t\nu_i)\ov \theta(v + \nu_i)}~.
\ee
Note that \eqref{pif0 def SU} is not invariant under large gauge transformations of $\nu_i$ or $\t \nu_j$, reflecting the non-zero $U(1)$-$SU(N_f)^2$ anomaly for $U(1)\subset U(N_c)$.

Similarly, the Bethe equation for the $SU(N_f-N_c)$ dual theory read:
\be\label{su BE SU 2}
\pif_0(v_\ba^D, -\lambda^D)=1~, \quad \ba=1, \cdots,N_f- N_c~, \qquad
\sum_{\ba=1}^{N_f-N_c} v_\ba^D = -\mu_B
\ee
where $\lambda^D$ is again a Lagrange multiplier, which is {\it a priori} unrelated to $\lambda$. Interestingly, we see that we can write the dual Bethe equations in terms of the same elliptic function \eqref{pif0 def SU} as in the $SU(N_c)$ theory.

The counting of Bethe vacua in the $SU(N_c)$ theory is more involved than in the $USp(2N_c)$ case. 
The first equation in \eqref{su BE SU 1} has $N_f$ solutions in $v_a$ for every choice of $\lambda$, which we denote by $\t v_k$, $k=1, \cdots, N_f$.  To construct a vacuum, we must assign the eigenvalues, $v_a$, to a size-$N_c$ subset of these solutions, and subsequently vary the parameter $\lambda$ until we are able to satisfy the second condition in  \eqref{su BE SU 1}. We must then further divide by the Weyl group $S_{N_c}$.

It is difficult to count the number of such solutions directly, however, we can indirectly arrive at the answer as follows. We will use the fact that the number of vacua, denoted by ${\bf N}_{N_c,N_f}$, satisfies the recursion relation:
\be \label{recrel} {\bf N}_{N_c,N_f} = {\bf N}_{N_c,N_f-1} + {\bf N}_{N_c-1,N_f-1}, \;\;\;\;\;\;\;\;\quad N_c,N_f>1 \ee
This can be derived by adding a complex mass for one flavor, as we will derive below in Section \ref{sec:degen}, but for now let us assume it is true.  Then, we will also need:

\be \label{recreliv} {\bf N}_{1,N_f} = 1~, \;\;\;\;\quad {\bf N}_{N_c,N_f < N_c} = 0~,  \;\;\;\;\quad {\bf N}_{N_c,N_f=N_c} = 0~,\ee
The first relation follows because for $N_c=1$ there is no gauge group, and a theory of chiral multiplets always has a single vacuum.  The second relation follows because in this case the first condition in \eqref{su BE SU 1} has fewer solutions than the number of eigenvalues, $v_a$, so it is impossible to take all the $v_a$ distinct.  And the last relation follows because when $N_f=N_c$, we must take the $v_a$ to lie among all of the solutions, $\tilde{v}_k$, to the first condition in \eqref{su BE SU 1}.  However, using the ellipticity of  \eqref{pif0 def SU}, one finds that
\be
\label{susv0} \sum_{k=1}^{N_f} \t v_k =0,
\ee
and thus it is impossible to satisfy the second condition for generic $\mu_B$, and thus there are no vacua.  Let us also formally define:

\be  \label{recrelnn} {\bf N}_{\ell,N_f} = 0, \;\;\; \ell \leq 0 \ee
which one can check is consistent with \eqref{recrel}.  Then, repeatedly applying \eqref{recrel}, one eventually obtains:
\be
 {\bf N}_{N_c,N_f} = \sum_{k=0}^{N_f-2} \binom{N_f-2}{k} {\bf N}_{N_c-k,2} 
 \ee
However, from \eqref{recreliv} and \eqref{recrelnn}, one sees the only non-zero term occurs at $k=N_c-1$, with ${\bf N}_{1,2}=1$. We thus find:
\be \label{sucount} 
\CZ_{T^4}= {\bf N}_{N_c,N_f} = |\cS_{BE}| = \binom{N_f-2}{N_c-1} 
\ee
Note that this formula agrees with the $USp(2)$ result for $N_c=2$, as required by  the isomorphism $SU(2) \cong USp(2)$. The existence of a single Bethe vacuum for $N_f=N_c+1$ is also consistent with the dual description in terms of a mesons and baryons only. More generally, \eqref{sucount} is invariant under $N_c \rightarrow N_f-N_c$, as required by Seiberg duality. 

The duality map for the Bethe vacua can be constructed  implicitly, as follows.  A vacuum of the first theory corresponds to a choice of $\lambda$ together with a subset $A$ of size $N_c$ of $\{\t v_k\}_{k=1}^{N_f}$, for that particular choice of  $\lambda$. We then claim that the duality map is:
\be \label{sudm}
 \cD\;:\; \; (\lambda, A) \quad\mapsto\quad (\lambda^D, A^D) = (-\lambda, A^c)~.
 \ee
Indeed, with $\lambda^D= -\lambda$, the first equation in \eqref{su BE SU 2} has the same solutions as that in \eqref{su BE SU 1}, and we are simply taking the $N_f-N_c$ eigenvalues of the dual theory to lie in the complement $A^c$ of $A$.  To ensure that we indeed have an $SU(N_f-N_c)$ vacuum---and not only a would-be $U(N_f-N_c)$ vacuum---we must verify that, given the second condition in  \eqref{su BE SU 1}, the second condition in \eqref{su BE SU 2} holds as well. This directly follows from  \eqref{susv0}. Therefore,  \eqref{sudm} provides a map from the Bethe vacua of the first theory to that of the second, and this map is clearly invertible, so is an isomorphism.

\subsubsection{Matching the flux, handle-gluing, and fibering operators}
Let us briefly discuss the duality relations for the flux and handle-gluing operators. The argument is analogous to the $USp(2N_c)$ case, and we refer to  Appendix \ref{app:suproof} for more details. 

 The flux operators for the $SU(N_f) \times SU(N_f)$ flavor symmetry appear as:
 \be
 \pif_{\rm flux}(v,\nu, \t\nu)= \prod_{a=1}^{N_c}\left[\prod_{i=1}^{N_f}  {1\ov \theta(v_a + \nu_i)^{\n_i}}  \prod_{j=1}^{N_f}  {1\ov \theta(-v_a + \t\nu_j)^{\t\n_j}} \right]
 \ee
 in the first theory, with $\n_i$, $\t\n_j$ the $SU(N_f)\times SU(N_f)$ background fluxes, subject to $\sum_i\n_i =0$ and $\sum_j \t\n_j=0$. Similarly, for the second theory, we have:
 \bea
& \pif^D_{\rm flux}(v^D, \nu, \t\nu)&=&\; \prod_{\ba=1}^{N_f-N_c}\left[\prod_{i=1}^{N_f}   \theta(-v^D_\ba - \nu_i)^{\n_i}  \prod_{j=1}^{N_f}  \theta(-v^D_\ba- \t\nu_j)^{\t\n_j} \right]\;  \cr
 &&&\qquad \times\prod_{i=1}^{N_f}\prod_{j=1}^{N_f} \left[{1\ov \theta(\nu_i + \t\nu_j)}\right]^{\n_i + \t\n_j}~.
 \eea
By an argument similar to that in the $USp(2N_c)$ case, one can prove that
\be
\label{suflux} \pif_{\rm flux}(\h v,\nu, \t\nu)=  \pif^D_{\rm flux}(\h v^D, \nu, \t\nu)
\ee
for any pair of dual vacua, with $A= \{\h v_a\}$ and $A^c = \{\h v^D_\ba\}$, and for andy $\lambda=-\lambda_D$.  In particular, this applies to the dual Bethe vacua in \eqref{sudm}.

Let us next consider the baryonic symmetry. The $U(1)_B$ flux operators of the first theory reads:
\be 
\pif_B(v, \nu, \t\nu) = \prod_{a=1}^{N_c} \prod_{i=1}^{N_f} \left[{\theta(-v_a + \t\nu_i)\ov \theta(v_a + \nu_i)}\right]^{1\ov N_c}~,
 \ee
The $U(1)_B$ flux operator of the dual theory reads:
 \be 
\pif_B^D(v^D, \nu, \t\nu) = \prod_{\ba=1}^{N_f-N_c} \prod_{i=1}^{N_f} \left[{\theta(v_\ba^D - \t\nu_i)\ov \theta(-v_\ba^D - \nu_i)}\right]^{1\ov N_f- N_c}~.
\ee
But using the Bethe equations, \eqref{su BE SU 1} and \eqref{su BE SU 2}, we find
\be
 \pif_B(\h v, \nu, \t\nu) = e^{- 2\pi i \lambda} , \;\;\;  \pif_B^D(\h v^D, \nu, \t\nu)= e^{2\pi i \lambda_D} 
\ee
and so these agree in dual vacua, \eqref{sudm}, as expected. 

We can similarly study the handle-gluing operator. We find:
\be
\CH(\h u, \nu)=(-1)^{N_f+ N_c N_f+(N_f+1)\sum_i r_i} \, \CH^D(\h u^D, \nu)~,
\ee
as discussed in Appendix \ref{app:suproof}.  Finally, just as in the $USp(2N_c)$ case, the matching of the fibering operators follows, for all $N_c$ and Bethe vacua, from the following identity of elliptic gamma functions:
\be \prod_{k=1}^{N_f} \prod_{i=1}^{N_f} \Gamma_0(\tilde{v}_k + \nu_i) \Gamma_0(-\tilde{v}_k + \tilde{\nu}_i) = \prod_{i,j}\Gamma_0(\nu_i + \tilde{\nu}_j)
 \ee
where $\tilde{v}_k$ runs over the $N_f$ solutions to:
\be \label{subef} 
\pif_0(v, \lambda) = e^{2 \pi i \lambda} \prod_{i=1}^{N_f} {\theta(-v + \t\nu_i)\ov \theta(v + \nu_i)}~ = 1 \ee
 and $\lambda$ is arbitrary.  We conjecture this as a true identity relating elliptic gamma functions.  Although we do not have an analytic proof, we have checked it numerically and perturbatively in $q$.

\subsection{Degeneration limits}
\label{sec:degen}

Seiberg dualities at different values of $N_f$ and $N_c$ can be related by various decoupling limits. For instance, one can decrease $N_f$ by decoupling flavors with superpotential mass terms. In the dual theory, this operation maps to a Higgsing mechanism, triggered by the deformation of the dual superpotential by a linear term in the mesonic singlets \cite{Seiberg:1994pq}.

By decoupling enough flavors (but keeping $N_f>0$, so that we preserve the $R$-symmetry), we reach a theory without any supersymmetric vacuum \cite{Affleck:1983mk}---that is the case for $N_f < N_c+1$ in the $USp(2N_f)$ theory, and for $N_f<N_c$  in the $SU(N_c)$ theory \cite{Seiberg:1994bz, Intriligator:1995ne}. In the limiting case---$N_f=N_c+1$ for $USp(2N_c)$ and $N_f= N_c$ for $SU(N_c)$---, quantum effects lift the origin of the classical moduli space without breaking supersymmetry  \cite{Seiberg:1994bz}.  In both cases, the Witten index \eqref{spcount} or \eqref{sucount}  vanishes. We should insist, however, that this is the result one obtains for {\it generic} values of the flavor fugacities. 
Here, we note that the index can also (formally) diverge at special values of the flavor parameters---it is thus a sort of $\delta$-function on parameter space. This certainly deserves further investigation. Closely related results have been obtained by studying the $S^3$ index \cite{Spiridonov:2014cxa}.~\footnote{We thank Z.~Komargodski for comments about this case.}

\paragraph{Mass deformations.}  Consider the $SU(N_c)$ theory with $N_f$ flavors. In the electric theory, we can decouple a flavor, say $Q_{N_f}$ and $\tilde{Q}_{N_f}$, by adding the mass term:
\be\label{mass term elec}
W= m Q_{N_f}\tilde{Q}_{N_f}~.
\ee
This reduces the flavor group from $SU(N_f) \times SU(N_f)$ to $SU(N_f-1) \times SU(N_f-1)$. Correspondingly, the new infrared theory should have $N_f-1$ solutions to the Bethe equation instead of $N_f$. This is easy to see on the Bethe equations \eqref{su BE SU 1}-\eqref{pif0 def SU}.  The superpotential \eqref{mass term elec} imposes the constraint:~\footnote{Here we have made a redefinition of $\mu_B$ to impose the second relation.}
\be\label{degeneration m}
\nu_{N_f} + \tilde{\nu}_{N_f}=0~, \qquad \quad \sum_{k=1}^{N_f-1} \nu_k = \sum_{k=1}^{N_f-1} \tilde{\nu}_k=0~,
\ee
instead of \eqref{sumasscon}. On this special subspace for the $2N_f$ parameters $\nu_i,\tilde{\nu}_i$, the Bethe equations degenerate to the Bethe equations of the $SU(N_c)$ theory with $N_f-1$ flavors.

In the dual magnetic theory, we expect this operation to map to a Higgsing of $SU(N_f-N_c)$ to $SU(N_f-N_c-1)$, however, since the Bethe equations of the two theories are the same, naively we obtain the $SU(N_f-N_c)$ theory instead.  To understand what happens here, it is useful to introduce a small formal parameter, $\epsilon$, and replace \eqref{degeneration m} by:

\be\label{degeneration m eps}
\nu_{N_f} = \nu+\epsilon, \;\;\;  \tilde{\nu}_{N_f} = -\nu+\epsilon ~, \qquad \quad \sum_{k=1}^{N_f-1} \nu_k =- \sum_{k=1}^{N_f-1} \tilde{\nu}_k=-\epsilon~,
\ee
Then the Bethe equations, \eqref{su BE SU 1}, can be rewritten:

\be \label{BEdegen} 1 = e^{2 \pi i \lambda} \frac{\theta(-v_a -\nu +\epsilon)}{\theta(v_a+\nu+\epsilon)} \prod_{k=1}^{N_f-1} \frac{\theta(-v_a + \tilde{\nu}_k)}{\theta(v_a + \nu_k)} \ee
This has two types of solutions: when $v_a$ is not close to $-\nu$, we may ignore the first factor, and one finds the Bethe equations for the $SU(N_c)$ theory with $N_f-1$ flavors, which has $N_f-1$ solutions, as noted above.  In addition, there is a solution at $v_a=-\nu+\delta$ for some small $\delta$.  Namely, near this point we may approximate \eqref{BEdegen} as:

\be \label{BEdegen2} 1 \approx e^{2 \pi i \lambda} \frac{-\delta+\epsilon}{\delta+\epsilon} \prod_{k=1}^{N_f-1} \frac{\theta(\nu + \tilde{\nu}_k)}{\theta(-\nu + \nu_k)}~, \ee
which has a single solution in $\delta$ of order $\epsilon$ for finite $\lambda$.  Thus when we construct Bethe vacua, there are two classes of vacua:

\begin{itemize}
\item If we take all the $v_a$ to lie among solutions of the first type, then the system of equations we are solving is identical to that of an $SU(N_c)$ theory with $N_f-1$ flavors.
\item Alternatively, we may take one of $v_a$ to lie at the special solution at $-\nu+\delta$.  Then one can check that the remaining eigenvalues solve the same system of equations as an $SU(N_c-1)$ theory with $N_f-1$ flavors, with a shifted value of $\mu_B$.
\end{itemize}
Since all vacua must lie in one of these two classes, we have shown
\be \label{recrel2} {\bf N}_{N_c,N_f} = {\bf N}_{N_c,N_f-1} + {\bf N}_{N_c-1,N_f-1} \ee
as claimed in \eqref{recrel} above. 

If we simply set $\epsilon=0$, as we did above, only the first class of vacua survive, however, for small but non-zero $\epsilon$, we find a contribution from both classes.  Specifically, one finds, for the electric theory:

\be (\cF^{(N_c,N_f)},\cH^{(N_c,N_f)}) \underset{\epsilon \rightarrow 0}{\longrightarrow} \;\left\{ \begin{array}{cc} 
(\cF^{(N_c,N_f-1)},\cH^{(N_c,N_f-1)}) & \text{for vacua in first class} \vspace{2mm}\\ 
(\cF^{(N_c-1,N_f-1)},\infty)& \text{for vacua in second class} \end{array} \right. \ee
Thus, for $g=0$, the second class of vacua are suppressed, and we indeed find a contribution only from the first class of vacua.  On the other hand, for $g=1$ we find a contribution from both classes, and for higher $g$ the index diverges as we take $\epsilon \rightarrow 0$.

Note that under the duality map, \eqref{sudm}, vacua in the first class map to those in the second.  Thus, in the $g=0$ case, the vacua that survive in the dual theory must come from the second class, and so we indeed obtain the Higgsed $SU(N_f-N_c-1)$ theory, as was noted in the case of the $S^3 \times S^1$ index, {\it e.g.}, in \cite{Rastelli:2016tbz}.  For higher genus the $\epsilon \rightarrow 0$ limit is more subtle, and one does not see a clean splitting of the Higgsed and un-Higgsed vacua.

\paragraph{The index for SQCD with a deformed moduli space.}
Consider now $USp(2N_c)$ with  $N_f=N_c+1$ fundamental chiral multiplets. The low-energy theory is described in terms of the gauge-invariant mesons  $M_{ij}= Q_i Q_j$ subject to the quantum-deformed constraint $\text{Pf}(M) = \Lambda^{2N_c+2}$ \cite{Seiberg:1994bz, Intriligator:1995ne}. We see from \eqref{spcount} that $Z_{T^4}= 0$ for  generic values of the $2N_c+2$ flavor parameters $\nu_i$. Let us now consider any arbitrary splitting of the $\nu_i$'s into two sets of $N_c+1$ parameters:
\be
\{\nu_i\}_{i=1}^{2N_c+2}= \{\mu_n, {\t \mu}_n\}_{n=1}^{N_c+1}~.
\ee
One can easily check that, on the special locus:
\be\label{codimNc locus}
\mu_n+ {\t\mu}_n =0~, \qquad \forall n~,
\ee
we trivially solve the Bethe equations \eqref{spvac} for any $u$, because:
\be
\pif_0(u)\Big|_{\mu_n +\t \mu_n =0} = 1~,
\ee
 identically. Therefore, in this case, the 4d $A$-model has a {\it continuum} of vacua---a quantum Coulomb branch---on any of the codimension-$N_c$ loci defined by \eqref{codimNc locus}. In such a case, the Witten index would formally diverge, instead of being zero. When studying the $S^3$ index, the degeneration locus \eqref{codimNc locus} has been interpreted in terms of chiral symmetry breaking from $SU(2N_c+2)$, at the origin of the classical moduli space, to $USp(2N_c)$ at the points of maximal symmetry on the quantum moduli space  \cite{Spiridonov:2014cxa}. As evidenced by these few examples, degeneration limits on the Bethe equations---and on supersymmetric indices---are rather subtle. This raises interesting questions for future work.

\section*{Acknowledgements} We would like to thank  Ofer Aharony, Benjamin Assel,  Lorenzo di Pietro, Zohar Komargodski, Victor Mikhaylov, Greg Moore,  Shlomo Razamat, Nathan Seiberg, Edgar Shaghoulian, Itamar Yaakov and Piljin Yi
for interesting discussions and comments.
This research was supported in part by Perimeter Institute for
Theoretical Physics. Research at Perimeter Institute is supported by the Government of Canada through Industry Canada and by the Province of Ontario through the Ministry of Economic Development \& Innovation. 
The work of HK was made possible through the support of a grant from the John
Templeton Foundation. The opinions expressed in this publication are those of the
author and do not necessarily reflect the views of the John Templeton Foundation.  BW was supported in part by the National
Science Foundation under Grant No. NSF PHY11-25915.


\appendix
\section{Supersymmetric background on $\Mgp \times S^1$}\label{app:  SUSY backg}

Consider four-dimensional $\CN=1$ theories on curved-space supersymmetric backgrounds \cite{Festuccia:2011ws}. For theories with an $U(1)_R$ symmetry, they correspond to new-minimal supergravity backgrounds, satisfying the generalized Killing spinor equations:~\footnote{We follow the notation and geometry conventions of \cite{Dumitrescu:2012ha}, with $A_\mu^{(R)}\equiv A_\mu - {3\ov 2}V_\mu$ for the $R$-symmetry gauge field. This leads to the conventions of \cite{Closset:2017zgf} when reducing to 3d along the second 4d coordinate, $x^2=y$, like in Appendix D of \cite{Closset:2012ru}.}
\be\label{KSE 4d}
(\nabla_\mu - i A_\mu^{(R)})\zeta= -{i\ov 2} V^\nu \sigma_\mu \t \sigma_\nu\zeta~, \qquad \qquad
(\nabla_\mu + i A_\mu^{(R)})\t\zeta= {i\ov 2} V^\nu \t\sigma_\mu  \sigma_\nu\t\zeta~.
\ee
In addition to the metric $g_{\mu\nu}$, we have the $U(1)_R$ gauge field $A_\mu^{(R)}$ and the additional background field $V_\mu$, which satisfies $\nabla_\mu V^\mu=0$. A supersymmetric background:
\be
\left(\CM_4~; \, g_{\mu\nu}~,\,  A_\mu^{(R)}~, \, V_\mu\right)~,
\ee
 is a choice of Riemannian manifold  $\CM_4$ together with a $U(1)_R$ line bundle over $\CM_4$ with connection $A_\mu^{(R)}$ and an auxiliary background field $V_\mu$, for which the Killing spinor equations \eqref{KSE 4d} have at least one non-trivial solution $\zeta$ or $\t \zeta$.
Such backgrounds were classified in \cite{Dumitrescu:2012ha}. 

The presence of a single supercharge, corresponding to $\zeta$ (or $\t \zeta$), implies that $\CM_4$ is an Hermitian manifold. We are interested in manifolds that preserve two supercharges of opposite chirality, corresponding to $\zeta$ and $\t \zeta$. From these two Killing spinors, one can construct the complex Killing vector:
\be\label{def K}
K^\mu \equiv  \zeta \sigma^\mu \t \zeta~.
\ee
If we further assume that $K$ commutes with its complex conjugate, $[K, \b K]=0$, the manifold must be a torus fibration over a two-dimensional base  \cite{Dumitrescu:2012ha}. We will further restrict ourselves to the case of a principal elliptic fibration over a Riemann surface $\Sigma_g$, with the $T^2$ fiber generated by $K^\mu$.  This is the $\Mgp\times S^1$ manifold discussed in the main text.

\subsection{Cohomology of $\Mgp \times S^1$ and line bundles}
The cohomology of $\CM_4 \cong \Mgp \times S^1$  is easily computed from the Gysin sequence for $\Mgp$ and from the K{\"u}nneth formula. We have:
\bea
&H^0(\CM_4, \Z) \cong H^4(\CM_4, \Z) \cong \Z~, \qquad H^1(\CM_4, \Z)\cong H^3(\CM_4, \Z) \cong \Z^{2g+1}~, \cr
& H^2(\CM_4, \Z) \cong \Z^{4g} \oplus \Z_p~.
\eea
The most important part, for our purposes, is:
\be\label{TorH2}
{\rm Tor}(H^2(\CM_4, \Z) ) \cong \pi^*\left(H^2(\Sigma_g, \Z)  \right)ï¿½\cong \Z_p~.
\ee
A few other interesting facts about the topology of $\CM_4$ can be found in \cite{Nishioka:2014zpa}.

All the supersymmetry-preserving line bundles $L$ appearing in this paper are pull-backs of line bundles over $\Sigma_g$, and are therefore torsion, with their first Chern class valued in \eqref{TorH2}.
As a complex line bundle, any such  line bundle $L$ comes in families indexed by complex moduli valued in the first Dolbeault cohomology $H^{0,1}(\CM_4, \C)$. By supersymmetry, the single modulus that enter the $A$-model is the one denoted by $\nu_F$ in the main text. It corresponds to a flat-connection along the $T^2$ fiber, for the component $A_{\b w}$ of the corresponding gauge connection.

\subsection{The supergravity background}
Let us discuss the $\Mgp\times S^1$ supergravity background in detail. The following discussion mostly follows from the results of  \cite{Dumitrescu:2012ha}.

\subsubsection{Real and complex coordinates}
Consider $\CM_4 \cong \Mgp\times S^1$ with coordinates $(\psi, t, z, \bz)$ on $T^2 \times \Sigma_g$. Here, $\psi\sim \psi+ 2\pi$ and $t\sim t +2 \pi$ are the angular coordinates on the torus fiber, with $T^2 \cong S^1_{\t\beta}\times S^1_{\beta}$.  We use the notation:
\be
x_1= \psi~, \qquad x_2 = t~, \qquad \qquad\beta_1 = \t\beta~,  \qquad \beta_2= \beta~,
\ee
for the coordinates and radii of the two circles. In the language of section \ref{sec: 3}, we constructed $\Mgp$ in the modular frame \eqref{p1 p2 to p}, using the first fibering operator, $\CF_1$.
The $\psi$ coordinate is the coordinate along the circle fiber in $\Mgp$, and 
the ``Euclidean time'' coordinate $t$ is the coordinate along the $S^1$ in $\Mgp\times S^1$. The complex coordinates $z, \bz$ are local coordinates on $\Sigma_g$. Let us define:
\be
\tau= \tau_1 + i \tau_2~, \qquad \tau_2 =\beta \t\beta^{-1}~,
\ee
the modular parameter on $T^2$. 
We choose the simple metric:
\be\label{MgpS1 metric 1}
ds^2(\Mgp\times S^1) = \beta^2 dt^2 + \t \beta^2 \big(d\psi + \tau_1 dt + \CC(z, \bz)\big)^2 + 2 g_{z\bz}\, dz d\bz~.
\ee
Here, $g_{z\bz}$ is the Hermitian metric on $\Sigma_g$, which we normalize to:
\be
\vol(\Sigma_g) = \pi~.
\ee
 The one-form $\CC$ is the connection of a principal circle bundle over $\Sigma$, with first Chern class $p$:
\be
{1\ov 2 \pi} \int_{\Sigma_g} d \CC = p \in \Z~.
\ee
It satisfies:
\be\label{rel C to g}
\d_z \CC_\bz- \d_\bz \CC_z =  p\, 2i g_{z\bz}~.
\ee
Let us choose the complex coordinates $(w, z)$ on $\CM_4$, with:
\be\label{w def gen}
w=  \psi + \tau t + f(z, \bz)~,
\ee
and $z$ the local coordinate on $\Sigma_g$.
Here, the complex function $f(z, \bz)$ is related to the one-form $\CC$ by:
\be
\CC_z = \d_z \b f~, \qquad \qquad \CC_\bz = \d_\bz f~.
\ee 
  It follows from \eqref{rel C to g} ï¿½that ${\rm Im}(f)/p$ is the K{\"a}hler potential on $\Sigma_g$, for $p \neq 0$, with $p g_{z\bz}= \d_z \d_\bz {\rm Im}(f)$. The real part of $f$, ${\rm Re}(f)$, is a gauge choice, which should be fixed so that $\CC$ is well-defined on each patch.~\footnote{See section \ref{app: S3S1} ï¿½for an explicit example.} 

\subsubsection{Background fields}
Using the complex coordinates $(w,z)$, the metric \eqref{MgpS1 metric 2} takes the standard form:
\be\label{MgpS1 metric 2}
ds^2(\Mgp\times S^1) = \t \beta^2 \big(dw + h(z, \bz) dz\big)\big(d\bw + \b h(z, \bz) d\bz \big)+ 2 g_{z\bz} \,dz d\bz~,
\ee
for a $T^2$ fibration, with:
\be
h(z, \bz)  = -2 i \d_z {\rm Im}\left(f(z, \bz)\right)~, \qquad \qquad
\b h(z, \bz)  = 2 i \d_\bz {\rm Im}\left(f(z, \bz)\right)~.
\ee
The complex structure ${J^\mu}_\nu$  compatible with the Hermitian metric \eqref{MgpS1 metric 2} takes the form ${J^i}_j = i {\delta^i}_j$ and ${J^{\b i}}_{\b j} = -i {\delta^{\b i}}_{\b j}$
in  the complex coordinates, $(z^i)\equiv (w, z)$, $\b z^{\b i}\equiv (\bw, \bz)$, with all other components vanishing. Let us define the holomorphic complex Killing vector:
\be
 K^\mu \d_\mu = {2\ov \t\beta} \d_w = {i \ov \beta} \left(\b \tau \d_\psi - \d_t \right)~,
\ee
which we can identify with \eqref{def K}.
The remaining supergravity background fields are given by:
\bea\label{V A}
&V_\mu &=&\; -\half \nabla_\nu {J^\nu}_\mu + \kappa K_\mu~,\qquad
&A^{(R)}_\mu  &=&\; A_\mu^{c}+ \half \nabla_\nu {J^\nu}_\mu + {i\ov 4}  {J_\mu}^\nu \nabla_\rho {J^\rho}_\nu~,
\eea
where we defined:
\be\label{Ac def}
A_\mu^{c}=  {1\ov 4}{J_\mu}^\nu \d_\nu \log\left(g_{z\bz}\right) + \d_\mu s~.
\ee
This last expression is only valid in the coordinate system $(w, z)$. The function $s$  in \eqref{Ac def} encodes the $U(1)_R$ gauge freedom, as we will discuss momentarily. Note the presence of a ``$\kappa$ ambiguity'' in the background field $V_\mu$ \eqref{V A}, which could be any function such that $K^\mu \d_\mu \kappa=0$. In this work, we choose:
\be\label{choice Kappa0}
\kappa= 0~,
\ee
like in the three-dimensional case \cite{Closset:2017zgf}. This is part of our definition of the background.  We then obtain:
\be\label{V and A explcit Mgp}
V_\mu = -\half  p \t\beta(K_\mu + \b K_\mu)~, \qquad A_\mu^{(R)}= A_\mu^c + {p \beta\ov 4} \big(K_\mu + 3 \b K_\mu\big)~.
\ee

As explained in \cite{Dumitrescu:2012ha}, on a complex four-manifold, we can rewrite the Killing spinor equations \eqref{KSE 4d} suggestively as:
\be
(\nabla_\mu^c -i A_\mu^{c}) \zeta=0~, \qquad \qquad (\nabla_\mu^c+i A_\mu^{c}) \t\zeta=0~,
\ee
where $\nabla_\mu^c$ is the Chern connection of \eqref{MgpS1 metric 2}.  
When $p=0$, the manifold $\CM_4 \cong \Sigma_g \times T^2$ is K{\"a}hler, in which case $\nabla_\mu^c = \nabla_\mu$ and $A_\mu^c=A_\mu^{(R)}$, and the supergravity background is a simple uplift of the $A$-twist on $\Sigma_g$. More generally, $\Mgp\times S^1$ is a non-K{\"a}hler  Hermitian manifold and the Chern connection has torsion proportional to $p$. In all cases, this supersymetric background is the pull-back of the $A$-twist background on $\Sigma_g$  \cite{Closset:2014pda} through the fibration $\pi: \CM_4 \rightarrow \Sigma_g$.

In particular, the $R$-symmetry background gauge field $A_\mu^{(R)}$---or, equivalently, $A_\mu^c$---is a connection over a line bundle:
\be\label{L eq K}
L^{(R)} \cong \b\CK^{-\half}~.
\ee
Here, the canonical line bundle $\CK$ over $\CM_4$ is the pull-back of the canonical line bundle over $\Sigma_g$.  It follows that $L^{(R)}$  is a torsion line bundle with first Chern class $g-1 \mod p$, as discussed in the main text.

\subsubsection{Spinors and spinor bilinears}
In the canonical complex frame 
\be
e^1 = \t \beta \big(dw + h(z, \bz) dz\big)~, \qquad \qquad e^2= \sqrt{2g_{z\bz}} \, dz~,
\ee
the Killing spinors are given explicitly by:
\be\label{KS explicit}
\zeta_\alpha=e^{i s}\mat{0\cr 1}~, \qquad\qquad \t\zeta^{\dot \alpha}=e^{-i s}\mat{0\cr 1}~.
\ee
From these Killing spinors, we can reconstruct the two commuting complex structures:
\be
J_{\mu\nu}={2i \ov |\zeta|^2} \zeta^\dagger \sigma_{\mu\nu}\zeta~, \qquad \quad
\t J_{\mu\nu}={2i \ov |\t\zeta|^2}\t \zeta^\dagger\t \sigma_{\mu\nu}\t\zeta~, 
\ee
and the Killing vector \eqref{def K}. The complex structure ${J^\mu}_\nu$ is the one associated to the coordinates $(w, z)$, while ${\t J^\mu}_{\phantom{\mu}\nu}$ corresponds to a choice of holomorphic coordinates $(w, \bz)$ instead. Another useful bilinear is the anti-holomorphic two-form:
\be
P_{\mu\nu}ï¿½\equiv \zeta \sigma_{\mu\nu}\zeta~, \qquad \qquad
 P_{\bw \bz} = \t \beta e^{2 i s} \sqrt{2 g_{z\bz}}~.
\ee
By construction, $P_{\bw \bz}$ is a globally-defined, nowhere-vanishing section of $\b \CK\otimes (L^{(R)})^2$, which leads to the identification \eqref{L eq K} \cite{Dumitrescu:2012ha}.

\subsection{``$A$-twist'' and ``physical'' gauge}
The $U(1)_R$ line bundle $L^{(R)}$ has a complex modulus determined by the gauge function $s$ in \eqref{Ac def}, namely:
\be
\nu_R=- 2 i \tau_2 \, \d_{\b w} s =-(\tau \d_\psi - \d_t)s~.
\ee
More precisely, this is a flat connection for the component $A_{\b w}^c$ of \eqref{Ac def}.
For the Killing spinors \eqref{KS explicit} to be well-defined, we must have:
\be
s= m \psi + n t~, \qquad m, n \in \Z~.
\ee
It follows that:
\be
\nu_R = -m \tau + n~.
\ee
The fact that $\nu_R=0 \mod 1, \tau$ is a consequence of supersymmetry. Any other value of $\nu_R$ would break supersymmetry explicitly, since fermions and bosons would acquire different phases when parallel-transported along $T^2$. 
In most of this work, we choose $s=0$, so that $\nu_R=0$. This is what we called the ``A-twist gauge'' in section \ref{sec: free R}. In that case, the constant Killing spinors \eqref{KS explicit}  are exactly the $A$-twist Killing spinors on $\Sigma_g$ pulled-back to $\CM_4$.

When $g-1=0 \mod p$ and $L^{(R)}$ is topologically trivial, it can be useful to consider the alternative gauge choice:
 \be
 s= {g-1\ov p} \psi~.
 \ee
This gives us the  ``physical gauge'' in \eqref{physical gauge}. For $S^3 \times S^1$, the physical gauge $s=- \psi$, corresponding to $\nu_R= \tau$, is the one considered implicitly in most of the literature. This was discussed more explicitly in \cite{Closset:2014uda}. As we explain in section \ref{sec: free R} (see also Appendix \ref{app: reg}), the physical gauge allows us to consider arbitrary $R$-charges, not only integer ones, effectively considering the $R$-symmetry group to be $\R$ instead of $U(1)$.

\subsection{The  $S^3 \times S^1$ background}\label{app: S3S1}
As an explicit example, consider $S^3\times S^1$. This complex four-manifold is the primary Hopf surface $\CM_4^{q,q}$ defined as a quotient $\C^2-(0,0)/\sim$, with:
\be\label{Hopf qt def}
(z_1, z_2) \sim (q z_1, q z_2)~, \qquad \qquad q= e^{2 \pi i \tau}~.
\ee
Let us  introduce the angular variables $\varphi, \chi$ of period $2\pi$, and $\theta \in [0, \pi]$. In the real coordinates $(t, \theta, \varphi, \chi)$, we have:
\be
z_1= e^{i \tau t} \cos{\theta\ov 2} e^{i \varphi}~, \qquad 
z_2= e^{i \tau t} \sin{\theta\ov 2} e^{i \chi}~,
\ee
which spans $\C^2$ for $t\in \R$. 
The identification \eqref{Hopf qt def} is equivalent to making the $t$ coordinate periodic, $t\sim t+ 2 \pi$.
To describe the Hopf surface in terms of the complex coordinates $(w, z)$ above, we also define:
\be
\phi = \chi-\varphi~, \qquad \psi = \varphi~.
\ee
The Hopf fibration $\pi: S^3 \rightarrow S^2$ is given by:
\be\label{def z N}
\pi : (z_1, z_2) \mapsto z\equiv {z_2\ov z_1} = \tan{\theta\ov 2}e^{i \phi}~.
\ee
Here, $(\theta, \phi)$ are the standard angular coordinates on $S^2$. The two sets of holomorphic coordinates are related by $z_1 = e^{i w}$, $z_2= z\, e^{iw}$.
In particular, the equation \eqref{w def gen} for the complex coordinate $w$ reads:
\be\label{w S3}
w= \psi+ \tau t - i \log \cos{\theta\ov 2}~.
\ee
This is on the ``northern'' patch $\theta \neq \pi$ spanned by the coordinate $z$ in \eqref{def z N}.~\footnote{On the southern patch $\theta \neq 0$, we have the complex coordinates $z'= {1\ov z}$ and $w' = w -i \log z$ instead.  This gives $w'= \psi' + \tau t - i \log \sin{\theta\ov 2}$, with $\psi'=\psi + \phi $ the Hopf fiber coordinate on that patch. Note that the branch cut ambiguity from the $\log$ in \protect\eqref{w S3} is accounted for by the periodicity of $\psi$.}
For  $2 g_{z\bz}= 1/(1+|z|^2)^2$, one can check that \eqref{MgpS1 metric 2} gives:
\be
ds^2= \beta^2 dt^2 + \t\beta^2 \big(d \psi + \tau_1 dt + \half(1- \cos \theta)\, d\phi\big)^2 + {1\ov 4}\big( d\theta^2+ \sin^2\theta \, d\phi^2\big)~,
\ee
 on $S^3 \times S^1$. For $\t\beta=1$ and $\tau_1=0$, this is the round metric on $S^3$ in Hopf coordinates. (The more usual Hopf fiber coordinate, of period $4\pi$, is $\h \psi \equiv 2 \psi$.)  
 The other background fields are:
 \bea
&A^{(R)}_\mu dx^\mu &=&\;  \t\beta^2 (d\psi + \tau_1 dt) + {i\ov 2} \beta\t\beta\, dt + \half (\t\beta^2 -1) (1-\cos\theta) \, d\phi~,\cr
&V_\mu dx^\mu &=&\;  -\t\beta^2 \big(d\psi + \tau_1 dt + \half (1-\cos\theta) \, d\phi\big)~.
 \eea
Note that $A_\mu^{(R)}$ is given on the northern patch of the $S^2$ base (with $A^{(R)}_S=A^{(R)}_N + d\phi$ on the southern patch), while $V_\mu$ is well-defined globally.
 
\paragraph{The round $S^3$.} On the round three-sphere with $\tau_1=0$, $\t\beta=1$, we can choose $\kappa$ in \eqref{V A} in such a way as to preserve four supercharges \cite{Dumitrescu:2012ha}. Namely, if we choose $\kappa=1$, while at the same time fixing the physical gauge:
\be
s= - \psi~,
\ee
then the background fields \eqref{V and A explcit Mgp} simplify to: 
\be
A_\mu^{(R)} dx^\mu = - \half V_\mu dx^\mu= {i\beta \ov 2} dt~.
\ee
With our choice \eqref{choice Kappa0}, on the other hand, we preserve only two supercharges. Whatever the choice of $\kappa$, we have $A^{(R)}= {i\beta \ov 2} dt$ in the physical gauge, corresponding to an imaginary chemical potential for $U(1)_R$ \cite{Festuccia:2011ws}. In the $A$-twist gauge, on the other hand, we have a non-zero (albeit flat) component of $A_\mu^{(R)}$ along the Hopf fiber, $A^{(R)}=d\psi + {i\beta \ov 2} dt$.

\section{Definitions and useful identities for quasi-elliptic functions}\label{app: el fct identities}
 Let us consider a torus $T^2$  with period $\tau \in \mathbb{H}$ and the complex variable $u\in \C$. We define the associated ``fugacities'':
\be
q \equiv e^{2\pi i \tau}~,  \qquad\qquad\qquad x\equiv e^{2 \pi i u}~.
\ee
In this Appendix, we collect various definitions and useful identities for the elliptic and quasi-elliptic functions that appear throughout this paper. We will denote by:
\be
S\big[f(u; \tau)\big]\equiv f\left({u\ov \tau}; -{1\ov \tau}\right)~, \qquad \qquad
T\big[f(u; \tau)\big]\equiv f\left( u; \tau+1\right)~,
\ee
the action of the $SL(2,\Z)$ generators $S$ and $T$ on any function $f(u; \tau)$.

\subsection{Eta, theta and elliptic gamma functions}

\paragraph{$\eta$-function:} Let us first recall the definition of the Dedekind eta function $\eta(\tau)$, and the associated Pochhammer symbol $(q,q)_\infty$, also known as the Euler function $\phi$:
\be
\eta(\tau) \equiv q^{1\ov 24}  \prod_{k=1}^\infty (1-q^k)~, \qquad \qquad \quad (q; q)_\infty = \phi(q) \equiv  \prod_{k=1}^\infty (1-q^k)~. 
\ee
The eta function transforms naturally under the modular group:
\be
S\big[\eta(\tau)\big]= \sqrt{-i \tau} \, \eta(\tau)~, \qquad \qquad T\big[\eta(\tau)\big]=e^{\pi i \ov 12} \eta(\tau)~.
\ee

\paragraph{$\theta$-functions:}  Let us define two closely related theta functions. The ``reduced theta function'':
\be\label{def theta0}
\theta_0(u; \tau) \equiv \prod_{k=0}^\infty (1-q^k x)(1- q^{k+1} x^{-1})~,
\ee
and the theta function:
\be\label{def theta ap}
\theta(u; \tau) \equiv  e^{\pi i\left({ \tau\ov 6}- u \right)}  \theta_0(u; \tau) =q^{1\ov 12}   x^{-\half} \prod_{k=0}^\infty (1-q^k x)(1- q^{k+1} x^{-1})~.
\ee
The theta function \eqref{def theta ap} has more natural elliptic and modular properties than the reduced theta function \eqref{def theta0}, but both appear naturally throughout this work.  Importantly, $\theta(u; \tau)$ is an odd function of $u$:
\be
\theta(-u; \tau)  = - \theta(u; \tau)~.
\ee
Under shifts of $u$ along the torus, $u \sim u+1\sim u+\tau$, we have:
\be
\theta(u+n + m\tau; \tau) = (-1)^{n+m} e^{-2 \pi i m u - \pi i m^2 \tau}\, \theta(u; \tau)~,\qquad \qquad \forall n, m \in \Z~.
\ee
Under modular transformations, we have:
\be
S\big[ \theta(u; \tau)\big]= -i \, e^{ {\pi i\ov \tau} u^2}\,  \theta(u; \tau)~, \qquad \qquad    T\big[ \theta(u; \tau)\big]= e^{\pi i \ov 6}  \theta(u; \tau)~.
\ee

\paragraph{Elliptic $\Gamma$-function:} 
The elliptic gamma function $\Gamma_e(u; \tau, \sigma)$ can be defined as the following converging product:
\be\label{Ge def}
\Gamma_e(u; \tau, \sigma)  = \prod_{j, k=0}^\infty {1- x^{-1} p^{j+1} q^{k+1}\ov 1- x p^j q^k}~.
\ee
with the two periods $\tau, \sigma$, and $q=e^{2\pi i \tau}$, $p=e^{2\pi i \sigma}$---see \cite{1999math......7061F} and references therein. In this work, we will only discuss the following specialization of \eqref{Ge def}, which we denote by $\Gamma_0(u; \tau)$:
\be\label{def Gamma0}
\Gamma_0(u; \tau) \equiv \Gamma_e(u+\tau ; \tau, \tau)  =  \prod_{n=0}^\infty  \left({1- x^{-1}q^{n+1}\ov 1- x q^{n+1}}  \right)^{n+1}~.
\ee
By abuse of language, we will often refer to \eqref{def Gamma0} simply as ``the elliptic gamma function''.
The function \eqref{def Gamma0} satisfies the reflection property:
\be
\Gamma_0(-u; \tau) = {1\ov \Gamma_0(u; \tau)}~.
\ee
It also satisfies  the difference equation:
\be
\Gamma_0(u-\tau; \tau)={1\ov \theta_0(u; \tau)}\, \Gamma_0(u; \tau)~,
\ee
with $\theta_0$ defined in \eqref{def theta0}. More generally, one can show that:
\be\label{shift m gamma}
\Gamma_0(u+n+ m \tau; \tau) =(-x)^{-{m(m+1)\ov 2}} q^{-{1\ov 6}m(m^2-1)} \, \theta_0(u; \tau)^m \, \Gamma_0(u; \tau)~,
\ee
for any $n, m \in \Z$.

\paragraph{Useful identities.}  Let us list a couple of useful results. First of all, we have the relations:
\be
\theta'_0(0; \tau) = - 2 \pi i \, (q; q)_\infty^2~, \qquad \qquad \theta'(0; \tau) = - 2 \pi i \, \eta(\tau)^2~,
\ee
with $f'(u; \tau)= \d_u f(u; \tau)$.
We can also show that:
\be
\oint_{u= -\tau}   {du \ov 2 \pi i} \,\Gamma_0(u; \tau) = -{1\ov 2 \pi i}  (q; q)_\infty^{-2}~,
\ee
for the residue of $\Gamma_0(u; \tau)$ at $u=-\tau$. More generally, $\Gamma_0(u)$ has poles of order $n$ at $u = -n \tau$, for $n \in \mathbb{N}$, and zeros of order $n$ at $u= n \tau$, for $n \in \mathbb{N}$. 
Using \eqref{shift m gamma}, one can prove the useful limit:
\be\label{limit u to n Gamma0}
\left(-{1\ov 2 \pi i}\right)^n \lim_{u \rightarrow n \tau} {\Gamma_0(u; \tau)\ov (u- n \tau)^n} = (-1)^{n(n+1)\ov 2} q^{-{1\ov 6} n(n^2-1)}  (q; q)_\infty^{2 n}~, 
\ee
for any $n \in \Z$.

\subsection{Chiral-multiplet flux operator $\pif^\Phi$}\label{app: Pif elliptic}
In the main text, we also define the function:
\be\label{pifPhi app}
\pif^\Phi(u; \tau) = e^{- {\pi i\ov \tau} u^2} \;  {1\ov \theta(u; \tau)}~.
\ee
It has the ellipticity properties:
\be
\pif^\Phi(u+n+ m \tau; \tau)  =   (-1)^{n+ m} \,e^{- {\pi i\ov \tau} \left(n^2 + 2 n u\right)}\, \pif^\Phi(u; \tau)~, 
\ee
for $n, m\in \Z$. It also transforms as:
\be
S\big[\pif^\Phi(u; \tau)\big] =  i \, e^{{\pi i \ov \tau} u^2}\, \pif^\Phi(u; \tau)~,\qquad \quad
T\big[\pif^\Phi(u; \tau)\big] =   e^{\pi i \left({u^2\ov \tau(\tau+1)}- {1\ov 6}\right)} \, \pif^\Phi(u; \tau)~,
\ee
under the modular transformations. The function $\pif^\Phi$ has a simpler transformation law under the element $\t T\in SL(2, \Z)$, with $\t T= - S T S$  defined as in \eqref{def t T}:
\be
\t T\big[\pif^\Phi(u; \tau)\big] = e^{-{\pi i \ov 6}}\, \pif^\Phi(u; \tau)~.
\ee
$S$ and $\t T$ give an equivalent presentation of the modular group.

\subsection{Chiral-multiplet fibering operators $\CF^\Phi_1$ and $\CF^\Phi_2$}\label{app: F1 F2 elliptic}
In section \ref{sec: 4dAtwist}, we defined the two functions:
\be\label{def F1ap}
\CF_1^\Phi(u; \tau) \equiv \exp\left(2 \pi i \left({u^3\ov 6 \tau^2}- {u\ov 12}\right) \right) \, \Gamma_0(u; \tau)~,
\ee
and
\be\label{def F2ap}
\CF_2^\Phi(u; \tau) \equiv \exp\left(2 \pi i \left({u^3\ov 6 \tau} - {u^2\ov 4}+ {u\tau\ov 12}+{1\ov 24} \right)\right)\,   \prod_{k=0}^\infty{ f_\Phi(u+ k \tau) \ov f_\Phi(-u+ (k+1) \tau)}~,
\ee
with the function $f_\phi(u)$ defined in \eqref{def fphi}. All these functions are meromorphic on the $u$ plane.  Note the reflection formulas:
\be
\CF_{1}^\Phi(-u; \tau) =\CF_{2}^\Phi(u; \tau)^{-1}~, \qquad \quad
\CF_{2}^\Phi(-u; \tau) =\CF_{2}^\Phi(u; \tau)^{-1}~.
\ee
The elliptic properties of \eqref{def F1ap}-\eqref{def F2ap}   are given in \eqref{el pr F2}-\eqref{el pr F1}, which we reproduce here for convenience:
\bea
&\CF^\Phi_1(u+n; \tau) &=&\;\; e^{-{\pi i n\ov 6}} e^{{2\pi i \ov \tau^2} \left({nu^2\ov 2} + {n^2 u\ov 2 }+ {n^3\ov 6 }\right)}  \CF^\Phi_1(u; \tau)~. \cr
&\CF^\Phi_1(u+m\tau; \tau) &=&\; \;  e^{-{\pi i \ov2} m^2} e^{-{\pi i \ov2} m}\, \pif^\Phi(u; \tau)^{-m}\, \CF^\Phi_1(u; \tau)~,\cr
&\CF^\Phi_2(u+n; \tau) &=&\;\; e^{-{\pi i\ov 2} n^2} e^{{2 \pi i \ov \tau} \left({n^2 u\ov 2}+ {n^3\ov 6}\right)} \, \pif^\Phi(u; \tau)^{-n}\, \CF^\Phi_2(u; \tau)~,\cr
&\CF^\Phi_2(u +m \tau; \tau) &=&\;\; e^{\pi i m\ov 6} \,\CF^\Phi_2(u; \tau)~,
\eea
The modular properties of \eqref{def F1ap}-\eqref{def F2ap} are also discussed in the main text.
We have:
\bea
&S\big[\CF^\Phi_1(u; \tau)\big] = e^{{ \pi i \ov 3 \tau} u^3} \CF^\Phi_2(u; \tau)^{-1}~, \qquad \quad
&& \t T\big[\CF^\Phi_1(u; \tau)\big] =\CF^\Phi_1(u; \tau) \CF^\Phi_2(u; \tau)~,\cr
&S\big[\CF^\Phi_2(u; \tau)\big] = e^{-{ \pi i \ov 3 \tau^2} u^3} \CF^\Phi_1(u; \tau)~, \qquad \quad
&& \t T\big[\CF^\Phi_2(u; \tau)\big] =\CF^\Phi_2(u; \tau)~,
\eea
for $S$ and $\t T$.
All these relations can be proven most easily from the definition of  $\CF_1^\Phi$, $\CF_2^\Phi$ in term of a twisted superpotential, but one can also check them directly from the explicit definition \eqref{def F1ap}-\eqref{def F2ap}.

\section{Regularized superpotential and one-loop determinants}\label{app: reg}
In this Appendix, we sketch the derivation of various key expressions through $\zeta$-function regularization. 

\subsection{Twisted superpotential}\label{app: subsec W}
Consider the twisted superpotential of a single chiral multiplet. The formal expression \eqref{3d sum W} can be massaged to \eqref{full Wphi} by splitting the sum over $n\in \Z$ in the middle. The second term in \eqref{full Wphi} converges. The cubic polynomial, on the other hand, originates from the formal sum:
\bea\label{W SUSY casimir inter}
& \CW_\Phi^{(0)}&=&\;{1\ov (2 \pi i)^2} \sum_{n=1}^\infty \left[\dilog(x q^{-n})+\dilog(x^{-1} q^{n})\right]\cr 
&&=&\; \sum_{n=1}^\infty \left[-{1\ov 12} -\half (-u + n \tau)^2 - {\epsilon \ov 2} (-u + n \tau) \right]~.
\eea
Here we used a dilogarithm identity, and  $\epsilon \in 2\Z+1$ corresponds to a choice of branch. We will set $\epsilon =0$ instead. We further manipulate the second line in \eqref{W SUSY casimir inter} to:
\be
 \sum_{n=1}^\infty \left[-{1\ov 12} \right] +\sum_{k=0}^\infty \left[-\half (-u + (k+1) \tau)^2\right]~.
\ee
The first term gives ${1\ov 24}$ using $\zeta(0)= -\half$, and the second term is regularized using the Hurwitz zeta function.~\footnote{Recall that $\zeta_H(-n, a) = -{1\ov n+1} B_{n+1}(a)$ for $n\in \Z_{>0}$,  with $B_n$ the $n$-th Bernoulli polynomial.} This directly leads to \eqref{full Wphi}.

Let us further comment on the {\it ad-hoc} choice $\epsilon=0$ in \eqref{W SUSY casimir inter}. At any fixed $n$,  \eqref{W SUSY casimir inter} is part of a three-dimensional superpotential, corresponding to three-dimensional Chern-Simons term. As explained in  \cite{Closset:2017zgf}, the term linear in $u$ leads to subtle signs in the partition function (through signs in the flux operators), which are necessary to preserve supersymmetry and gauge invariance. Once we sum over $n$ to obtain a four-dimensional theory, it is reasonable to posit that such signs only lead to other signs in four dimensions. This is possible only if we set $\epsilon=0$ by hand.

\subsection{One-loop determinants}\label{App: oneloop dets}
In this section, we very briefly discuss the derivation of the $A$-model operators from the path integral on $\Mgp \times S^1$. The one-loop determinants around a general (geometric and gauge) supersymmetric background are the building blocks for the localization computation of section \ref{sec: integral form and BE}.

\subsubsection{Chiral multiplet}
Consider a chiral multiplet on $\Mgp\times S^1$, of $R$-charge $r\in \Z$ and with charge $1$ under a background $U(1)$ gauge field chemical potential and background flux $(u, \m)$.  
By a standard argument---see {\it e.g.} \cite{Pestun:2007rz, Hama:2011ea, Closset:2013sxa}, most of the field modes on $\CM_4$ are paired by supersymmetry and cancel out between bosons and fermions. In the present case, the modes that contribute non-trivially to the one-loop determinant are in one-to-one correspondence with holomorphic sections on $\Sigma_g$ \cite{Ohta:2012ev, Nishioka:2014zpa, Closset:2015rna, Closset:2017zgf}. Schematically, we have:
\be
Z^\Phi = {\det_{{\rm coker}(D_{\b z})} D_\bw \ov \det_{{\rm ker}(D_{\b z})} D_\bw}~,
\ee
where the operator:
\be
D_\bz: \CH_{r\ov 2} \rightarrow \CH_{r-2\ov 2} 
\ee
maps fields of two-dimensional twisted spin ${r\ov 2}$ to fields of 2d twisted spins ${r-2\ov 2}$. This gives:
\be\label{ZPhi Atwist}
Z^\Phi_{\Mgp\times S^1} = \prod_{n, m\in \Z} \left[{1\ov  u +m \tau + n}\right]^{p m + \m + (g-1)(r-1)}~. 
\ee
This infinite product simply corresponds to the product over the full KK tower of $A$-twisted chiral multiplets on $\Sigma_g$, which have twisted masses $u+ m \tau +n$.
More generally, the $A$-model partition function for $\Phi$ reads:
\be
Z^\Phi = \prod_{n, m\in \Z} \left[{1\ov  u +m \tau + n}\right]^{p_1 m+ p_2 n + \m + (g-1)(r-1)}~. 
\ee
In the $A$-model language, we have:
\be
Z^\Phi = (\CF^\Phi)_1^{p_1} \, (\CF_2^\Phi)^{p_2} (\pif^\Phi)^{\m + (g-1)(r-1)}~.
\ee
This gives the following formal expression for the flux operators:
\be\label{pif formal}
\pif^\Phi=  \prod_{n, m\in \Z} {1\ov  u +m \tau + n}~,
\ee
and for the fibering operators:
\be
\CF_1^\Phi=  \prod_{n, m\in \Z} \left[{1\ov  u +m \tau + n}\right]^{m}~, \qquad 
\CF_2^\Phi=  \prod_{n, m\in \Z} \left[{1\ov  u +m \tau + n}\right]^{n}~.
\ee
Formally, all these expressions are gauge- and modular-invariant (more precisely, $\CF_1$ and $\CF_2$ mix under modular transformation), but the chiral anomaly forbids a regularization that fully preserve these symmetries. By  $\zeta$-function regularization, we can derive the expression $\pif^\Phi = e^{- {\pi i\ov \tau} u^2}  \theta(u; \tau)^{-1}$ from \eqref{pif formal}, and similarly for the fibering operators. This is consistent with the derivation of these operators from the twisted superpotential.

\subsubsection{Vector multiplet}
Consider the vector multiplet one-loop determinant on this supersymmetric background. The $W$-bosons contribute like chiral multiplets of $R$-charge $2$ and gauge charges $\alpha^a$. For each abelian vector multiplet in the Cartan of $\GG$, we have:
\be\label{U1 mn expr}
Z_{U(1)}= \prod_{\substack{n, m \in \Z \\ (n,m) \neq (0,0)}} \left[{1\ov m\tau + n}\right]^{p m + (g-1)}
\ee
where we removed the zero-mode $m=n=0$. The expression  \eqref{U1 mn expr} contributes trivially to the fibering operator, since the $p$ dependence cancels out. Upon regularization, we obtain:
\be\label{ZU1 Atwist res}
Z_{U(1)}= \eta(\tau)^{2-2g}~.
\ee

\subsubsection{One-loop determinants in the physical gauge}\label{app: subsec phys g}
The one-loop determinant \eqref{ZPhi Atwist} was computed in the $A$-twist gauge \eqref{Atwist gauge}. Consider instead an arbitrary gauge for $U(1)_R$, with parameters $(\nu_R, \n_R)$.
The chiral determinant one-loop determinant is similarly given by the formal product:
\be\label{prod Phi gen nR}
Z^\Phi = \prod_{m, n} \left[{1\ov  u + m \tau+ n + \nu_R (r-1)}\right]^{pm + \m + \n_R(r-1)}~.
\ee
Note that, in general, this expression only makes sense for $R$-charges that respect the Dirac quantization $r\, \n_R \in \Z$ on $\Sigma_g$. 
For $\CM_4$ such that $g-1=0 \mod p$, we can consider  the physical gauge \eqref{physical gauge}. This gives:
\be\label{prod Phi gen nR}
Z^\Phi_{\rm phys} =(\CF_{\rm phys}^\Phi)^p\,  (\pif_{\rm phys}^\Phi)^{\m}~,
\ee
with the ``physical'' fibering and flux operators:
\bea
& \CF_{\rm phys}^\Phi= \prod_{m, n}  \left[{1\ov  u + m \tau+ n + \nu_R (r-1)}\right]^m~,  \cr
& \pif_{\rm phys}^\Phi= \prod_{m, n}  {  1\ov  u + m \tau+ n + \nu_R (r-1)}~.
\eea
We can similarly consider the one-loop determinant of the vector multiplet in the physical gauge. The contribution from the W-bosons is again the same as from chiral multiplets of $R$-charge $2$ and gauge charges $\alpha^a$. Finally, for every $U(1)$ along the Cartan, we have:
\be\label{U1 mn phys}
Z_{U(1)}^{\rm phys}= \prod_{\substack{n, m \in \Z \\ (n,m) \neq (0,-l_R)}} \left[{1\ov m\tau + n+ \nu_R}\right]^{p m}=  \left(\CF^{\rm phys} _{U(1)}\right)^p~,
\ee
similarly to \eqref{U1 mn expr}. To compute this, we consider the limit:
\be
 \CF_{U(1)}^{\rm phys} \propto \lim_{u\rightarrow \nu_R}  \CF_1^\Phi(u)^p
\ee
 of the regularized fibering operator $\CF_1^\Phi$. Since $\nu_R= l_R \tau$, with $l_R = {1-g\ov p}\in \Z$, this limit diverges (or vanishes), but we can remove the corresponding bosonic (or fermionic) zero-modes by hand. This gives:
 \be
 Z_{U(1)}^{\rm phys} = {1\ov (-2 \pi i)^{l_R}}  \lim_{u\rightarrow \nu_R} {1\ov (u-\nu_R)^{l_R} }\, \CF_1^\Phi(u)~,
 \ee
where the overall factor has been chosen for convenience.
Using the limit \eqref{limit u to n Gamma0}, we directly obtain \eqref{CF U1}, namely:
\be
\CF_{U(1)}(\nu_R; \tau) = (-1)^{l_R(l_R+1)\ov 2}  \; \eta(\tau)^{2 l_R}~.
\ee
This is equivalent to the $A$-twist gauge result \eqref{ZU1 Atwist res}, except for a sign.
This sign contributes to the relative sign in \eqref{G to H sign}, and it is therefore interpreted as the result of a $U(1)_R$ 't Hooft anomaly contribution when changing gauge.


\section{Further details on the duality checks}
\label{app: sp dual}
In this appendix we present some further details of the proofs of matching of  $\cM_{g,p} \times S^1$ partition functions for the dual theories considered in section \ref{sec:dualities}.  
  
\subsection{$Sp(2N_c)$ duality}
\label{app:spproof}

The matching of the flux operators across the $USp(2N_c)$ duality was shown in the main text, and follows from the identity \eqref{sp id 01}, namely:
\be\label{sp id 01 app}
 \prod_{i=1}^{2N_f}\theta(-u + \nu_i) -  \prod_{i=1}^{2N_f}\theta(u + \nu_i)  = C(\nu, \tau) \,\theta(2 u) \, \prod_{l=1}^{N_f-2} \theta(u+ \h u_l) \theta(-u+ \h u_l)~,
\ee
To derive the matching of handle-gluing operators, we will need another relation which can be obtained by differentiating \eqref{sp id 01 app} with respect to $u$:

\bea 
&- \prod_{j=1}^{2N_f} \theta(-u+\nu_j) \sum_{j=1}^{2N_f} \frac{\theta'(-u+ \nu_j)}{\theta(-u+ \nu_j)} - 
\prod_{j=1}^{2N_f} \theta(u+ \nu_j)\sum_{j=1}^{2N_f} \frac{\theta'(u+ \nu_j)}{\theta(u+ \nu_j)} \\
 &\qquad= C(\nu,\tau)  \theta(2u)\prod_{l=1}^{N_f-2} \theta(u \pm  {\hat{u}_l} )\bigg(2 \frac{\theta'(2u)}{\theta(2u)} + \sum_{l=1}^{N_f-2} \sum_{\pm}\frac{\theta'(u\pm {\h u}_l)}{\theta(u \pm \h u_l)} \bigg)~.
  \eea
Substituting $u=\hat{u}_l$ and using the Bethe equation, \eqref{spvac}, we find:
\be \label{sphident} h(\hat{u}_l)\prod_{j=1}^{2N_f} \theta(\hat{u}_l +  \nu_j) = - C(\nu,\tau) \eta(\tau)^2 \theta(2{\hat{u}_l})^2\prod_{j \neq i} \theta(\hat{u}_l \pm  {\hat{u}_j}) \ee
where we defined:
\be \label{sphdefA} h(u) =- \frac{1}{2\pi i}\sum_{j=1}^{2N_f}\bigg( \frac{\theta'(u + \nu_j)}{\theta(u + \nu_j)} + \frac{\theta'(-u + \nu_j)}{\theta(-u + \nu_j)} \bigg)~. \ee
The handle-gluing operator of the electric theory is given by:
\bea\label{sph}
& \cH = H \; \eta(\tau)^{-2N_c} \prod_{a=1}^{N_c} \bigg[ \prod_{j=1}^{2N_f} \big( \theta(\nu_j \pm  {u_a})^{1-r_j} \big) \theta(\pm 2 {u_a})^{-1} \bigg] \cr
 &\qquad\times\prod_{a \neq b} \theta(u_a + u_b)^{-1}\theta(u_a -  {u_b})^{-1}~.
 \eea
 Similarly, in the magnetic theory we have:
\bea \label{sphd} 
&\cH^D = H^D \; \eta(\tau)^{-2(N_f-N_c-2)} \prod_{{\bar a}=1}^{N_f-N_c-2} \bigg[ \prod_{j=1}^{2N_f} \big( \theta(-\nu_j  \pm u^D_{\bar a})^{r_j} \big)  \theta(\pm 2 u^D_{\bar a})^{-1} \bigg]\cr
&\qquad\times \prod_{{\bar a} \neq {\bar b}} \theta(u^D_{\bar a} + u^D_{\bar b})^{-1}\theta(u^D_{\bar a}- u^D_{\bar b})^{-1} \times \prod_{i<j} \theta(\nu_i +\nu_j)^{1-r_i-r_j}~.
 \eea
 The last factor in \eqref{sphd} corresponds to the gauge-singlets $M_{ij}$.
The $R$-charges are mapped under the duality as in Table \ref{tab:sp}. The Hessian determinants,
\be H =  \prod_{a=1}^{N_c} h(u_a)~, \qquad \qquad
 H^D  = (-1)^{N_f-N_c} \prod_{{\bar a}=1}^{N_f-N_c-2} h(u^D_{\bar a})~,
 \ee
are given in terms of $h(u)$ in  \eqref{sphdefA}. 
Let us pick a Bethe vacuum, assigning a dimension-$N_c$ subset of the non-trivial Bethe roots $\h u_l$ to the $u_a$'s in the original theory, and the complement $A^c$ to the $\h u_{\b a}^D$ of dual theory.
  Then, using \eqref{sp id 01 app}, \eqref{sphident}, and the Bethe equations, one can massage the ratio of \eqref{sph} and \eqref{sphd} to:
\be\label{res app HHD}
 \frac{\cH(\h u)}{\cH^D(\h u^D)} = (-1)^{N_f+N_c+1}~.
 \ee
Here, it is important to impose the anomaly cancellation condition \eqref{Rs Sp AF cond}; in particular, the dependence on the unknown function $C(\nu, \tau)$ cancels out because of it. This completes the proof the matching the handle-gluing operators. One can also easily check \eqref{res app HHD} numerically for low values of $N_f$ and $N_c$.

\subsection{$SU(N_c)$ duality}
\label{app:suproof}

The arguments here are completely analogous to the $USp(2N_c)$ case, so we will be brief.  First we have the following analogue of \eqref{sp id 01 app} for the $SU(N_c)$ case:

\be \label{sunid app}  e^{2 \pi i \lambda} \prod_{j=1}^{N_f} \theta(\tilde{\nu}_j -  v) - \prod_{j=1}^{N_f} \theta(\nu_j + v) = C \prod_{k=1}^{N_f} \theta(\t v_k  -v )~,
\ee
for some $C$ independent of $v$, and where $\t v_k$ run over the solutions to the Bethe equation, \eqref{subef}.  Substituting $v=-\nu_j$ and $v=\tilde{\nu}_i$, we straightforwardly derive \eqref{suflux}.  Note this argument holds for arbitrary $\lambda$, and does not rely on imposing the trace condition for $SU(N_c)$.

Next, consider the handle-gluing operators.  As above, this will require differentiating \eqref{sunid app}, which gives the identity:
\be  \label{sunhid} h(\tilde{v}_k)  \prod_{j=1}^{N_f} \theta(\nu_j + \tilde{v}_k) = C \eta(\tau)^2 \prod_{j \neq k} \theta(\tilde{v}_k -  {\tilde{v}_j})~,
\ee
for every Bethe root $\t v_k$. Here we defined:
\be \label{hdefsu} h(v) =- \frac{1}{2\pi i}\sum_{j=1}^{N_f} \bigg( \frac{\theta'(\tilde{\nu}_j -v )}{\theta(\tilde{\nu}_j  - v)}  + \frac{\theta'(\nu_j + v)}{\theta(\nu_j + v)} \bigg)~.
\ee
The handle-gluing operators for the two theories are given by:
\be \cH = \eta(\tau)^{-2(N_c-1)} H \; \prod_{j=1}^{N_f} \bigg[ \prod_{a=1}^{N_c} \theta(\nu_j +  v_a)^{1-r_j} \theta(\tilde{\nu}_j  - v_a)^{1-\tilde{r}_j} \bigg] \prod_{a \neq b} \theta(v_a  - v_b )^{-1}~,
\ee
and:
\bea
& \cH^D = \eta(\tau)^{-2(N_f-N_c-1)}  H^D \; \prod_{j=1}^{N_f} \bigg( \prod_{{\bar a}=1}^{N_f-N_c} \theta(- \nu_j  -  v^D_{\bar a})^{r_j-\Delta} \theta( - {\tilde{\nu}_j} + v^D_{\bar a})^{\tilde{r}_j+\Delta} \bigg) \cr
&\qquad\times \prod_{{\bar a} \neq {\bar b}} \theta(v^D_{\bar a}  - {v^D_{\bar b}})^{-1} \times 
\prod_{i,j} \theta(\nu_i +\tilde{\nu}_j)^{1-r_i-\tilde{r}_j}~,
\eea
where $r_j,\tilde{r}_j$ are the R-charges of the chirals, which we have mapped under the duality as in Table \ref{tab:su}, and $\Delta=\frac{1}{2(N_f-N_c)}\sum_j (r_j-\tilde{r}_j)$ assumed to be integer. The Hessian determinants  and $H$ and $H^D$ are given by:
\be
 H = \prod_{a=1}^{N_c} h(v_a) , \;\;\;\; H^D = (-1)^{N_f-N_c}  \prod_{{\bar a}=1}^{N_f-N_c} h(v^D_{\bar a})~.
 \ee
Evaluating these at dual Bethe vacua, and using  \eqref{sunid app}, \eqref{sunhid}, and the Bethe equations, we find:
\be \frac{\cH}{\cH^D}  = (-1)^{N_f + N_c N_f + (N_f+1) \sum_i \tilde{r}_i}
\, C^{2N_c-2N_f +\sum_j (r_j+\tilde{r}_j)} z^{(N_f-N_c)(1+\Delta) - \sum_j r_j}~.
\ee
We then use the fact that the non-anomalous R-charges must satisfy:
\be 
\sum_j r_j = (N_f-N_c)(1+\Delta), \;\;\;\;\qquad  \sum_j \tilde{r}_j = (N_f-N_c)(1-\Delta) 
\ee
to find:
\be
 \cH = (-1)^{N_f+N_f N_c +(N_f+1) \sum_i r_i}\, \cH^D~,
\ee
as claimed in the main text.

\bibliographystyle{utphys}
\bibliography{bib4dfibering}{}

\end{document}